\documentclass[final,5p,times,twocolumn,authoryear]{elsarticle}

\usepackage{amssymb}
\usepackage{amsmath}
\usepackage{amsthm}

\usepackage[switch]{lineno}
\usepackage{hyperref}
\usepackage{multirow}
\usepackage{subcaption}
\journal{Icarus}

\begin{document}

\begin{frontmatter}



\title{In-flight calibration of ESA Hera’s HyperScout-H imager} 

\author[iac,ull]{G. P. Prodan}
\author[iss,ucv]{M. Popescu}
\author[iac,ull]{J. de Le\'{o}n}
\author[mogi]{G. Kov\'{a}cs}
\author[esa]{M. K\"{u}ppers}
\author[esa2]{B. Grieger}
\author[esa3]{C. Guerbuez}
\author[iac,ull]{E. Tatsumi}
\author[iac,ull]{J. Licandro}
\author[esa4]{A. Escalante L\'opez}
\author[mogi]{V. Nagy}
\author[padova]{A. Farina}
\author[iss]{B. A. Dumitru}
\author[arcetri]{G. Poggiali}
\author[dlr]{J. B. Vincent}
\author[turku,aalto,cze]{T. Kohout}
\author[ucv]{I. Petri\c{s}or}
\author[iaps]{E. Palomba}
\author[cosine]{M. Esposito}
\author[cosine]{N. Vercruyssen}
\author[uot]{S. Sugita}
\author[padova]{M. Lazzarin}
\author[jsc]{P. Abell}
\author[oca]{P. Michel}

\affiliation[iac]{organization={Instituto de Astrofísica de Canarias}, addressline={C/Vía Láctea s/n}, city={La Laguna}, postcode={E-38205}, state={Tenerife}, country={Spain}}
\affiliation[ull]{organization={Department of Astrophysics, University of La Laguna}, addressline={C/Astrofísico Francisco Sánchez, s/n}, city={La Laguna}, postcode={E-38205}, state={Tenerife}, country={Spain}}
\affiliation[iss]{organization={Institute of Space Science – INFLPR Subsidiary}, addressline={409 Atomi\c{s}tilor Street}, city={M\u{a}gurele}, postcode={077125}, state={Ilfov}, country={Romania}}
\affiliation[ucv]{organization={University of Craiova}, addressline={Alexandru Ioan Cuza 13}, city={Craiova}, postcode={200585}, country={Romania}}
\affiliation[mogi]{organization={Department of Mechatronics, Optics and Mechanical Engineering Informatics, Faculty of Mechanical Engineering, Budapest University of Technology and Economics}, addressline={Muegyetem rkp. 3.}, city={Budapest}, postcode={H-1111}, country={Hungary}}
\affiliation[esa]{organization={European Space Agency (ESA), European Space Astronomy Centre (ESAC)}, addressline={Camino Bajo del Castillo s/n}, city={Villanueva de la Cañada}, postcode={28692}, state={Madrid}, country={Spain}}
\affiliation[esa2]{organization={Aurora Technology B.V. for the European Space Agency (ESA), European Space Astronomy Centre (ESAC)}, addressline={Camino Bajo del Castillo s/n}, city={Villanueva de la Cañada}, postcode={28692}, state={Madrid}, country={Spain}}
\affiliation[esa3]{organization={ESA/ESOC (European Space Operations Centre)}, addressline={Robert-Bosch-Str. 5}, city={Darmstadt}, postcode={64293}, country={Germany}}
\affiliation[esa4]{organization={Starion Group for the European Space Agency (ESA), European Space Astronomy Centre (ESAC)}, addressline={Camino Bajo del Castillo s/n}, city={Villanueva de la Cañada}, postcode={28692}, state={Madrid}, country={Spain}}
\affiliation[padova]{organization={Dipartimento di Fisica e Astronomia, Padova University}, addressline={Vicolo dell'Osservatorio 3}, city={Padova}, postcode={35122}, country={Italy}}
\affiliation[arcetri]{organization={INAF-Osservatorio Astrofisico di Arcetri}, addressline={Largo Enrico Fermi, 5}, city={Firenze}, postcode={50125}, country={Italy}}
\affiliation[dlr]{organization={DLR Institute of Space Research}, addressline={Rutherfordstraße 2}, city={Berlin}, postcode={12489}, country={Germany}}
\affiliation[turku]{organization={Department of Physics and Astronomy, University of Turku}, addressline={Yliopistonmäki (Vesilinnantie 5), Quantum}, city={Turku}, postcode={20014}, country={Finland}}
\affiliation[aalto]{organization={Department of Electronics and Nanoengineering, Aalto University}, addressline={Maarintie 8}, city={Espoo}, postcode={02150}, country={Finland}}
\affiliation[cze]{organization={Institute of Geology of the Czech Academy of Sciences}, addressline={Rozvojova 269}, city={Prague}, postcode={16500}, country={Czech Republic}}
\affiliation[iaps]{organization={INAF-IAPS}, addressline={Via del Fosso del Cavaliere 100}, city={Rome}, postcode={00133}, country={Italy}}
\affiliation[cosine]{organization={cosine Remote Sensing BV}, addressline={Warmonderweg 14}, city={Sassenheim}, postcode={2171}, country={Netherlands}}
\affiliation[uot]{organization={Department of Earth and Planetary Science, University of Tokyo}, addressline={7-3-1, Hongo, Bunkyo}, city={Tokyo}, postcode={113-0033}, country={Japan}}
\affiliation[jsc]{organization={NASA Johnson Space Center}, addressline={2101 NASA Parkway, Mail Code XI}, city={Houston}, postcode={TX 77058-3696}, country={USA}}
\affiliation[oca]{organization={Observatoire de la C\^{o}te d'Azur, CNRS, Laboratoire Lagrange}, addressline={96 Bd de l'Observatoire}, city={Nice}, postcode={06304}, country={France}}

\begin{abstract}
ESA's Hera space mission is on its way to the mission target, the binary asteroid (65803) Didymos. HyperScout-H, one of the instruments onboard Hera, is a hyperspectral imager operating in the visible and near-infrared regions between 0.65 and 0.95 $\mu$m. HyperScout-H will enable a detailed assessment of the composition of both objects, Didymos and its satellite Dimorphos, the characterization of space weathering effects, and the possible presence of exogenous material on their surfaces. To monitor instrument functionality, calibration exposures are acquired regularly. This article describes the in-flight calibrations carried out for HyperScout-H during the commissioning and cruise phases. Bias and dark exposures, as well as stellar field observations, were acquired several times after launch. We update the calibration data and monitor instrument performance in the space environment. In addition, images of Earth and Moon were acquired from distances of $1.5 \times 10^{6}$ to $2.0 \times 10^{6}\,\mathrm{km}$, and Mars and its satellite were imaged during the flyby. In five images, the surface of Mars fills the entire field of view, enabling cross-validation of HyperScout-H results with those reported by other Mars missions. We characterize the detector under in-flight operational conditions. The calibration data indicate that the bias pattern is stable, the dark current remains negligible for short exposures, and the detector response is highly linear. We quantify the field-of-view alignment and geometric distortion, and evaluate the point spread function based on the stellar field observations. Stellar observations and Mars swing-by data provide updated radiometric calibration constants, suggesting that in-flight conditions have slightly modified the detector’s spectral response. In-flight calibrations are essential to ensure data quality and reliability. The results obtained for HyperScout-H demonstrate that the instrument can achieve its scientific goals in observations of the Didymos-Dimorphos system.
\end{abstract}




\end{frontmatter}




\section{Introduction}
\label{sec:intro}
HyperScout-H (HS-H) is one of the scientific instruments onboard Hera~\citep{Michel2025}, a European Space Agency (ESA) spacecraft targeting the binary asteroid (65803) Didymos. The primary objective of HS-H is to obtain close-range hyperspectral images that deliver both spectrophotometric and geomorphological information~\citep{Popescu2025hyperscout}, particularly in the aftermath of NASA's Double Asteroid Redirection Test (DART) impact~\citep{2024PSJ.....5...49C}. The HyperScout family of instruments~\citep{Esposito2019}, developed by cosine Remote Sensing with ESA and Netherlands Space Office support, was originally designed for Earth observation~\citep{2024IJRS...45.2488B}. Building on the versatile HyperScout-2 platform, HS-H represents a fully miniaturized hyperspectral instrument optimized for planetary missions, offering high efficiency and reduced data volume.

ESA's Hera spacecraft was launched on 7 October 2024 from the Kennedy Space Center (Cape Canaveral, Florida) and is currently on trajectory toward its rendezvous at (65803) Didymos. The expected arrival is in October 2026~\citep{Michel2025}. Hera executed a gravity-assist flyby of Mars on 12 March 2025. This event provided an opportunity to test the HS-H, AFC, and TIRI instruments by performing observations of Mars and its moons, Phobos and Deimos. These measurements provide a valuable data set for assessing instrument performance in the space environment. In particular, new high-resolution data were obtained for the anti-Mars side of Deimos~\citep{Popescu2025EPSC,Sugita2025EPSC}. The Martian satellites are also key targets of the JAXA Martian Moons eXploration (MMX) mission~\citep{2022EP&S...74...12K}. Consequently, Hera’s in-flight calibration activities contribute not only to mission readiness but also to the broader landscape of Solar System exploration.

Over the past decades, a succession of space missions has transformed our understanding of small bodies of the solar system. The first dedicated asteroid mission, NASA’s NEAR-Shoemaker, successfully orbited and landed on (433) Eros in 2001, demonstrating the feasibility of in-situ asteroid exploration~\citep{2002aste.book..351C, PROCKTER2002491}. Since then, missions such as Hayabusa at Itokawa conducted by JAXA~\citep{2006Sci...312.1341S, 2019Sci...364..268W, TATSUMI2019153}, or Dawn at Vesta~\citep{Russell2007Dawn} by NASA, expanded the contribution to asteroid exploration and provided valuable insights into the composition and internal structure of these small rocky bodies.

The risk of potential impacts posed by near-Earth asteroids has motivated increasingly detailed physical characterization campaigns, both from Earth-based facilities and spacecraft~\citep{Mainzer2017, Baum2018AsteroidRisk}. NASA’s DART mission represented the first kinetic-impact experiment to intentionally alter the orbit of a small body, demonstrating that momentum transfer can be an effective means of deflection~\citep{2023Natur.616..457C}. The post-impact evolution of the Didymos–Dimorphos system highlighted the need to fully understand the nature of asteroids to be able to predict the outcomes of such impacts~\citep{2022PSJ.....3..158A, Richardson2024DidymosDynamicalState}. The ESA Hera mission is designed to provide the follow-up of the impact, allowing a full assessment of the outcome of the DART mission and establishing the framework for future deflection strategies~\citep{michel2022psj}.

The binary near-Earth asteroid (65803) Didymos, composed of a primary body (Didymos) and a secondary component (Dimorphos), follows an orbit with eccentricity $e$ = 0.383, with perihelion and aphelion distances of 1.013 au and 2.272 au, respectively. A combination of ground-based studies~\citep[e.g.][]{2022PSJ.....3..175P, 2024PSJ.....5...35M, 2024PSJ.....5...17S, 2024Icar..41816138P}, data from NASA’s DART camera and images acquired by the LICIACube CubeSat~\citep{2023Natur.616..443D, 2024NatAs...8..445R, 2024Natur.627..505D, 2023Natur.616..457C} offer the first characterization of its properties. The Didymos–Dimorphos system is classified as an S-type based on its visible–near-infrared (VNIR) reflectance spectra, which show the characteristic 1 $\mu$m and 2 $\mu$m silicate absorption bands typical of ordinary chondritic material~\citep{2006AdSpR..37..178D, 2022PSJ.....3..183I,2023PSJ.....4..214R,2023PSJ.....4..229P}.

Operating over the 0.65–0.95 $\mu$m range, HS-H will produce spatially resolved hyperspectral measurements that enable detailed mapping of surface composition, space-weathering effects, and the distribution of freshly exposed material from Dimorphos' interior. These data will be combined with the observations obtained by Asteroid Framing Cameras (AFC) -- \cite{Vincent2024}, and Thermal InfraRed Imager (TIRI) -- \citet{Okada2025}, to obtain a comprehensive geological and geophysical characterization of the Didymos - Dimorphos system.

The ground calibration of HS-H took place at the premises of the manufacturer, {\it cosine Remote Sensing BV}, and at the European Space Research and Technology Centre (ESTEC). It included measurement of the bias and dark level, flat-fielding, linearity tests, radiometric calibration and distortion correction. The results are described in~\citet{Popescu2025hyperscout}.

The first in-flight HS-H observations were taken approximately three days after the launch, when all the instruments acquired images of the Earth-Moon system. Then, the commissioning phase continued by collecting calibration frames, biases and darks, several frames exposing Vega with the purpose of performing initial radiometric assessments, and various star fields to characterize the camera alignment and the camera field distortion. As a regular periodic check, additional bias and dark frames are acquired to monitor the stability of the bias pattern, dark current, and hot pixels. The Mars swing-by has been an opportunity to evaluate the spectral response of the detector.

The in-flight observations have enabled us to fine-tune the calibration across all critical aspects of the instrument's performance~—~including bias subtraction, geometric distortion correction, spectral response, and optical alignment~—~under actual space environment conditions that differ significantly from laboratory settings. Whereas pre-flight characterization provided baseline parameters, factors such as launch stresses \citep{verissimostructural}, thermal cycling in space \citep{Gilmore2002SpacecraftTC}, radiation exposure, zero-gravity conditions can lead to contamination and degradation~\citep{SHI20243993, BenMoussa_2013, Ono1996} and subtly affect instrumental behavior in ways that cannot be fully replicated on the ground. 

In this paper, we present the in-flight observations acquired by HS-H and their role in refining the instrument calibration. Section 2 introduces the HS-H instrument, briefly describes the decoding of telemetry packets, and as well as the data reported by temperature sensors inside the instrument. Section 3 presents the in-flight calibration data acquired with HS-H. Section 4 provides a detailed analysis aimed at evaluating and fine-tuning the instrument calibration. The final section is devoted to additional discussion and conclusions.


\section{HyperScout-H instrument}
\label{sec:hsh}
The HS-H instrument is a miniaturised configurable imager based on a 2D sensor. It is optimized for high-resolution, VNIR multi-band imaging, with onboard filtering and flexible acquisition parameters tailored for planetary surface characterization. The main technical specifications of the HS-H instrument are summarized in Table~\ref{tab:hsh_specs}. It features a nominal focal length of 41.25~mm and an entrance pupil diameter of 10.31~mm (corresponding to an f-number of 4.0), providing a field of view of approximately 16$^\circ$ × 10$^\circ$.

The instrument operates in the VNIR spectral range of 0.65~–~0.95$~\mu$m, divided into 25 discrete bands with full width at half maximum (FWHM) values ranging from 8 to 27~nm. The detector, an AMS CMV2000 CMOS sensor, is equipped with a spectral filter mosaic directly integrated on the sensor surface, allowing simultaneous spatial and spectral imaging. The pixels are grouped as $5\times 5$ patches that are called macropixels, where each pixel corresponds to a different channel (narrow band filter). 

\begin{table}[ht]
\centering
\begin{tabular}{@{}ll@{}}
\hline
\textbf{Parameter}                & \textbf{Value} \\
\hline
Focal length                     & 41.25~mm \\
Entrance pupil diameter          & 10.31~mm (F:4.0) \\
Field of view                    & 15.97$^\circ$~$\times$~9.85$^\circ$ \\
Spectral range                   & 0.650~–~0.975 $\mu$m \\
Bandwidth (FWHM)                 & 8~–~27 nm \\
 
Detector pixels                  & 2048~$\times$~1088 \\
Pixel size                       & 5.5~$\times$~5.5~$\mu$m \\
A/D Resolution                   & 12-bit (saved as 16-bit) \\
Integration time                 & 0.1~ms to 9900~ms \\
Readout frequency                & 0.002~Hz \\
Full well capacity                 & 11,000~e$^{-}$\\
Gain                & 3.2~e$^{-}$/DN \\
Quantum efficiency               & 9~–~18~\% (hyperspectral)\\
\hline
\end{tabular}
\caption{HS-H specification data}\label{tab:hsh_specs}
\end{table}

\subsection{Telemetry packages decoding}

The application protocol for HS-H is directly derived from the Space Packet Protocol, CCSDS 133.0-B-1 standard\footnote{\url{https://ccsds.org/Pubs/133x0b2e2.pdf}} (Consultative Committee for Space Data Systems, Space Packet Protocol, Blue Book). The communication uses the space packet encapsulation as a data wrapper. The space packet protocol consists of multiple parts; the primary header, secondary header, application data and packet error control field. 

The HS-H science team has access to the observations acquired by the instrument during the flight through the EGOS (ESA Ground Operations System) Data Dissemination System (EDDS). To ensure fast (as soon as it became available) and reliable use of the data acquired by the hyperspectral imager, we developed a software tool to decode telemetry packets. These packets include both the image data and information from the various onboard sensors, commonly referred to as housekeeping (HK) parameters. In particular, the temperatures measured by the sensors within HS-H are essential for understanding the instrument’s behaviour and for applying bias and dark-frame corrections. 

The telemetry packages are received as hexadecimal characters, following the standard outlined in the HS-H Communication Interface Control Document. The images are stored as 16-bit FITS (Flexible Image Transport System) files, along with all corresponding HK parameters.

The final step consists of using the SPICE (Spacecraft, Planet, Instrument, C-matrix, Events) kernels~\citep{Acton1996NAIF} to add information in the header of each FITS image regarding the observation geometry, target, and spacecraft location and orientation. These are provided and updated regularly by the ESA SPICE team\footnote{\url{https://www.cosmos.esa.int/web/spice/hera}} \citep{ESA_HERA_Operational_SPICE_2024}. 


\subsection{Temperature sensors}
Temperature is a key parameter that influences the behavior of the Complementary Metal–Oxide–Semiconductor (CMOS) AMS CMV2000 sensor, particularly the dark current and electronic properties. Establishing the relationship between dark current values and temperature is a required step for the calibration process. Additionally, temperature monitoring is essential for ensuring the instrument's safety.

Nine temperature sensors are employed to monitor the thermal conditions at different locations within the HS-H instrument. The instrument housekeeping (HK) data include readings from eight temperature sensors: ICU1, ICU2, HSSP, BEE, T1, T2, T3, and T4. The ICU1 and ICU2 are located near the Instrument Control Unit (ICU), while T1, T2, T3, and T4 monitor the telescope section. The HSSP and BEE sensors are positioned near the electronic boards, namely the HSSP (High-Speed Serial Protocol) board and the Backend Electronics (BEE). The temperature of the Focal Plane Array (FPA) is reported separately via spacecraft telemetry. In addition, the detector temperature is included in the header of each acquired image.

The ESA/European Space Operations Centre (ESOC) and the instrument manufacturer designated the FPA sensor as the reference for monitoring the instrument temperature and for characterizing the sensor behaviour. The additional temperature sensor integrated within the CMOS detector remains uncalibrated. The difference between the temperatures measured by the CMOS detector sensor and the FPA sensor has a mean value of $7.5^\circ$~C, median of $8.0^\circ$~C, and standard deviation of $1.4^\circ$~C. Because all thermal calibrations are strictly tied to the designated FPA sensor, this hardware offset does not impact the dark current correction pipeline.

\begin{figure}[h]
    \centering
    \includegraphics[width=0.8\linewidth]{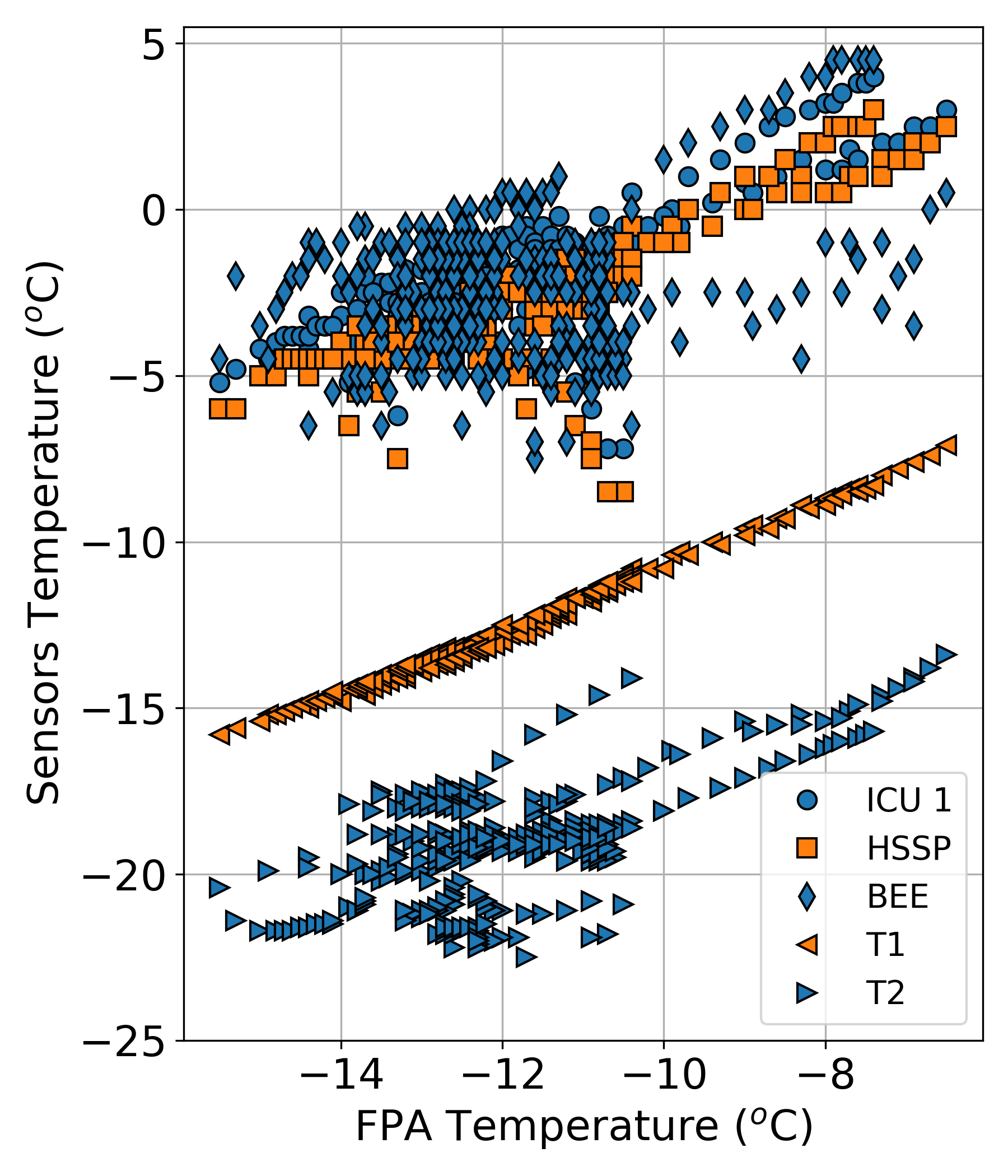}
    \caption{The dependence of different sensor temperatures with respect to FPA temperature during the cruise. The dependences of five HS-H sensors are shown with different marker shapes.}
    \label{fig:sensors}
\end{figure}

The temperatures inside the instrument vary significantly. The average temperature recorded by sensor T2 at the telescope level is $-19.1 \pm 1.8^\circ$~C, while sensor ICU2 reports values around $-1.7 \pm 1.7^\circ$~C at the instrument control unit level. Although all sensors generally follow the same trend, there is no significant correlation between the values reported by different sensors (Fig.~\ref{fig:sensors}), except for the relationship between the FPA sensor and the T1 sensor (this is understandable, as both are at the telescope level). The temperature difference between these two sensors is $0.69 \pm 0.15^\circ$~C.


\section{In-flight calibration observations}
\label{sec:cal}

The in-flight calibration images consisted in the commissioning phase, periodic checks, and opportunity observations. A summary of these activities is presented in Table~\ref{tab:activities}. The commissioning activities included imaging the Earth–Moon system as early as possible after launch, acquiring bias and dark frames, observing Vega as a standard star, and imaging several stellar fields. All planned commissioning activities were completed by November 10, during which a total of 167 images were acquired. The periodic checks, planned to be performed approximately every six months, include the acquisition of bias and dark frames, observations of Vega as a standard star, and imaging of several stellar fields. Mars swing by offered us the opportunity to acquire additional observations.

\begin{table*}
\centering
\begin{tabular}{@{\extracolsep{\fill}}l p{4.2cm} p{6.3cm} c c@{}}
\hline
\textbf{Timeline} &
\textbf{Objective} &
\textbf{Observing constraints} &
\textbf{\#} &
\textbf{Exposure~(s)} \\
\hline \\

\multicolumn{5}{@{}l@{}}{\textbf{Earth \& Moon}} \\ \hline

Oct 2024 &
\textbf{Earth \& Moon} &
Earth and Moon simultaneously in the FoV &
41 &
0.005~-~0.050 \\
\\

\multicolumn{5}{@{}l@{}}{\textbf{Standard calibration frames}} \\ \hline

Oct 2024 &
\textbf{Dark} &
No bright objects &
5 &
0.10~-~0.50 \\

Nov 2024 &
\textbf{Bias} &
Lowest exposure allowed; no bright objects &
8 &
0.0001 \\

Dec 2024 &
\textbf{Dark} &
No bright objects &
36 &
0.01~-~9.00 \\

Jan 2025 &
\textbf{Bias} &
Lowest exposure allowed; no bright objects &
11 &
0.0001 \\

Sep 2025 &
\textbf{Bias} &
Lowest exposure allowed; no bright objects &
11 &
0.0001 \\ \\

\multicolumn{5}{@{}l@{}}{\textbf{Stellar calibration}} \\ \hline

%
Oct 2024 &
\begin{tabular}[c]{@{}l@{}}\textbf{Standard star (Vega):}\\ \hspace{10pt}saturation test\end{tabular} &
Vega in the FoV &
5 &
0.10~-~2.00 \\

Oct 2024 &
\textbf{Alignment} &
Different star field pointings &
22 &
3.00~-~5.00 \\

Nov 2024 &
\begin{tabular}[c]{@{}l@{}}\textbf{Star fields:}\\ \hspace{10pt}geometric and \\ \hspace{10pt}radiometric calibration\end{tabular} &
High density of 3~-~5 mag stars; 3×3 raster &
40 &
0.50~-~5.00 \\

Nov 2024 &
\begin{tabular}[c]{@{}l@{}}\textbf{Standard star (Vega):}\\ \hspace{10pt}radiometric calibration\end{tabular} &
Vega in FoV; 3×3 raster &
27 &
0.50 \\

Feb 2025 &
\begin{tabular}[c]{@{}l@{}}\textbf{Standard star (Vega):}\\ \hspace{10pt}radiometric calibration\end{tabular} &
Vega in FoV; 3×3 raster  &
27 &
2.00 \\

Feb 2025 &
\begin{tabular}[c]{@{}l@{}}\textbf{Standard star (16 Cyg):}\\ \hspace{10pt}radiometric calibration\end{tabular} &
16 Cyg in the FoV; 3×3 raster &
27 &
9.00 \\

Sep 2025 &
\begin{tabular}[c]{@{}l@{}}\textbf{Star fields:}\\ \hspace{10pt}geometric and \\ \hspace{10pt}radiometric calibration\end{tabular}&
High density of 3~-~5 mag stars; 3×3 raster &
9 &
5.00~-~9.00 \\ \\

\multicolumn{5}{@{}l@{}}{\textbf{Mars swing-by}} \\ \hline

Feb 2025 &
\textbf{Mars rehearsal} &
Dark-like exposures taken to rehearse the Mars swing &
37 &
0.010~-~0.025 \\

Mar 2025 &
\begin{tabular}[c]{@{}l@{}}\textbf{Mars (pre-Deimos):}\\ \hspace{10pt}radiometric calibration\end{tabular} &
Full-frame Mars near image center &
22 &
0.010 \\

Mar 2025 &
\textbf{Deimos} &
One high-resolution image &
3 &
0.025 \\

Mar 2025 &
\begin{tabular}[c]{@{}l@{}}\textbf{Mars (post-Deimos):}\\ \hspace{10pt}radiometric calibration\end{tabular} &
Mars covers nearly the entire FoV &
5 &
0.010 \\

Mar 2025 &
\textbf{Phobos} &
Phobos detected in FoV &
7 &
0.025 \\

\hline
\end{tabular}
\caption{Summary of the in-flight calibration and science observations, detailing the objectives, observational constraints, exposure configurations, and number of images acquired for each activity. Most of the star fields were centered around $\alpha$ Tau (Aldebaran) or $\delta$ CMa (Wezen).}
\label{tab:activities}
\end{table*}

Due to the rapidly increasing distance between the spacecraft and Earth, priority was given to imaging the Earth-Moon system. These images were first captured on October 10, 2024, at 14:12 UT, just three days after launch, following the instrument's activation. After observing the Earth-Moon system, the next step was to acquire bias frames, followed by observations of Vega. The in-flight observing campaign was continued by observing several star fields and acquiring bias frames. In March 2025, the Mars swing-by provided an opportunity to observe an object that fully covered the instrument’s field of view. All of these observations are described in this section.

\subsection{Earth - Moon observations}
The planned activities began with imaging the Earth-Moon system. The strategy involved capturing exposures by dithering the camera pointing in a $3 \times 3$ raster pattern, with a $1^\circ$ offset
between consecutive pointings. Thus, there were 9 different pointings. This approach helped mitigate potential issues related to malfunctioning pixels or unexpected detector behavior. Additionally, it ensured that all wavelength channels were spatially covered, allowing the same surface area sampled by a pixel to be imaged at all wavelengths.

The exposure times were calculated based on the range of light intensity, from the brightest Earth clouds to the darkest regions of the Moon. Given that the dynamic range of HS-H is limited by its 12-bit ADC and the varying sensitivity across different channels, images were planned to be captured with exposure times between 0.5 and 50~ms.

\begin{figure}[ht]
    \centering
    \includegraphics[width=\columnwidth]{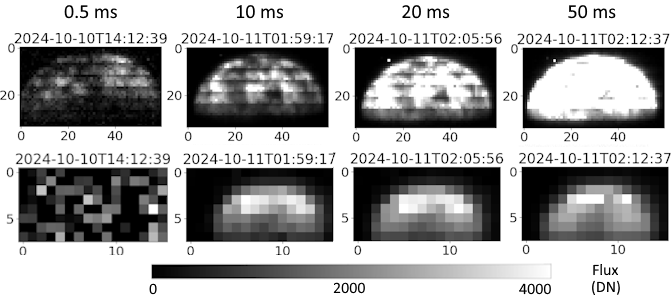}
    \caption{Earth (top row) and Moon (bottom row) observations with HS-H. Saturation is observed for the images at 20 and 50 ms. At 0.5 ms the S/N is very low, making the Moon indistinguishable. }
    \label{fig:earth_moon}
\end{figure}

These preliminary acquisitions showed that HS-H instrument operates in space in accordance with its specifications and enabled the first in-flight linearity test to be performed. The shortest exposures ($\leq$10~ms) are used to characterize the Earth, while the longest exposures result in most of Earth's pixels being saturated as presented in Fig. \ref{fig:earth_moon}. The best signal-to-noise (S/N) ratio for the Moon is achieved with exposure times of 20~ms and 50~ms. Various false color images were generated for public outreach \footnote{\url{https://www.esa.int/ESA_Multimedia/Images/2024/10/Spooky_Earths_seen_by_Hera_s_HyperScout}}. However, the illumination geometry of the Earth--Moon system, observed at a phase angle of approximately $\sim 90^\circ$, together with the low spatial resolution, make these images unsuitable for straightforward radiometric calibration.

Earth and Moon observations were utilized for the initial in-flight functional verification of the instrument and to perform the first linearity test.

\subsection{Bias and dark frames}

The bias frames were acquired with the lowest exposure time allowed by the camera, with the additional constraint that no bright object was present in the field of view (FoV). A total of 8 bias frames were taken in November 2024, 11 frames in January 2025, and 11 more frames in September 2025, summing up to 30 bias exposures. These were obtained at an exposure time of 0.1~ms.

The dark frames were taken under two sets of conditions. The first set consists of 5 frames acquired in October 2024 at short exposure times between $0.1$ and $5$~s, with no bright objects present in the FoV. A more extensive set of dark-like exposures was recorded in December 2024, totaling 36 frames with exposures of $0.015$, $0.1$~s and one half-frame acquired with an exposure time of $9$~s. 

In addition to the dedicated dark datasets, a third set of dark-like exposures is available from the Mars swing-by rehearsal operations. So, 37 exposures were acquired in February 2025 with exposure times between $0.01$ and $0.025$~s. These images were taken under dark conditions and are thus suitable for use as supplementary dark frames. 

\subsection{Standard stars} 

Exposures were obtained for the well-known standard $\alpha$ Lyr (Vega), an A0Va spectral type star. The test included five images taken at 0.1~s, 0.3~s, 0.5~s, 1.0~s and 2.0~s. Their purpose is to evaluate the detector's linearity regime and the point spread function (PSF) distribution. Then, a set of 27 images was acquired for the same target on 7th November 2024 at an exposure time of 0.5~s. The frames captured a ROI of $1024 \times 1088$ pixels (i.e. only the central half of the image was saved, thus reducing the image size by half). A dithering pattern was applied, after three exposures, the pointing was moved by 3 degrees in a square pattern.  Another set of 27 images was acquired again in February at an exposure time of 2~s, to optimize the S/N ratio.

Additionally, other stars have been detected across the star field frames used for alignment and the calibration of optical distortions.  The brightest two stars are $\alpha$ Tau (Aldebaran), a K5III spectral type star,  and  $\delta$ CMa (Wezen), F8Ia spectral type. Observations were obtained also for 16 Cyg star system to obtain spectra for the two G2V spectral type components.  The apparent magnitudes of these two stars (5.96 V mag, and 6.2 V mag)  allow only for obtaining data with $ S/N \leq 5$ -- even at 9 s exposure time (9.9 s is the maximum allowed by the instrument), unsuitable for radiometric, spectral, or PSF evaluations.

The camera is based on a CMOS sensor, where the pixel pitch has a fill factor of 42~\%.  These characteristics must be taken into account when evaluating the radiometric characterization using point-like sources. The goal of these observations was to validate the instrument's functionality and cross-check the calibration performed in the laboratory. Secondary goals included evaluating the PSF distribution and performing linearity tests.

\subsection{Star fields} \label{data-ss}

The first objective of the star field observations was to evaluate the instrument alignment. To this end, four different pointings were observed, with five images acquired for each star field. The second objective, aimed at assessing optical distortions, involved observations of two star fields using a dithering pattern with a $3^\circ$ offset arranged in a $2 \times 2$ raster and acquiring multiple frames for each FoV pointing. The exposure times were varied between 0.5~s and 9~s in order to cover a wider stellar magnitude range. Our data showed that stars are detected above an exposure of 1.0~s, whereas very bright stars in near infrared (e.g. $\alpha$ Tau) could be detected at 0.5~s. 

A total of 71 frames of star fields were acquired. The star fields include:
\begin{itemize}
    \item Five alignment fields observed with different exposure times (3, 4, and 5 seconds)
    \item Two fields around $\alpha$ \textit{Tau} (Aldebaran) and $\delta$ \textit{CMa} (Wezen) were observed with varying exposure times from 0.5 to 9 seconds and a 3$^\circ$ dithering applied between consecutive pointings in a $2\times2$ raster pattern.
\end{itemize}

An automatic pipeline to detect stars was implemented, relying on their expected near-infrared flux at wavelengths corresponding to the HS-H channels. The detections were filtered to keep only the sources with $S/N \geq 5$ to avoid possible mismatches between stars expected at low signal and "bright" groups of noisy pixels that happen to form near the expected location of such stars. Additionally, several artifacts that might have been caused by cosmic rays were detected but discarded after comparing multiple frames with the same pointing. In total, 312 detections have been registered with the help of the HYG catalog, which includes all stars listed in the \textit{Hipparcos}~\citep{1997A&A...323L..49P}, \textit{Yale Bright Star}~\citep{yale}, and \textit{Gliese}~\citep{gliese} catalogs. 

The star fields were used for geometric calibration, linearity tests, and radiometric calibration through the extraction of stellar profiles.

\subsection{Mars swing-by}

The Mars swing-by phase was used to validate the instrument functionality and the execution of the command sequence under strict timing constraints. This phase also enabled verification and refinement of the end-to-end data processing chain, from telemetry packet decoding to the generation of calibrated FITS products, ensuring that the image headers contain all information required for scientific analysis. The first HS-H observations of a planetary surface were successfully acquired, including images in which the Martian surface fully fills the instrument FoV. These data are essential for the assessment and validation of the instrument performance, in particular:
\begin{itemize}
    \item radiometric calibration (a key objective)
    \item geometric distortion correction
    \item conduct a scientific investigation of the far side of Deimos
    \item support public outreach by showcasing the capabilities of the HS-H instrument
\end{itemize}

A total of 22 images of Mars were acquired prior to the three images of Deimos (two at low spatial resolution and one at higher spatial resolution), followed by five images of Mars in which the planet fully covers the FoV. Seven images targeting Phobos were attempted, with the moon detected in three of them as a faint crescent. The instrument operated nominally, except for a minor line loss in a single image, which does not affect the scientific usability of the data. The temperature measured at the FPA detector remained stable at $-12.7 \pm 0.3^\circ$~C. The data were downlinked, decoded, and pre-processed in near real time, immediately following the reception of telemetry packets on Earth. The commanding sequence was successfully validated under the imposed strict time constraints.

The initial observations capture the full disk of Mars, with spatial resolutions ranging from 134~km/macropixel down to 25~km/macropixel. Images in which the Martian surface covered the entire FoV were acquired at spatial resolutions ranging from 11.5~km/macropixel down to 4.3~km/macropixel. These primarily cover southern latitudes, where the terrain is dominated by impact craters. Notable landforms include Schiaparelli and Huygens craters, Hellespontus Montes, and Noachis Terra. The illumination conditions vary across the images as the phase angles for the Mars images start from around $15^\circ$, increasing progressively up to approximately $65^\circ$ in the final frame. The images obtained of Mars serve as key input for validating the laboratory calibrations. This marks the first time the HS-H instrument has been used to image a planetary surface under reasonable illumination conditions. The flat-field calibration was conducted in the laboratory at 20$^\circ$~C, whereas the instrument operates in space at -12$^\circ$~C. Therefore, additional checks are required to validate the spectral response of each channel.


\section{Results}
The observations described above are used to validate and fine-tune the calibrations performed under laboratory conditions. They also allow the evaluation of the instrument performance in the space environment.

\subsection{Bad pixels}

The bad pixels are those that exhibits abnormal characteristics compared to surrounding pixels under identical illumination and operating conditions. To identify pixels with anomalous behaviour, bias frames were used. The detected bad pixels were classified into two categories: noisy pixels and outlier pixels. Noisy pixels are identified by an abnormally high temporal standard deviation of their digital number (DN) values, indicative of elevated and unstable read-out noise (RON). These pixels do not necessarily exhibit high signal levels. In contrast, outlier pixels are characterized by a persistently high mean DN value across all frames, significantly deviating from the global DN distribution, and typically correspond to pixels with elevated dark signal and limited temporal variability. The two pixel types are analyzed separately.

To identify the outlier pixels, we normalize the bias frames by subtracting the mean value and then dividing by the standard deviation of each image. Pixels deviating beyond a threshold of $5\sigma$ in all the frames are flagged as outliers. We cross-check these results with a second method involving image smoothing through convolution by $5\times 5$ and $10\times 10$ averaging kernels. The smoothed background is subtracted from the original data to highlight localized pixel deviations. Pixels with deviations beyond the same threshold of $5\sigma$ are flagged as outliers. These masks are then updated iteratively across all available bias frames.

Two sets of bias images were acquired during the early cruise phase. The first set, consisting of eight frames, was obtained in October 2024 during the commissioning phase, followed by a second set of 11 frames acquired in January 2025. An additional bias data set of 11 frames was collected later, in September 2025. For all bias acquisitions, the entire FoV was exposed for 100~$\mu$s. These three bias data sets are used to characterize the distribution of bad pixels.

\begin{figure*}[ht]
    \centering
    \includegraphics[width=0.8\linewidth]{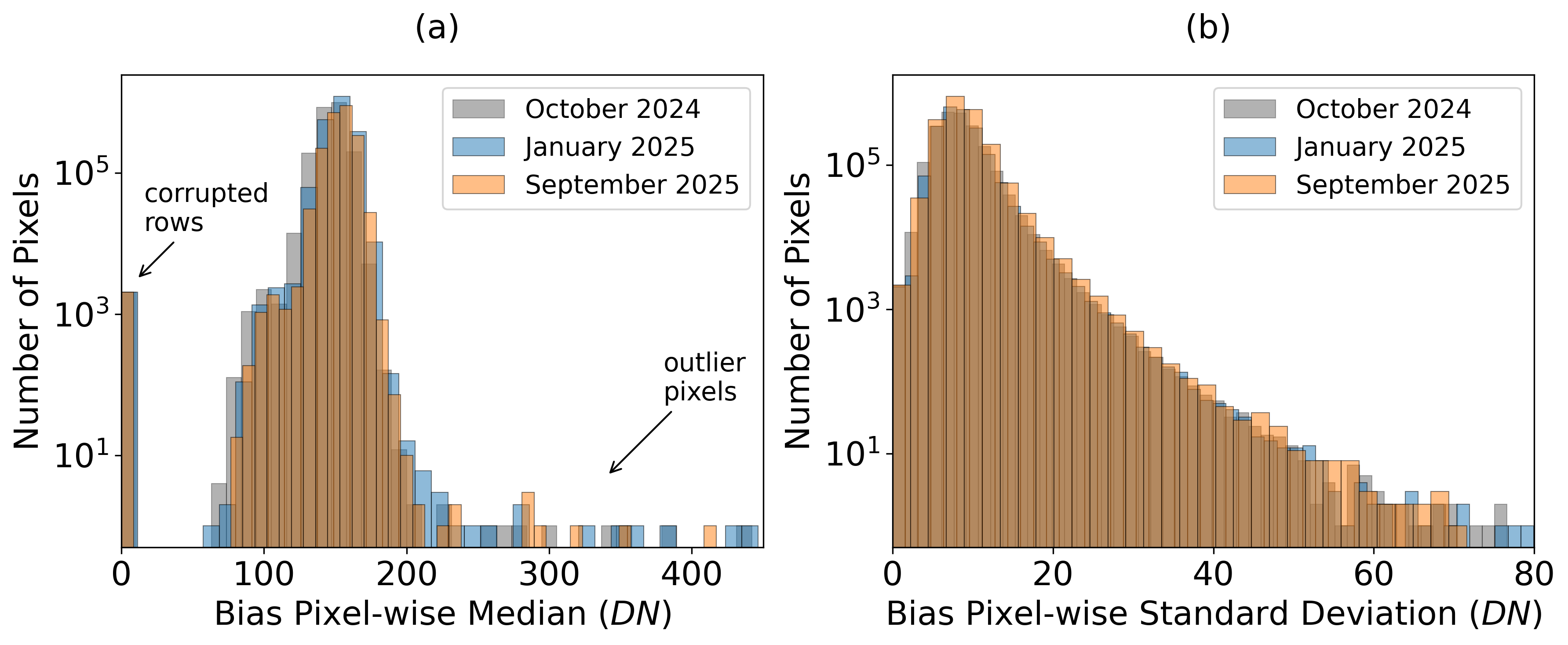}
    
    \caption{(a) Distribution of pixel-wise mean of the global bias level, highlighting outliers and corrupted rows. (b) Histogram of standard deviation values per pixel across bias frames, showing baseline readout noise and the presence of highly variable pixels. Both plots compare data from October 2024, January 2025 and September 2025, with logarithmic y-scales.}
    \label{fig:hot_pixel_stats}
\end{figure*}

In October 2024, a total of 231 pixels were identified as noisy, with $\sigma>50$ DN across all acquired bias frames, while 16 individual pixels were flagged as outliers. In January 2025, 221 noisy pixels and 21 individual outlier pixels were detected, whereas in September 2025, there were 191 noisy pixels and 9 outliers. The number of common noisy pixels between the three observational sessions was 71, and all 16 outlier pixels from October were also present in January. No new outliers were detected in September. In addition to the individual outlier pixels, three additional corrupted rows were detected across all frames, each containing 2048 outlier pixels (the first two rows and the last row). In Fig.~\ref{fig:hot_pixel_stats}a, it is shown how individual pixels deviate from the median bias level. The few outlier pixels mentioned before can be noticed, and also, the prominent negative spike at zero corresponds to the group of corrupted rows.

Thus, for October 2024, the total number of bad pixels was 6.389, with no overlap between the noisy pixels and the outliers. In contrast, the January session exhibited 6.380 hot pixels, with a small intersection of 4 pixels classified as both noisy and outliers. We notice that the number of individual outlier pixels increased from 16 to 21 from October 2024 to January 2025, however, in September 2025 the number went down to 9 pixels.  

The histogram in Fig.~\ref{fig:hot_pixel_stats}b illustrates the distribution of pixel-wise standard deviations derived from the two bias frame datasets. The similarity between the three histograms suggests that the detector’s overall noise characteristics remained stable over time. However, a long tail of warm pixels is present in both cases, extending beyond 50~DN and representing a small population of highly variable pixels. The noise level, derived from the most probable values of standard deviation histograms, corrected from statistical artifacts, which approximates the detector’s RON, ranged from 8.76~DN in October 2024 to 8.66~DN in January 2025 and 8.91~DN in September 2025. 

To correct the bad pixels in the other calibration and science images, we use an interpolation method based on the values of neighboring pixels within the same wavelength channel. For each corrupted pixel, its value is replaced with the mean of the four functioning pixels in its local neighborhood. This correction was applied to all images acquired with HS-H as an initial step, prior to any further calibration procedures.

\subsection{Bias frames}

The bias frames are first corrected using the hot-pixel maps characterized in the previous section. Subsequently, the mean value of each frame is subtracted to remove the offset introduced by temporal fluctuations in the bias level. These variations in the bias baseline are driven by random telegraph noise; physically, it arises from the stochastic trapping and subsequent release of individual charge carriers by localized defects within the silicon lattice or the oxide interface of the readout amplifiers~\citep{10.1088/978-0-7503-1272-1}, so the bias level is affected by groups of pixels having irregular jumps in brightness, as was observed in other CMOS detectors as well~\citep{Alarcon_2023}.The median bias level varies between 144~DN and 156~DN. The overall median bias level is 151~DN, with a standard deviation of 3.4~DN. Following the procedure described above, the resulting frames are combined into a single frame using pixel-wise median values. This process provides the master bias for each acquired bias data set.

To evaluate the stability of the bias pattern over time, we compared the master bias frames from October 2024 and January 2025. Visual inspection shows the two maps are nearly identical, with similar low-frequency structures across the sensor. The difference image is shown in the right panel of Fig.\ref{fig:master_bias_comparison}. No significant large-scale structures are observed to have evolved, suggesting good stability of the bias structure over the three-month interval. However, minor local variations are visible. This residual pattern is known as row noise and is described in the next subsection.

\subsubsection{Row noise}
\begin{figure*}
    \centering
    \includegraphics[width=0.98\linewidth]{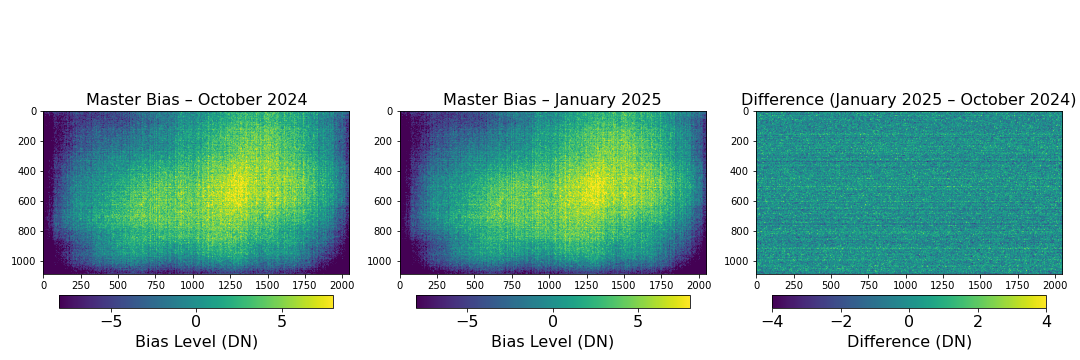}
    \caption{
    Comparison of the master bias frames acquired in October 2024 and January 2025. Each master bias was median-combined from a stack of corrected bias images. The left and middle panels show the master biases derived from October 2024 data, and from January 2025 data respectively. The right panel displays the pixel-wise difference between the two. }
    \label{fig:master_bias_comparison}
\end{figure*}

The difference between the different master biases showed that there is no long-term structural drift of the bias pattern (see Fig.\ref{fig:master_bias_comparison}). However, the observed residual pattern, i.e. the horizontal stripes, was not averaged out over time when combining the individual bias frames, meaning that these stripes are variable from frame to frame. Therefore, to capture this row noise, we compute the difference between the individual bias frames and the master bias estimated within the same session.

The higher variance in the row direction is a direct result of the sensor's row-by-row readout architecture. The detector digitizes all pixels in a single row simultaneously using a shared global analog reference voltage. As this reference voltage fluctuates over time due to normal power supply ripple and electronic noise~\citep{Mikkonen2014VerificationOC, Shao2023TheIO}, each row receives a slightly different baseline offset. This temporal voltage drift during the frame readout creates horizontal stripes. In contrast, the much lower column variance confirms that the parallel analog-to-digital converters are uniformly matched and introduce minimal spatial noise.

\begin{figure}
    \centering
    \includegraphics[width=0.95\linewidth]{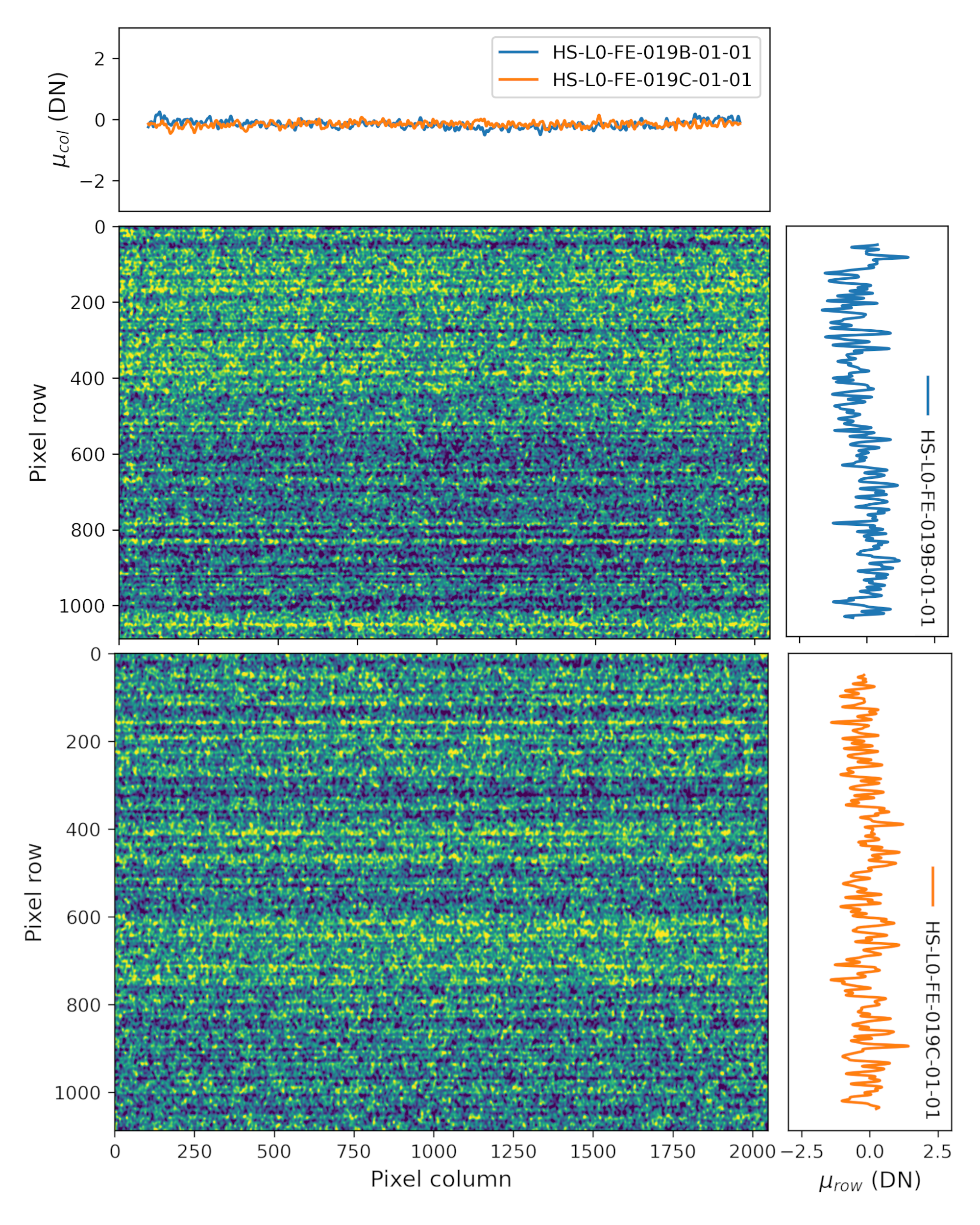}
    \caption{Residual bias patterns after subtracting the October master bias from two consecutive individual bias frames. A $5\times 5$ Gaussian filter was applied to suppress pixel-level noise and highlight the residual faint structures. The top and right panels show the column-wise ($\mu_{\mathrm{col}}$) and row-wise ($\mu_{\mathrm{row}}$) means, respectively, for each frame. The colormap spans from -2~DN to 2~DN.}
    \label{fig:residual_bias_structure}
\end{figure}

Two residual patterns derived from two consecutive individual bias frames acquired in the same session (October 2024) are shown in Fig.\ref{fig:residual_bias_structure}. A Gaussian kernel with a $5\times 5$ window was applied to the residuals to suppress high-frequency noise and emphasize the residual signal. The results reveal a faint horizontal banding, primarily visible as structure in the pixel row direction. This is also confirmed by the row-wise mean signal ($\mu_{\mathrm{row}}$), which exhibits variations with a standard deviation of approximately 2.1~DN (0.5~DN after applying the filter), significantly larger than the standard deviation of the column-wise mean ($\mu_{\mathrm{col}}$), which is only 0.25~DN (0.1~DN after applying the filter). These results suggest a residual low-amplitude signal not fully removed by the standard bias correction procedure.

\begin{figure}
    \centering
    \includegraphics[width=0.95\linewidth]{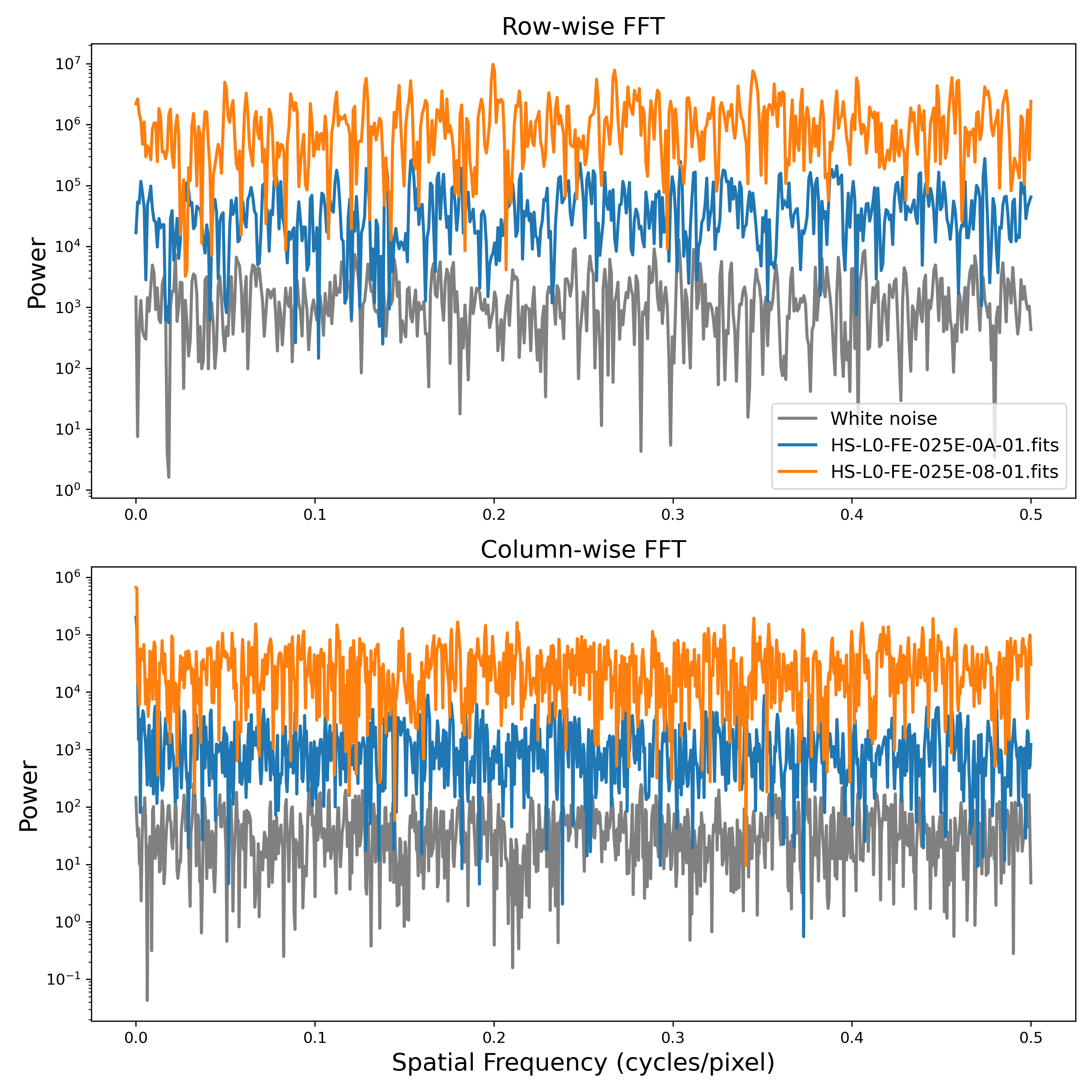}
    \caption{Spatial power spectra of residual bias frames. Top and bottom panels display the row-wise (horizontal) and column-wise (vertical) FFTs, respectively. Detector residuals are compared against a simulated Gaussian white noise baseline (bottom line), with successive spectra vertically offset for clarity.}
    \label{fig:fft}
\end{figure}

One-dimensional Fast Fourier Transforms (FFTs) were computed for both row-wise and column-wise mean residuals to isolate horizontal and vertical banding, respectively. The resulting spatial power spectra of the physical bias residuals were then compared against a synthetic, purely uncorrelated Gaussian white noise profile, scaled to match the empirical standard deviation of the detector's row and column variations. As shown in Fig.~\ref{fig:fft}, the power spectra of the actual residuals closely mirror the flat, featureless continuum of the simulated white noise across all spatial frequencies up to the Nyquist limit (0.5 cycles/pixel). The similarity of the power spectra with the white noise baseline suggests that the frame row or column noise is mostly caused by thermal fluctuations and is free from other types of periodic electronic artifacts.

\subsection{Dark frames}

\subsubsection{Short-exposure dark frames}
First, we studied the behavior of short-exposure dark frames (between 10 and 100~ms). These exposure times covers the range of exposures which will be used during the asteroid phase. Three datasets were analyzed, comprising 4 frames at 100~ms, 16 frames at 15~ms (both acquired in December 2024), and 27 frames at 10~ms acquired in February 2025. For each dataset, we calculated the pixel-wise standard deviation across frames to estimate the RON value as the most probable value of the distribution and identify noisy/outlier pixels in the same manner we proceeded before with the bias frames. Table~\ref{tab:dark_frame_summary} summarizes the results obtained. For each dark frame we compute the median of all the pixel values in that frame. Then, across all the dark frames in a dataset (e.g. the 4 frames with 100~ms exposure), we compute the mean and standard deviation of those medians. As seen in Table~\ref{tab:dark_frame_summary}, the median stays in the range 150~-~153~DN. The value is comparable to the one determined for the bias frames, showing that the dark current at short-exposure time is negligible with respect to the bias fluctuations due to the detector's electronics.

The estimated RON values measured at 10 and 15~ms are comparable to those measured for the bias frames. There is no significant increase in the number of noisy and outlier pixels. A number of 53 noisy pixels are the same in each dataset, of which 38 of them were classified as noisy also for the two bias frame datasets. Two new outlier pixels were detected across the dark frames acquired at 100~ms. However, more data are needed to confirm this, as they might be noisy pixels. Only three individual outlier pixels were detected across the dark frame acquired in February, which means that most of the outlier pixels, which were detected in the previous sessions, behaved as normal or noisy pixels during the February session.

\begin{table*}
\centering
\begin{tabular}{ccccccc}
\hline
\textbf{Session     }                   & \textbf{\#} & \textbf{Exposure time} \textbf{(ms)} & \textbf{Median (DN)} & \textbf{Noise (DN)} & \begin{tabular}[c]{@{}c@{}}\textbf{Noisy pixels}\\ $\sigma > 50~DN$\end{tabular} & \begin{tabular}[c]{@{}c@{}}\textbf{Outliers}\\ $>5\sigma$ deviations\end{tabular} \\ \hline
\multirow{2}{*}{December 2024} & 4           & 100                & $150.5\pm4.3$            & $8.22$        & 251                                                                     & 13                                                                             \\
                               & 16          & 15                 & $153.0\pm3.5$            & $8.13$        & 235                                                                     & 10                                                                             \\
February 2025                  & 27          & 10                 & $150.0\pm1.0$            & $8.10$        & 184                                                                     & 3                                                                              \\ \hline
\end{tabular}

\caption{Summary of short-exposure dark frame statistics. For each session and exposure time, we report the number of frames processed ($\#$), the average median signal per frame (with standard deviation), the estimated RON derived as the most probable value of the pixel-wise standard deviation across the stack, and the number of noisy and outlier pixels detected. A pixel is considered noisy if its temporal standard deviation exceeds 50~DN, and an outlier if it exhibits values exceeding 5$\sigma$ from the mean across all frames of the datasets.}
\label{tab:dark_frame_summary}
\end{table*}


\subsubsection{Long-exposure dark frames}

To monitor the dark current as a function of temperature and exposure time, we analyzed long-exposure dark frames. The HS-H camera does not have a shutter. Therefore, the 62 frames that were included in this analysis contained signal from stars or were affected by cosmic rays. The exposure times varied from one to five seconds

By stacking all the frames obtained at a specific exposure time, we cancel the incoming signal from different external sources, which is exposed by less than 1-2\% of the pixels. In this way, we obtained a different bad pixel map and master dark frame at each exposure time.

\begin{figure}
    \centering
    \includegraphics[width=0.49
    \textwidth]{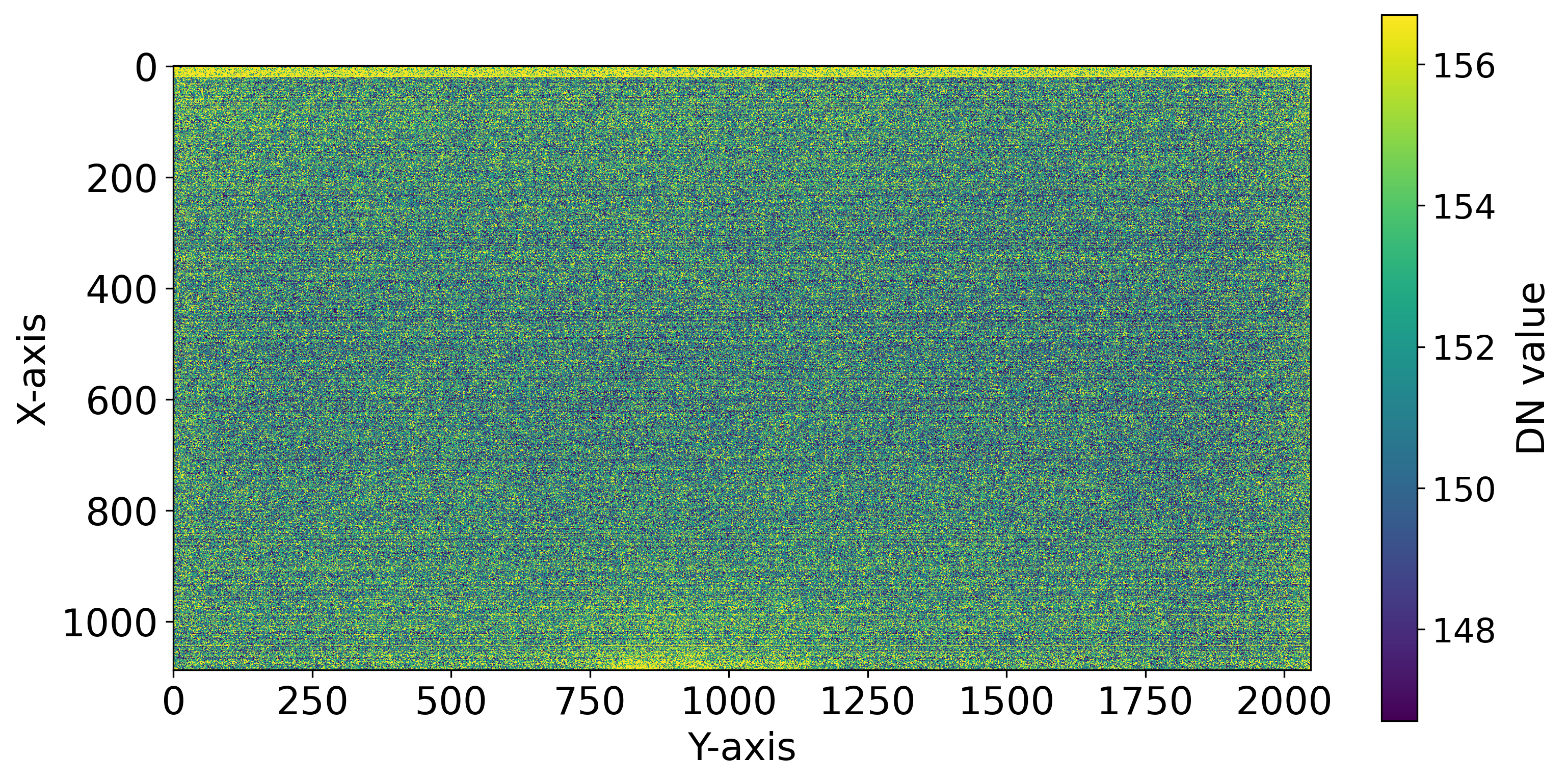}
    \caption{Master dark frame acquired at an exposure time of 5~s. }
    \label{fig:masterdark5}
\end{figure}

Our analysis showed that the number of these common outliers above 5~$\sigma$ is increasing with respect to the exposure time reaching almost 100 outliers at 5~s. So, the bad pixels become more prominent at longer integrations. In addition to individual bad pixels, some patterns are identifiable in the dark frames. These are clearly observed in Fig.~\ref{fig:masterdark5}, where the master dark frame was derived from 5~s dark exposures. At the top of the images, faint horizontal bands are visible, suggesting a systematic effect. Similarly, a low-intensity cloud-like structure near the bottom indicates a region with slightly larger DN level.

These patterns are characteristic signatures of amplifier glow, a well-known phenomenon in CMOS detectors. This is primarily driven by an electroluminescence phenomenon associated with hot carrier generation within the in-pixel source follower transistors, where impact ionization emits photons and excess minority carriers that are parasitically collected by the photosensitive area during prolonged integrations~\citep{10.1117/12.476792}. Similar macroscopic glow patterns have been recently observed in comparable CMOS cameras~\citep{apergis2025highprecisionphotometryscientificcmos}. Because this parasitic gradient scales proportionately with exposure time, it only becomes prominent during extended integrations. 

For the nominal target observations of the asteroid, which require sub-second exposures, this amplifier glow will remain suppressed below the RON floor. However, these artifacts can be successfully corrected by subtracting the master dark from the raw frame.

While the FPA temperature fluctuated between $-13.5^\circ$C and $-12^\circ$C, this window is too narrow to establish a statistically significant correlation between the dark current rate and the detector temperature. Within such a limited range, any intrinsic temperature-dependent variations in the dark current are hidden under bias fluctuations. However, we can evaluate the dark current contribution at this temperature level by analyzing the variation with respect to the exposure time.
\begin{figure}
    \centering
    \includegraphics[width=0.49
    \textwidth]{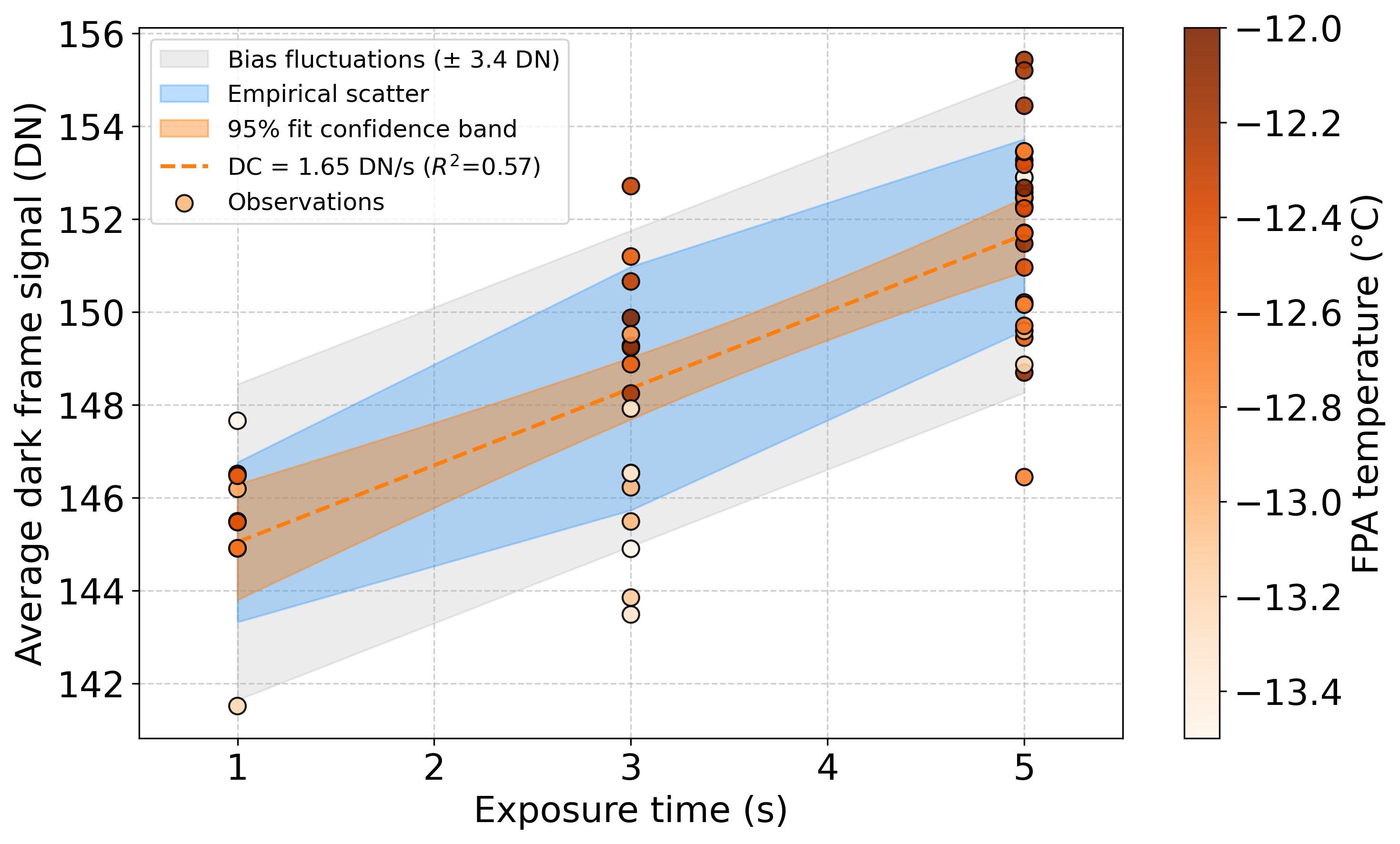}
    \caption{Mean values of dark frames as a function of exposure time in a narrow range of FPA temperatures. }
    \label{fig:darkexp}
\end{figure}

The average dark frame signal at the pixel level and at a specific exposure time can be modeled as the linear sum of the static baseline, the time-dependent dark current, and the bias fluctuations, expressed as $D(t) = B_0 + DC \times t + \Delta B$. The term $B_0$ defines the fundamental, static bias level of the detector, while $DC$ represents the intrinsic dark current generation rate. The bias fluctuations are captured by $\Delta B$, which acts as a stochastic noise component, as observed during the bias frames analysis discussed earlier. This term is modeled as a random variable sampled from a normal distribution, $\Delta B \sim \mathcal{N}(0, \sigma_B^2)$, where the standard deviation is empirically determined as $\sigma_B \approx 3.4$~DN. Because $\Delta B$ is normally distributed and operates independently of the integration time, it introduces a stochastic vertical offset to the data. By performing a linear regression, we cancel the bias temporal fluctuations. Thus, we extract the dark current rate $DC=1.65\pm 0.20$~DN/s. 

The linear regression within the stable operational regime, $t \geq 1$~s, yields $R^2 = 0.57$. Under these environmental conditions, the accumulated dark signal remains small relative to the bias fluctuations. The maximum possible $R^2$ is constrained, as the bias variance is comparable to the dark current increase. Theoretically, $R^2$ is bounded by the ratio of the true signal variance to the total combined variance, which is approximately $0.45$ for this exposure sequence given an underlying white noise of $\sigma_B \approx 3.4$~DN. The empirical value of $0.57$ therefore exceeds this threshold, which means that the bias fluctuations are lower at exposures beyond $1$ s compared to the bias exposures at 100 $\mu$s. This decrease can be observed also visually in Fig.~\ref{fig:darkexp}, where most of the observations fall inside the $1\sigma_B$. 

The measured temporal noise, which is dominated by RON, exhibits a linear increase from approximately $8.20$~DN at $1$~s to $8.36$~DN at $5$~s. This slight elevation is an expected consequence of the accumulating dark signal. Taking into account the empirically established dark current rate of $1.65$~DN/s, the theoretical contribution of the dark current shot noise over this interval precisely accounts for the observed noise growth.

In the preliminary on-ground characterization of the instrument~\citep{Popescu2025hyperscout}, the dark signal accumulation was modeled as following an exponential growth curve, and the baseline RON was reported at a significantly higher $12.57$~DN. However, that initial analysis was heavily influenced by the higher ambient temperatures of the test environment.

\subsection{Linearity tests}

We report the linearity of the HS-H detector based on inflight observations of Earth~-~Moon system and stellar observations. The sensor response is supposed to be linear with respect to the incident flux or exposure time.

\subsubsection{Earth~-~Moon system}

The linearity of the signal is analyzed using 39 frames of the Earth–Moon system. The Earth shows significant saturation for 50 ms exposure time, and even the 10 ms and 20 ms exposures contain saturated regions. To ensure a consistent comparison across exposure times, we restricted the analysis to the continuously illuminated core region of the Earth's disk. Because the Earth was rotating and its atmospheric cloud cover was rapidly changing between frames, applying a static spatial mask would have introduced severe photometric errors. Instead, by dynamically isolating this consistently sunlit area---which corresponds to approximately the brightest 25\% of the field of view---we established a stable photometric reference. This approach safely excludes the dark space background and the sharp brightness gradients near the terminator. The resulting trend shows a clear linear response up to the onset of saturation, with a slope of 188.42 DN/ms, and the deviation at 50 ms aligns with the expected clipping due to full-well saturation.

The Moon images remain fully unsaturated for all tested exposure times. Their signal increases linearly with exposure, with a slope of 48.93 DN/ms, reflecting the lower intrinsic brightness of the target. The linearity fit for both bodies confirms that the detector response is linear within the unsaturated regime up to around 3800~DN.

\begin{figure}[ht]
    \centering
    \includegraphics[width=0.49
    \textwidth]{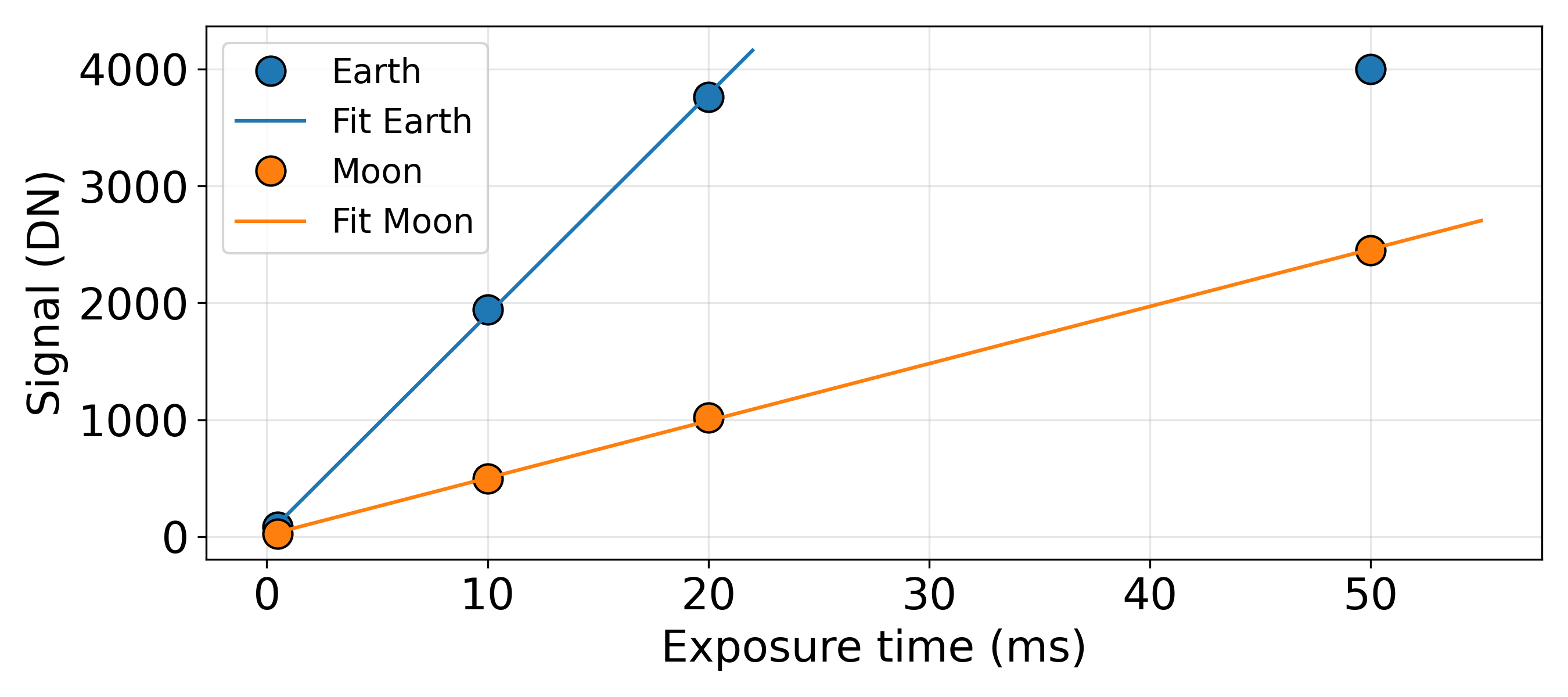}
    \caption{Linearity test of the Earth–Moon system. The plot shows the measured signal as a function of exposure time for both Earth and Moon images, together with their corresponding linear fits. The error bars are too small to be distinguished from the data markers. Earth reaches saturation.}
    \label{fig:lin_earth}
\end{figure}

\subsubsection{Standard stars}

The linearity test of Vega consisted of five exposures with the exposure time increasing from 0.1~s to 2.0~s. The acquired signal was not saturated in any image. On the other hand, for Aldebaran there are five exposures between 3.0~s and 5.0~s, and it almost reaches saturation on the brightest pixel at an exposure time of 5.0~s. We checked that the flux of Vega and Aldebaran is linear within different apertures. For Aldebaran, the slopes increase with aperture size, ranging from 2445 DN/s for the $3\times 3$ px aperture to 3199 DN/s for the $9\times 9$ px aperture, with the corresponding values of the coefficient of determination exceeding 0.96, indicating a good fit to a linear model. Similarly, Vega exhibits a linear trend, with slopes increasing from 1001 DN/s at $3\times 3$ px to 1529 DN/s at $9\times 9$ px. 

\begin{figure}[ht]
    \centering
    \includegraphics[width=0.49
    \textwidth]{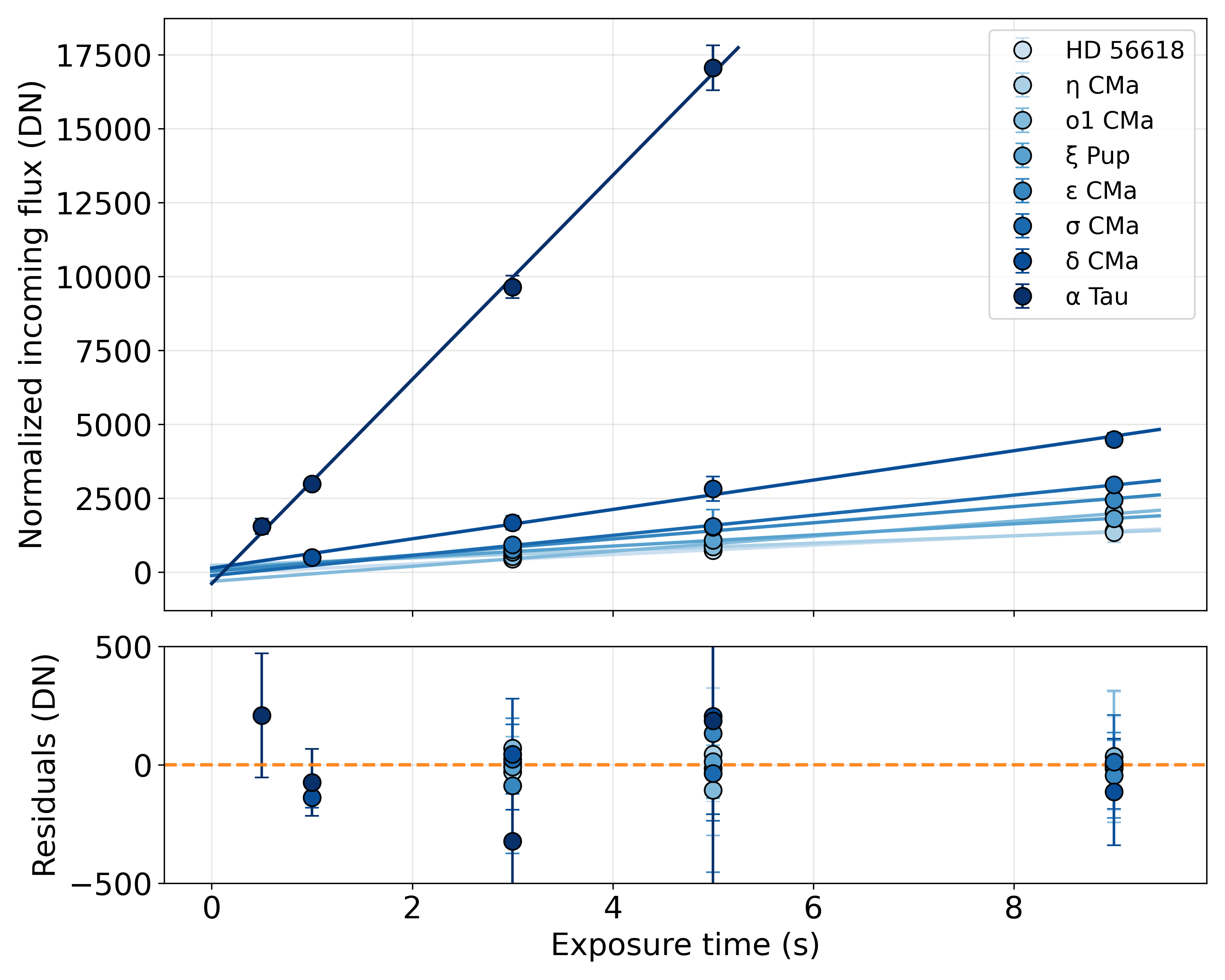}
    \caption{Linearity test with most frequent star detections. Error bars show the standard deviation across all frames with identical exposure time. The bottom panel illustrates the residual analysis. }
    \label{fig:linearity_stars}
\end{figure}

We continued our analysis with data gathered from the most frequently detected stars across the star field acquisitions. We filter them considering the SNR of the detections, exposure time, and the condition of having at least three detections for each exposure time group. The plot in Fig.~\ref{fig:linearity_stars} presents the normalized incoming signal as a function of exposure time for a large aperture of  $9\times 9$ px. The fluxes were then normalized to the median channel.

To validate the linear regime and removal of electronic offsets, the dependences were modeled using linear regressions. The resulting free intercepts yielded a mean of $-40 \pm 70$~DN across the targets (inside an aperture of $9\times 9$ px), which is statistically indistinguishable from zero. This validates the applied bias and background subtraction. Furthermore, a quantitative analysis of the residuals demonstrates no systematic structure, showing that the deviations from linearity are consistent with the camera's noise.

\subsection{Point spread function}

The distribution of the point spread function (PSF) is characterized using stellar observations of Vega at 0.5 and 2.0~seconds. The PSF covers multiple pixels. Thus, the flux is distributed to several wavelength channels. The position of the star corresponds to the peak of the PSF distribution, which depends on the pointing of HS–H. 

If the distribution peak overlaps with the center of a pixel, then we will observe a symmetrical distribution of the PSF which is modulated by the spectral behavior of each channel. On the other hand, if it is located somewhere inside the pixel, not so close to the center, we can expect a non–symmetrical PSF distribution. Numerical simulations characterizing the effect of the detector filling factor and centroid positions on the PSF are detailed in \ref{app:psf_sim} together with the description of the model used in processing the stellar profiles. The simulations show that the signal recorded by the center pixel of the profile is between 10.8 and 26.6\% of the total measured flux. These values were the result of noise-free simulations. Therefore, some of the observations might not respect the constraints derived above.

The stellar observations allowed us to analyze the observed PSF distribution. We present an analysis of the PSF distribution using data acquired for Vega.

\begin{figure}
    \centering
    \includegraphics[width=0.49
    \textwidth]{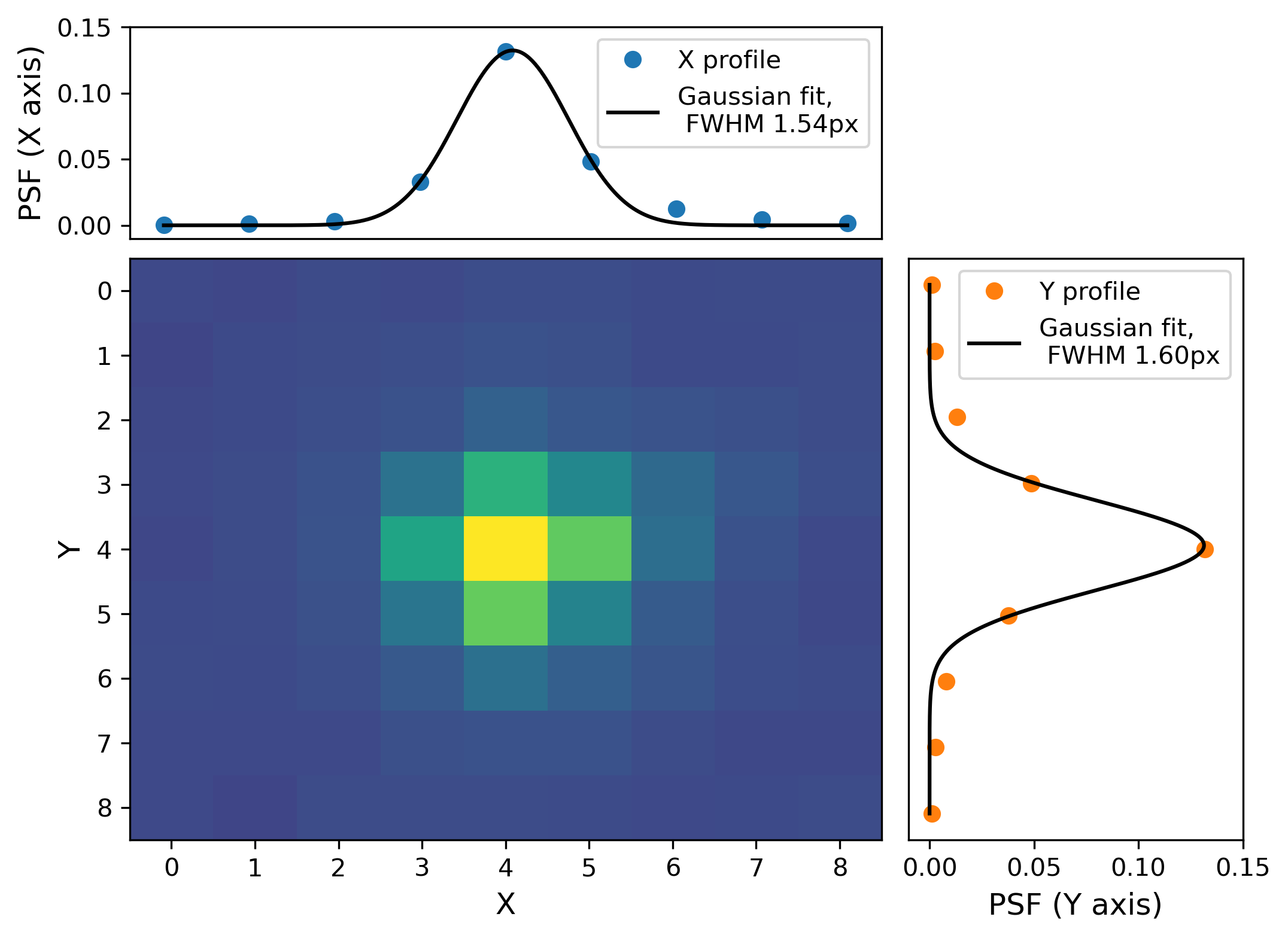}
    \caption{Two-dimensional PSF map obtained from all stellar profiles of $\alpha$ Lyr (Vega), along with its 1D projections on the X and Y axes. The Gaussian fits yield FWHM values of 1.54~px in the X direction and 1.60~px in the Y direction. }
    \label{fig:psf_map}
\end{figure}

First, the PSF structure is analyzed in Figure~\ref{fig:psf_map}, where we visualize the 2D flux distribution and its corresponding maximum projections along the detector axes. Gaussian functions were fitted to the X and Y axis profiles, yielding FWHM values of 1.54~px and 1.60~px, respectively. 

\begin{figure}
    \centering
    \includegraphics[width=0.45\textwidth]{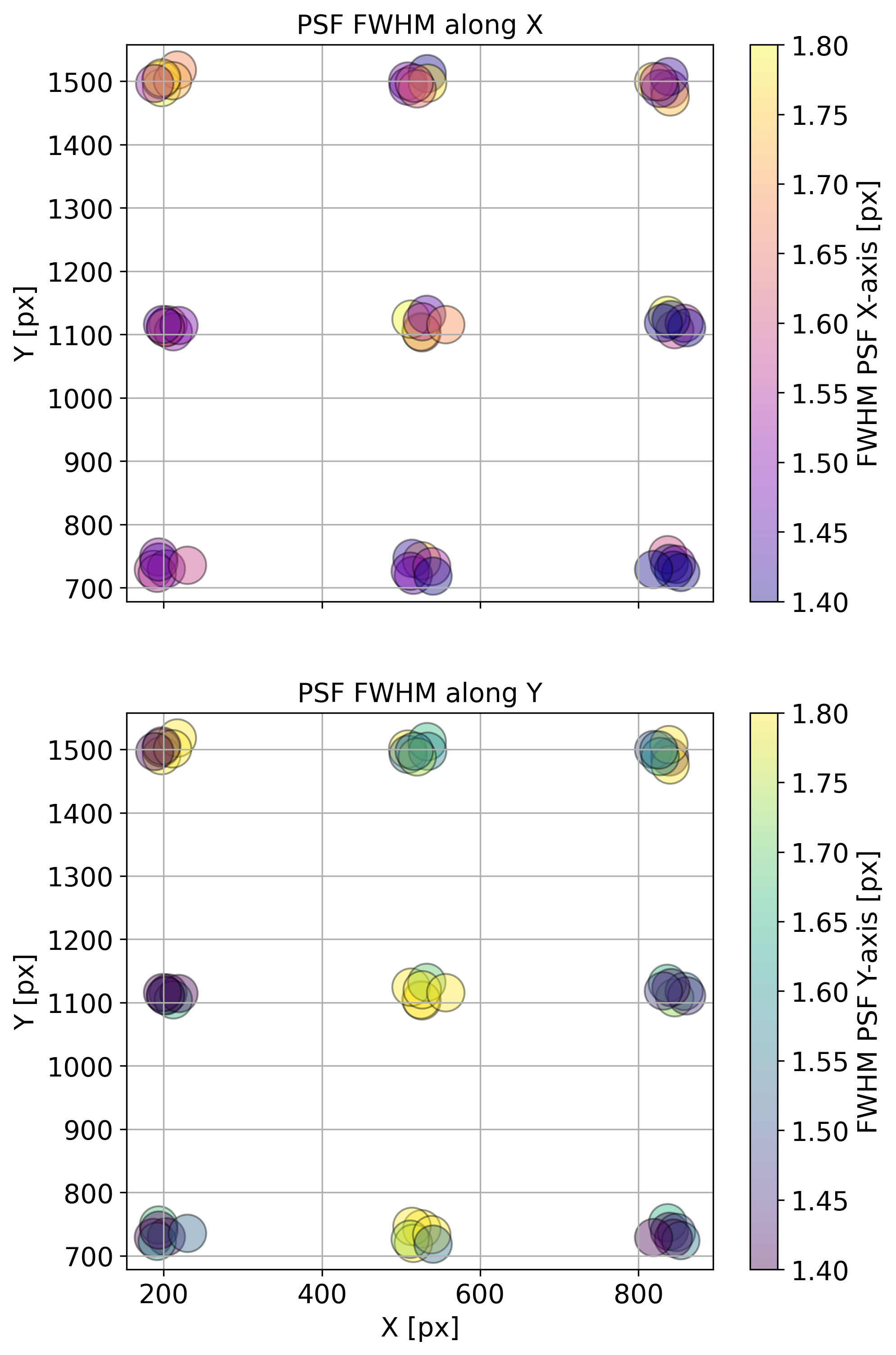}
    \caption{Top: Scatter map of PSF FWHM along the X-axis across the detectors.
    Bottom: Same for the Y-axis. A small random position shift was added to reduce marker overlap improving the visualization clarity. Color scales indicate FWHM in pixels, ranging from 1.4 to 1.8~px.}
    \label{fig:psf_fwhm_map}
\end{figure}

Then, to evaluate the spatial variation of the PSF width across the FoV, we compare the FWHM along both the X and Y detector axes. Figure~\ref{fig:psf_fwhm_map} presents two spatial scatter plots of the PSF FWHM values. We notice a few regions that exhibit slightly elevated or reduced PSF widths, suggesting minor spatial dependence, possibly due to optical distortions. The FWHM values are spread between 1.2 and 2.2~px. However, the average values are around 1.6~px as determined before, being in good agreement with previous pre-flight measurements provided by the producer.

\subsection{Radiometric calibration}

The purpose of this section is to quantify the radiometric constants using in-flight data and to compare the results with the values measured in the laboratory. Data were pre-processed using the flat-field correction established on-ground prior to flight \citep{Popescu2025hyperscout}. No dedicated flat-fielding exposures were taken during the cruise due to a lack of appropriate targets. On the Martian surface, we did not observe the presence of significant variations above the noise level as the cause of potential flat-field effects. Our in-flight determination of the radiometric constants is based on multiple observations, with regions of interest analyzed at different locations on the sensor. Thus, by averaging, we considerably reduce the effect of potential variations in the flat-field patterns. The pre-flight flat fields showed variations within a 10\% range, the primary causes being the sensor electronics and filters.

Different methods for evaluating the sensitivity constants based on laboratory images yield slightly varying results due to factors such as temperature, illumination conditions, etalons, and the instruments used in the laboratory. The results and the uncertainties of these calibrations are described by \citep{Popescu2025hyperscout}. Below, we discuss the radiometric calibrations performed in flight using data from standard stars and the Mars images.

\subsubsection{Cross-calibration with standard stars}
We used the frames acquired for $\alpha$ Lyr (Vega) and all the star field frames to detect stellar profiles with $S/N\geq 5$. The Vega frames were acquired in two sessions - first session included 27 acquisitions at an exposure time of 0.5~s and the second session another 27 acquisitions, but at a longer exposure time of 2.0~s. As mentioned in Section~\ref{data-ss} and detailed in Table~\ref{tab:activities}, there are 71 star field frames with varying exposure times from 0.5 to 9.0~s.

Stellar profiles were extracted using a $9\times9$ pixel aperture, a size sufficient to capture the full profile according to the PSF analysis presented earlier. To quantify the total incoming signal, the value of each pixel count was scaled to the gain level of one reference channel, then divided by the normalized spectrum of the star at the wavelength of the same reference channel. The reference channel was chosen at 0.806~$\mu$m, corresponding to channel number 12. The sum of the resulting values is the channel-normalized incoming signal $I_{ref}$, as would have been measured by a pixel area of the detector that operates in the regime of reference channel 12. We used the calibration data provided by the Mars swing-by to retrieve updated information about the relative gain levels of different channels. The entire processing pipeline is detailed in \ref{app:psf_sim}.

\begin{figure}
    \centering
    \includegraphics[width=0.49
    \textwidth]{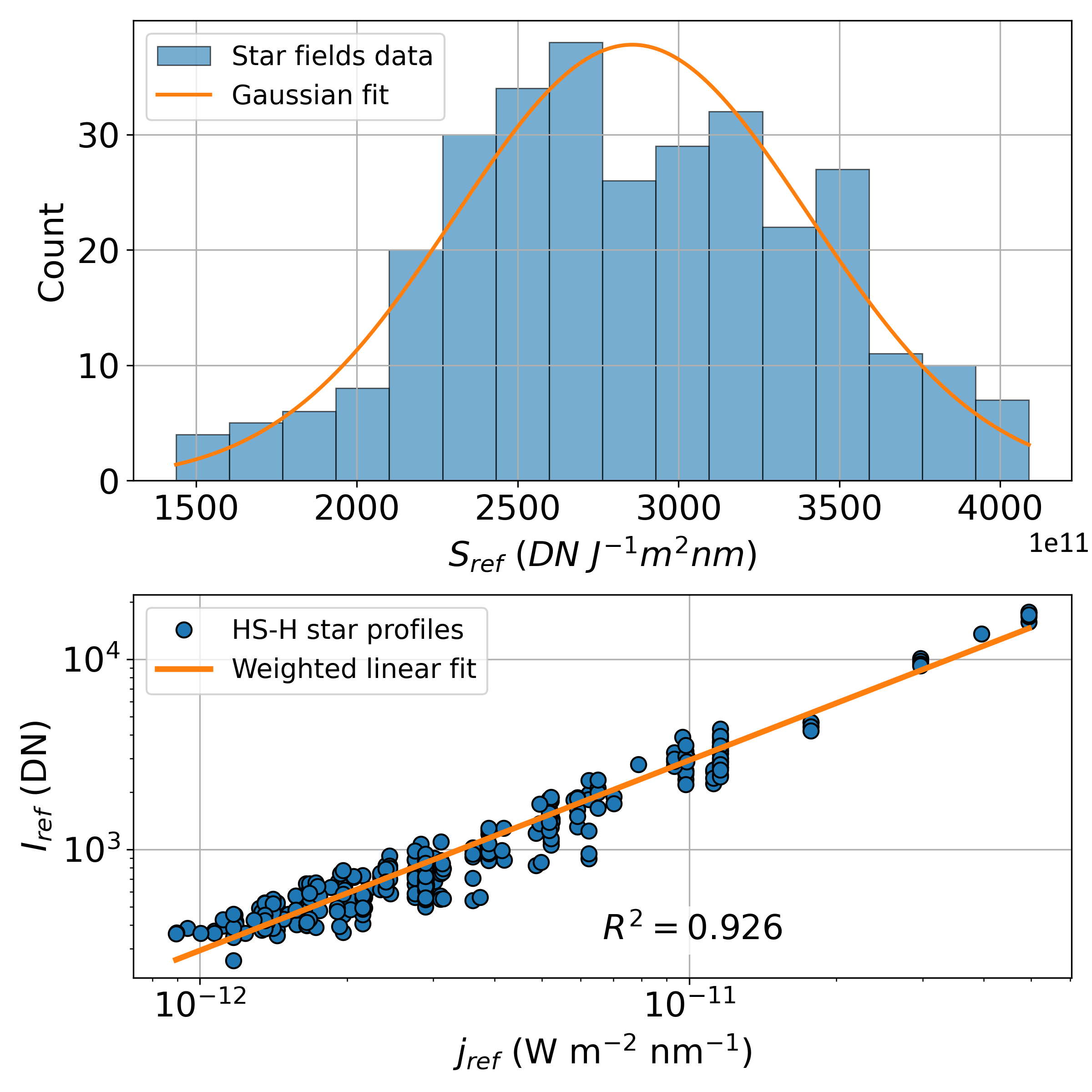}
    \caption{\textit{Top}: Distribution of the reference-channel–normalized stellar signals. A Gaussian model is fitted to the histogram. \textit{Bottom}: Comparison between measured reference-channel signals and convolved stellar irradiances. A weighted linear regression is applied using weights proportional to $1/I_{ref}$, reducing the input of noisy detections. The resulting best-fit line and coefficient of determination are shown. In both subplots, the outlier values leading to very high sensitivity were discarded to avoid large model errors.}
    \label{fig:detections_fit}
\end{figure}

The $I_{ref}$ values were computed for 367 star profiles. The results are summarized in Table~\ref{tab:stellar_obs}, which lists all stars with more than 5 detections. The last column shows the sensitivity value of the reference channel for point-like sources, calculated as the average of all detections. The sensitivity calculation takes into account the filling factor of the detector's geometry,
\begin{equation}
S_{12} = \frac{I_{ref}}{j_{12}\times FF}
\end{equation}
The distribution of the sensitivity constant derived from the measured reference-channel signals is shown in the upper panel of Fig.~\ref{fig:detections_fit}. The measurements are fitted with a Gaussian distribution. The peak value is found at  $\mu$ = (3030$\pm$40) DN/(10$^{-11}$ Jm$^{-2}$nm$^{-1})$, which provides the first in-flight estimate of the effective response of the detector in the reference channel. The measurement uncertainties combined with the systematic errors of the theoretical model are outlined in a standard deviation of the distribution of $\sigma$ = 730 DN/(10$^{-11}$ Jm$^{-2}$nm$^{-1})$. This is an upper limit of the measurement uncertainties, and the value is consistent with the condition imposed on the lower limit of the S/N ratio.

The lower panel of Fig.~\ref{fig:detections_fit} compares the measured channel-normalized incoming signal with the convolved stellar irradiance derived from the HYG catalog. Because low-signal stars are strongly affected by readout noise, the regression is performed with weights proportional to the inverse signal, thereby limiting the influence of noisy data points. The resulting weighted fit yields a correlation coefficient of $R^2=0.926$, indicating good agreement between the instrument response and catalog-based expectations. Also, this validates the use of a single linear sensitivity coefficient for the reference channel. Neither gain compression nor non-linear digitization effects play a significant role within the flux range sampled by the stellar fields data. The high-signal stars align closely with the fitted relation. 

The model uncertainties are not negligible when deriving the reference-channel flux using a simple Planck-law approximation. The cool giants (e.g. M2–M3) are an illustrative example for this kind of error as the real spectral energy distribution departs significantly from a blackbody due to the complex molecular opacity in their atmospheres~\citep{1994A&AS..105..311F}. As a result, the calculations can be biased depending on how the real spectrum is shaped by absorption features. These biases lead to large sensitivity values as one can see in Table~\ref{tab:stellar_obs} for $o$ Ori and HD 56618, which are red giants on the asymptotic giant branch of spectral type M3Sv and M2III, respectively. In their case, the model spectral density across the visible band in underestimated. This makes Eq.~\eqref{eq:jlambda} unreliable, leading to sensitivity values around $\mu+2.5\sigma$ and $\mu+7\sigma$. On the other hand, $\delta$ CMa (Wezen) and $\alpha$ Lyr (Vega) exhibit an almost featureless spectrum. Thus, the model errors are much smaller. This is confirmed by the calculated sensitivity values, which are much closer to the distribution parameter $\mu$, inside $\pm\frac{1}{3}\sigma$.

\begin{figure}
    \centering
    \includegraphics[width=0.49
    \textwidth]{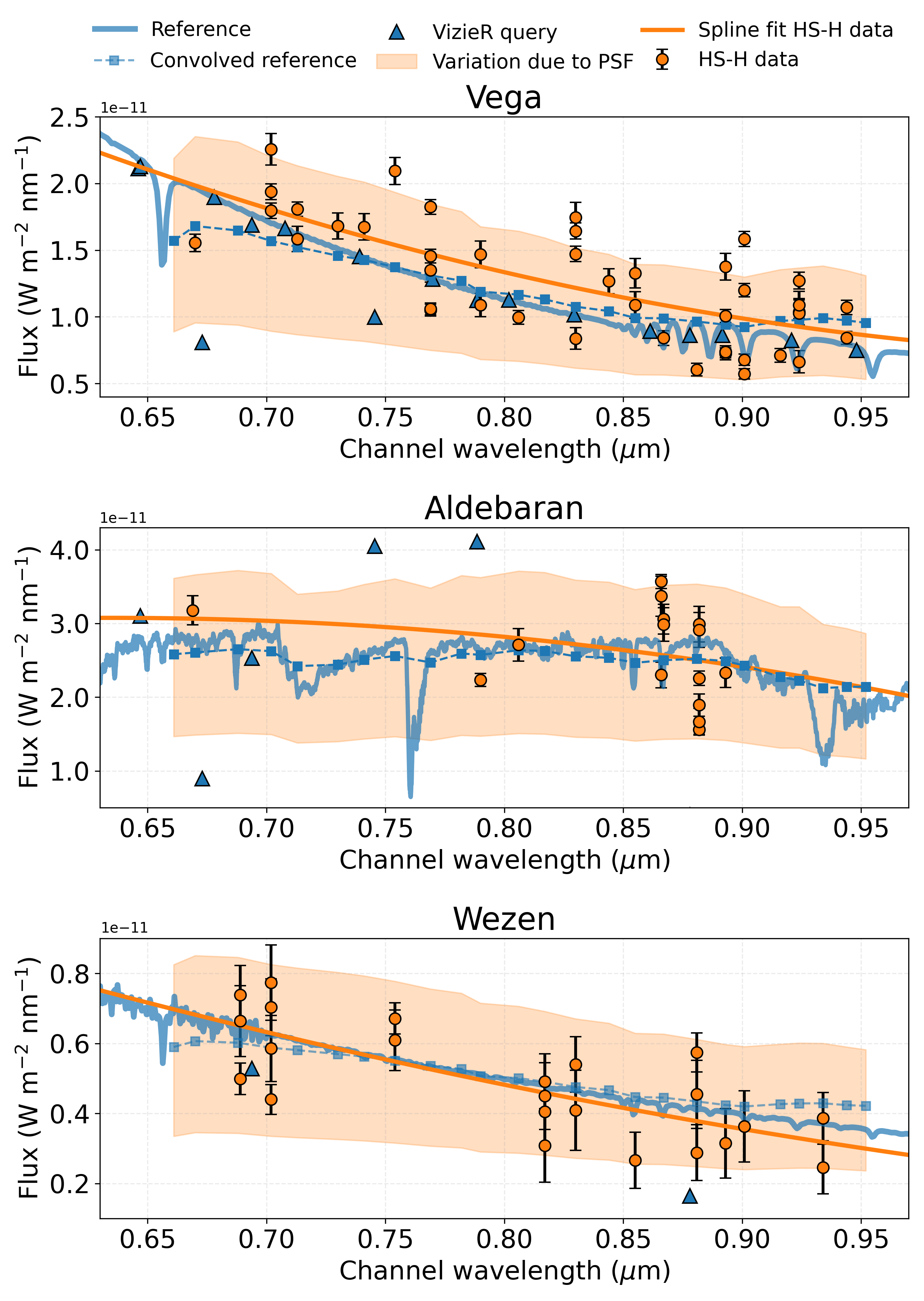}
    \caption{Comparison between the measured HS-H channel fluxes (data points along with error bars) and reference spectra stars: Vega (A0V), Aldebaran (K5 III), and Wezen (F8 Ia). The lines show the high-resolution reference spectrum, and the dashed lines with squared data points show the same spectrum after convolution with the HS-H transfer functions. The triangles indicate independent photometric fluxes retrieved from VizieR service. The featureless curves represent spline fits of the HS-H observations, and the shaded bands illustrate the expected variation due to the PSF distribution. All spectra are shown as flux density versus the effective wavelength of each HS-H channel. Aldebaran spectrum is smoothed only for improving visualization}.
    \label{fig:stellar_spectra}
\end{figure}

The spectral fluxes shown in Fig.~\ref{fig:stellar_spectra} were obtained using only the detector pixel located at the maximum of each stellar PSF. This “peak-pixel” allows a direct comparison between detected signals and real physical spectra. The DN values measured at the PSF peak were converted into spectral flux density using the radiometric constant derived earlier from the Gaussian fit to the sensitivity distribution, specifically the mean value $\mu$. The HS-H measurements closely follow the convolved reference spectra, validating the updated spectral and photometric calibration.

To robustly evaluate the instrument's spectral response, we compared our reconstructed stellar spectra against multiple data sources. The spectrum of \textit{$\delta$ CMa} was estimated as the spectrum of a standard star with a similar spectral type, C26202, retrieved from the CALSPEC database ~\citep{2014PASP..126..711B}. Data products including spectra of \textit{$\alpha$ Tau} ~\citep{1977RMxAA...2...71J} and  \textit{$\alpha$ Lyr} \citep{Rieke2011AbsoluteSpectrum} were retrieved with the help of VizieR catalog ~\citep{vizier}. Additional photometric data were retrieved using the same service. 

The spectrum of C26202 was scaled according to the difference between the visual magnitudes of the two stars. No scaling was applied to the spectra of \textit{$\alpha$ Tau} and \textit{$\alpha$ Lyr}. The spectrum of Aldebaran was smoothed through a rolling average (boxcar filter with a window size of 301 data points) with the sole purpose of improving data visualization.

High-resolution spectroscopic references were utilized to establish a precise ground truth, while photometric spectral energy distribution data retrieved via VizieR provided a necessary baseline for understanding uncertainties inherent to broad-bandpass measurements. 

The shaded PSF band in Fig.~\ref{fig:stellar_spectra} represents the range of possible flux values associated with variations in the PSF across the detector. Because the width and shape of the PSF change slightly depending on the exact landing position of the diffraction figure on the detector chip, the fraction of the total stellar flux that falls into the peak pixel is not constant. PSF simulations show that the central pixel captures between 5.4\% and 13.3\% of the total incoming flux, with an average of 9.4\%. This spread directly translates into an uncertainty on the expected peak-pixel flux, which is visualized as the shaded band in the plot. 

While our empirical data generally trace the theoretical model we developed, some reconstructed points fall outside the shaded variance regions. This indicates that our initial error estimations, which accounted only for Poisson noise, dark current, and read-out noise, slightly underestimate the total uncertainties. These deviations are likely driven by unmodeled factors such as random noise from temporary warm pixels, background fluctuations (which particularly impact the low signal-to-noise observations of Vega at 0.5~s and Wezen), and residual systematic errors in the PSF modeling.

As HS-H is based on filter with a significant band width, it lacks the spectral resolution to fully resolve narrow stellar features, such as the deep absorption band in Aldebaran's spectrum. As demonstrated by the HSH-convolved theoretical spectrum included in Fig.~\ref{fig:stellar_spectra}, the instrument registers only a slight flux drop in this region, comparable to the local noise level. This challenge of reconstructing a spectrum from a single pixel across multiple frames served as a rigorous stress test that strongly validated our calibration pipeline. Future observations of the extended mission target will be comparatively straightforward, as they will not be required to account for such complex PSF corrections.

\subsubsection{Mars cross-calibration}

The accuracy of stellar calibration is insufficient for characterizing the fine features of asteroid spectra. Thus, Mars cross-calibration is crucial to better characterize the relative gains of HS-H channels. We are performing a cross-calibration of the laboratory-based radiometric calibration of the HS-H instrument by comparing it with Mars surface data acquired during the cruise phase. Specifically, we have analyzed reflectance measurements from several regions such as Huygens, Schiaparelli, Schroeter craters or brighter terrains of the Mars surface. The spectral data acquired from such images have been compared to CRISM spectral data~\citep{CRISM} to validate and refine the in-flight calibration.

For each set of CRISM measurements, we retrieve manually the median spectra from each HS-H exposure within a ROI that has the same spread as the area from which CRISM data was gathered. The official interactive map of high-resolution observations was used to retrieve the data~\footnote{\url{http://crism-map.jhuapl.edu/}}. To compare the data, we compute the response of CRISM irradiance spectra within the HS – H filters. However, CRISM data is accurately provided only for the spectral range above 0.70~$\mu$m. So, we fit a cubic spline over the CRISM irradiance spectrum, extrapolate the missing interval between 0.65~$\mu$m and 0.70~$\mu$m, and convolve with HS – H transmission functions. For this comparison, we did not take into account the HS – H channels with wavelengths under  0.70~$\mu$m to avoid artifacts. Validation against observed spectra indicates a maximum extrapolation error of 5\%. Because the out-of-band contribution in the 0.65--0.70~$\mu$m range represents a minor fraction of the total cumulative transfer function of each channel (peaking at $\sim$30\% for the 0.702~$\mu$m channel and falling to 2--10\% for longer wavelengths), the systematic error induced by this extrapolation is strictly bounded at $\sim$1.5\% for the worst-case channel and $<$0.5\% for all others. These uncertainties have been included in the sensitivity error calculations.

\begin{figure}
    \centering
    \includegraphics[width=0.49
    \textwidth]{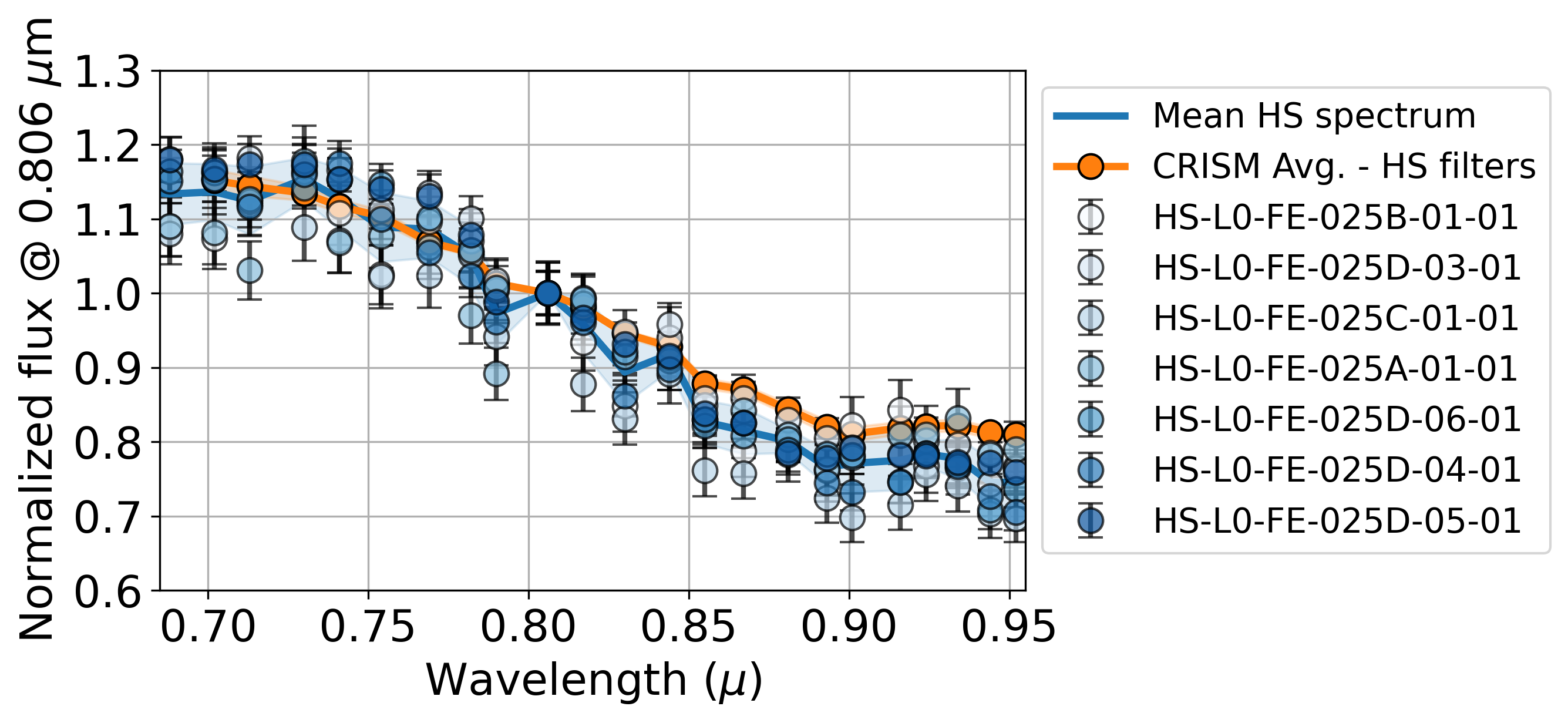}
    \caption{Comparison between the mean HS-H reflectance spectrum of a subregion of Huygens crater (line) and the CRISM average spectrum convolved with the HS-H filter responses (line with data points). Individual HS-H observations from different Mars frames are shown as semi-transparent points with error bars, each corresponding to a distinct pointing of Mars and phase angle. All spectra are normalized at 0.806 $\mu$m. }
    \label{fig:huygens_spectra}
\end{figure}

An example of the comparison between HS-H observed and CRISM spectra is illustrated in Fig.~\ref{fig:huygens_spectra}, where we analyze one subregion of the Huygens crater. The figure presents normalized irradiance spectra from multiple HS-H exposures within the selected ROI, alongside the corresponding HS-H convolved CRISM spectrum. The results demonstrate a good agreement in spectral shape between multiple HS-H measurements. However, we observe systematic offsets between CRISM and HS-H data that are consistently present across the spectral comparisons derived for all the regions we have investigated so far. These discrepancies guide further refinement of the radiometric calibration by adjusting the gains for each filter. 

\begin{figure}
    \centering
    \includegraphics[width=0.49
    \textwidth]{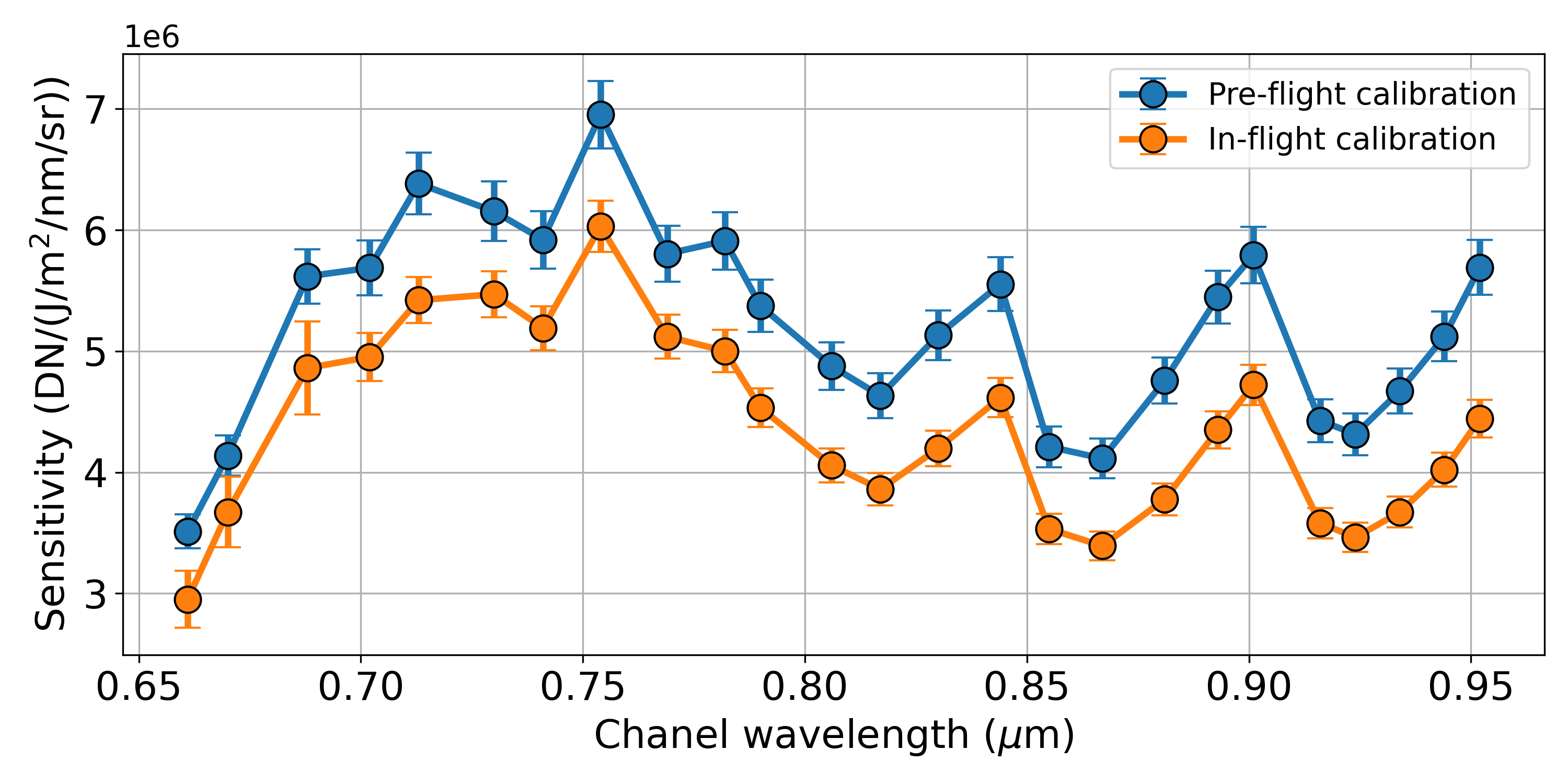}
    \caption{Comparison between the pre-flight and in-flight radiometric sensitivity curves of HS-H as a function of channel wavelength. The larger in-flight sensitivities are derived from Mars surface observations acquired during the cruise phase, while pre-flight values are retrieved from laboratory calibration.}
    \label{fig:sensitivity}
\end{figure}

To refine the radiometric calibration, we first average the HS-H observations within each region of interest to reduce random noise and increase the S/N ratio. We then compute the ratio between these averaged spectra and the corresponding CRISM-derived spectra (convolved with the HS-H spectral responses). These ratios define a preliminary set of radiometric correction coefficients across the given spectral range. Fig.~\ref{fig:sensitivity} shows the resulting channel-by-channel sensitivity curve obtained from the in-flight calibration, compared directly with the pre-flight laboratory measurements. The detailed values for the in-flight sensitivity factors, along with their associated uncertainties, are listed in Fig.~\ref{tab:hsh_inflight_sensitivities}. The in-flight spectral responses take into account the radiometric constant determined from stellar observations. The relative gain factors across the spectral channels are determined exclusively from Mars observations.

Direct cross-calibration was not possible for the three shortest-wavelength channels ($< 0.70~\mu$m) due to a lack of spectral overlap. We estimated the values by linearly extrapolating the derived correction coefficients, defined as the channel-averaged ratios between the CRISM-convolved reflectances and the corresponding HS-H reflectances. , the extrapolation introduces an additiona uncertainty for these three channels, which has been incorporated into their final errors. 

Comparison of the calibration datasets indicates that the in-flight spectral response was altered by the lower operating temperatures in space. This shift in the thermal environment caused channel-dependent sensitivity deviations of approximately 10\%. These deviations are systematic and exceed the calibration uncertainties estimated during ground-based testing.

The observed deviations of the sensitivity with respect to the pre-flight baseline are likely driven by a combination of two physical mechanisms. The first potential cause is the temperature dependence of the silicon detector's band-to-band absorption coefficient~\citep{sturm1992silicon, nguyen2014temperature}. At colder operating temperatures, the semiconductor lattice contracts and the bandgap widens, which reduces the CMOS detector's intrinsic quantum efficiency~\citep{velichko2019intrinsic}. The second hypothesis is that temperature variations affect the transmission profiles of the monolithic Fabry--Pérot filters of HS-H because of the thermal contraction of the filter substrate and temperature-dependent refractive indices, which can alter the optical behaviour of the narrow-bandpass filters~\citep{Takashashi:95}. So far, these coupled effects can be quantified holistically by cross-calibrating our Mars surface data with those from other instruments. To decouple and characterize these deviations, additional laboratory tests are intended to be conducted in the future using a sensor identical to that of HS-H.

\subsubsection{Spectral behaviour of different Sun literature spectra}

To investigate the potential origin of the systematic correction coefficients observed in our radiometric calibration, we examined whether uncertainties in the solar reference spectrum used in our pipeline could contribute to the discrepancy. We compared our adopted baseline solar spectrum~\citep{2001AJ....122.2118B} to several alternative literature spectra, including the Kurucz model~\citep{1993yCat.6039....0K}, Air Mass 0 and 1.5 solar analogs~\citep{2004SoEn...76..423G} and MODTRAN extraterrestrial solar spectra~\citep{osti_5671682} that include multiple versions based on different source combinations. Each spectrum was normalized and convolved through the HS-H filter responses to isolate spectral shape differences within HS–H operational range.

We measured deviations from the baseline spectrum and they remained below $2\%$, and most spectra stayed within $\pm 1\%$ of the baseline. This analysis indicates that differences among solar reference spectra are too small to account for the magnitude of the correction coefficients derived from cross-checking with CRISM data.

Consequently, we retain the \citet{2001AJ....122.2118B} spectrum as our radiometric baseline. The band-integrated solar irradiance values derived from this spectrum, which serve as the basis for our Reflectance factor ($I/F$) conversion, are detailed in Table~\ref{tab:hsh_inflight_sensitivities}.

\begin{table}[ht]
    \centering
    \begin{tabular}{c c c}
    \hline 
    \textbf{Wavelength} & \textbf{Sensitivity} & \textbf{Solar Irradiance} \\
    $\mu$m & $10^6$~DN/(J/(m$^{2}$sr\ nm)) & W m$^{-2}$ nm$^{-1}$ \\ 
    \hline
    0.661 & $2.95 \pm 0.16$ & 1.324 \\
    0.670 & $3.67 \pm 0.19$ & 1.375 \\
    0.688 & $4.86 \pm 0.26$ & 1.364 \\
    0.702 & $4.95 \pm 0.11$ & 1.323 \\
    0.713 & $5.42 \pm 0.09$ & 1.306 \\
    0.730 & $5.47 \pm 0.09$ & 1.290 \\
    0.741 & $5.19 \pm 0.09$ & 1.272 \\
    0.754 & $6.03 \pm 0.10$ & 1.251 \\
    0.769 & $5.12 \pm 0.09$ & 1.214 \\
    0.782 & $5.00 \pm 0.09$ & 1.197 \\
    0.790 & $4.53 \pm 0.08$ & 1.153 \\
    0.806 & $4.06 \pm 0.07$ & 1.138 \\
    0.817 & $3.86 \pm 0.07$ & 1.116 \\
    0.830 & $4.20 \pm 0.07$ & 1.083 \\
    0.844 & $4.62 \pm 0.08$ & 1.063 \\
    0.855 & $3.53 \pm 0.06$ & 1.012 \\
    0.867 & $3.39 \pm 0.06$ & 1.009 \\
    0.881 & $3.78 \pm 0.07$ & 0.983 \\
    0.893 & $4.35 \pm 0.08$ & 0.959 \\
    0.901 & $4.72 \pm 0.08$ & 0.948 \\
    0.916 & $3.58 \pm 0.06$ & 0.956 \\
    0.924 & $3.46 \pm 0.06$ & 0.963 \\
    0.934 & $3.67 \pm 0.06$ & 0.962 \\
    0.944 & $4.02 \pm 0.07$ & 0.945 \\
    0.952 & $4.44 \pm 0.08$ & 0.933 \\
    \hline
    \end{tabular}
    \caption{In-flight HS-H radiometric parameters. The table lists the effective wavelength, the measured sensitivity with associated uncertainties, and the band-averaged solar irradiance at 1 au derived from the \citet{2001AJ....122.2118B} solar spectrum used to normalize radiance to reflectance ($I/F$).}
    \label{tab:hsh_inflight_sensitivities}
\end{table}

\subsection{Geometric calibration}
 
 The telescope within the HS-H instrument follows a three-mirror anastigmat configuration. This design is subject to barrel distortion, a phenomenon in which the focal length decreases towards the edges of the field.  Another effect observed is related to the optical anisotropy; the axes X and Y of the detector are characterized by different pixel scales due to focal length variation. Together, these effects cause the image to appear slightly stretched in one direction and curved toward the center, requiring correction through anisotropic scaling and distortion modeling.
 
 An estimate of the instrument's FoV was obtained in the laboratory using backward ray tracing, which involved constraining the position of the spot on the image plane and measuring the corresponding field angle. This method determined that the largest rectangular FoV observed is $15.9^\circ \times 9.9^\circ$. These effects are characterized below using in-flight observations of various star field images.

\subsubsection{Alignment with the spacecraft}
 
The instrument kernel provided by ESA SPICE incorporates the FoV of HS-H. Several star-field frames were acquired to test the alignment of the instrument. It was identified a boresight misalignment that generates pointing errors of up to 100 pixels. The cause of these errors is the offset between the actual optical axis (as followed by the instrument in flight) and the nominal direction defined in the instrument kernel.

The boresight misalignment was estimated following an optimization approach to minimize the pointing offset by reducing the mean squared error between the observed pixel coordinates of the stars and those obtained from SPICE (after converting the celestial equatorial coordinates to pixel coordinates). The alignment corrections derived from this process do not take into account the effects of distortion. However, they significantly improve the precision of the pointing.

The celestial equatorial coordinates, right ascension and declination, are included in the header of the HS-H images, and they correspond to the location of the spacecraft pointing.  The optimized projection computed for these coordinates is at pixel $(529, 1115)$.  On the other hand, the camera pointing corresponds to the center of the image at pixel $(x_0, y_0)=(544, 1024).$


\subsubsection{Optical anisotropy}

The average focal lengths along the X and Y axes differ, indicating anisotropy in the optical system, i.e. the image scale is different in the horizontal and vertical directions. 

The in-flight focal length of the imaging system can be derived from the astrometric World Coordinate System (WCS) fitted to the star field observations corrected for barrel distortion, such that WCS worked with undistorted coordinates only, $(x_u, y_u)$. For this procedure, we used only the star fields with more than 3 stars detected, meaning a total of 31 frames. 

Plate scales $s_X=28.3$ arcseconds/px and $s_Y=34.0$ arcseconds/px were calculated as Euclidean norms of the respective column vectors of the WCS CD matrix~\citep{2005ASPC..347..491S}. As barrel distortion was corrected, these scales will correspond to focal lengths at the centers of the X and Y axes, in the paraxial approximation, $f=\frac{l_{px}}{pixel \ scale}$, where $l_{px}=5.5 \ \mu m$ is the pixel size. Thus, we get $f_X=39.9 \pm 0.1$ mm and $f_Y=33.27 \pm 0.05$ mm. These values are in very good agreement with the measurements provided by the instrument manufacturer. The FoV without any distortion would be in this case $10.2^\circ \times 16.0^\circ$. Additional mathematical details are provided in \ref{app:processing-geometric}.

\subsubsection{Barrel distortion}
The focal length varies with field position along both axes, which is characteristic of radial (barrel) distortion: points farther from the optical center are projected closer to the center than they would be in an ideal pinhole camera. 

\begin{figure}
    \centering
    \includegraphics[width=0.95\linewidth]{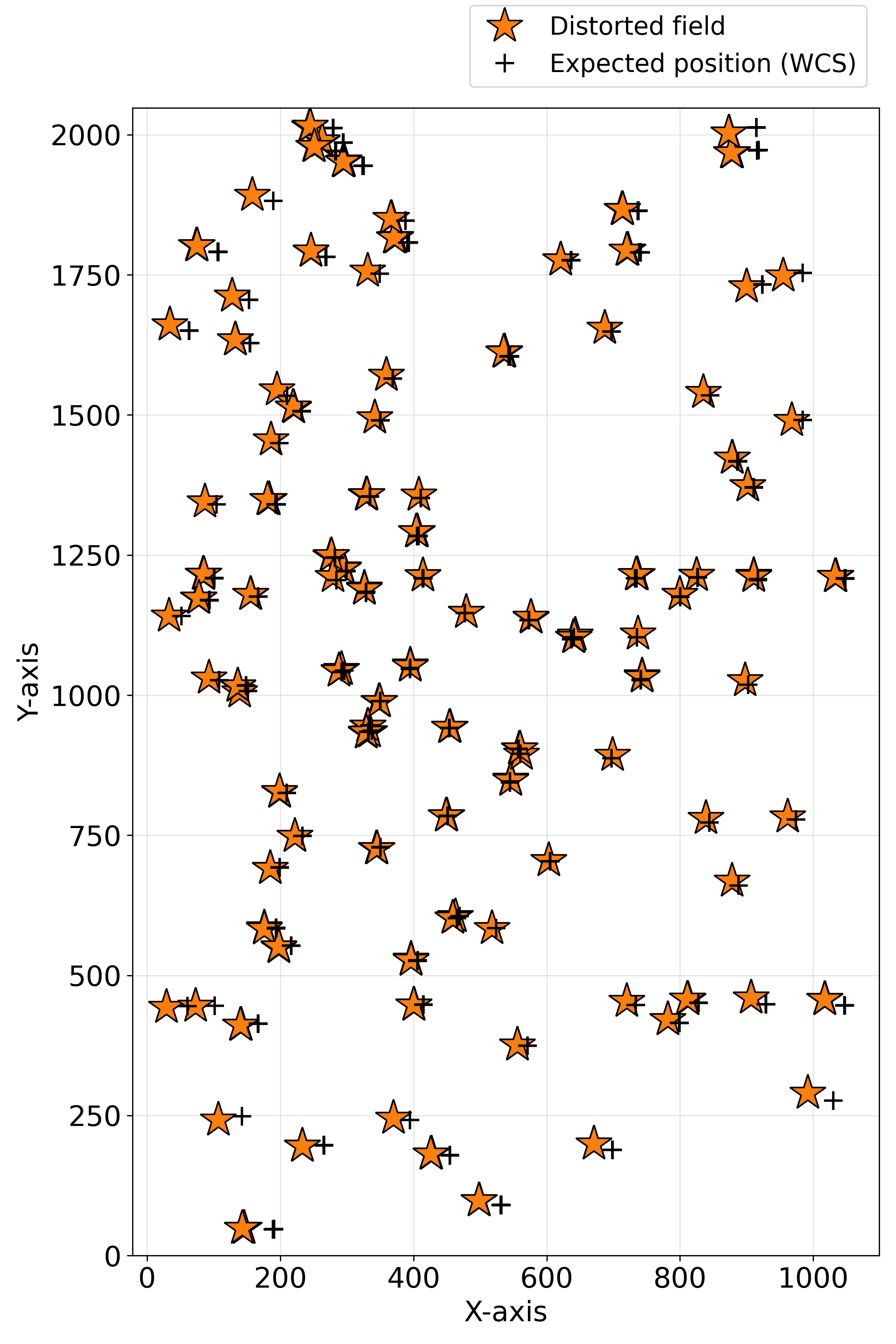}
    \caption{Comparison between the distorted stellar field (star signs) and the expected undistorted positions derived from the WCS solution (cross signs).}
    \label{fig:geom_star_pos}
\end{figure}

A single star field frame is not sufficient to evaluate the field distortion because the number of detected stars used to derive the correction polynomials would be low, providing only one or two reference points at the edges of the field. To address this, we combine the data retrieved from multiple frames and evaluate the distortion once. A total of 312 detections of stars were made across 71 frames. Each star was identified using the HYG catalog, as described in Section~\ref{data-ss}. The detections cover most of the  field of view (FoV), as illustrated in Fig.~\ref{fig:geom_star_pos}. Future observations of stellar fields will further extend this spatial coverage.

\begin{figure}[ht]
    \centering
    \includegraphics[width=\linewidth]{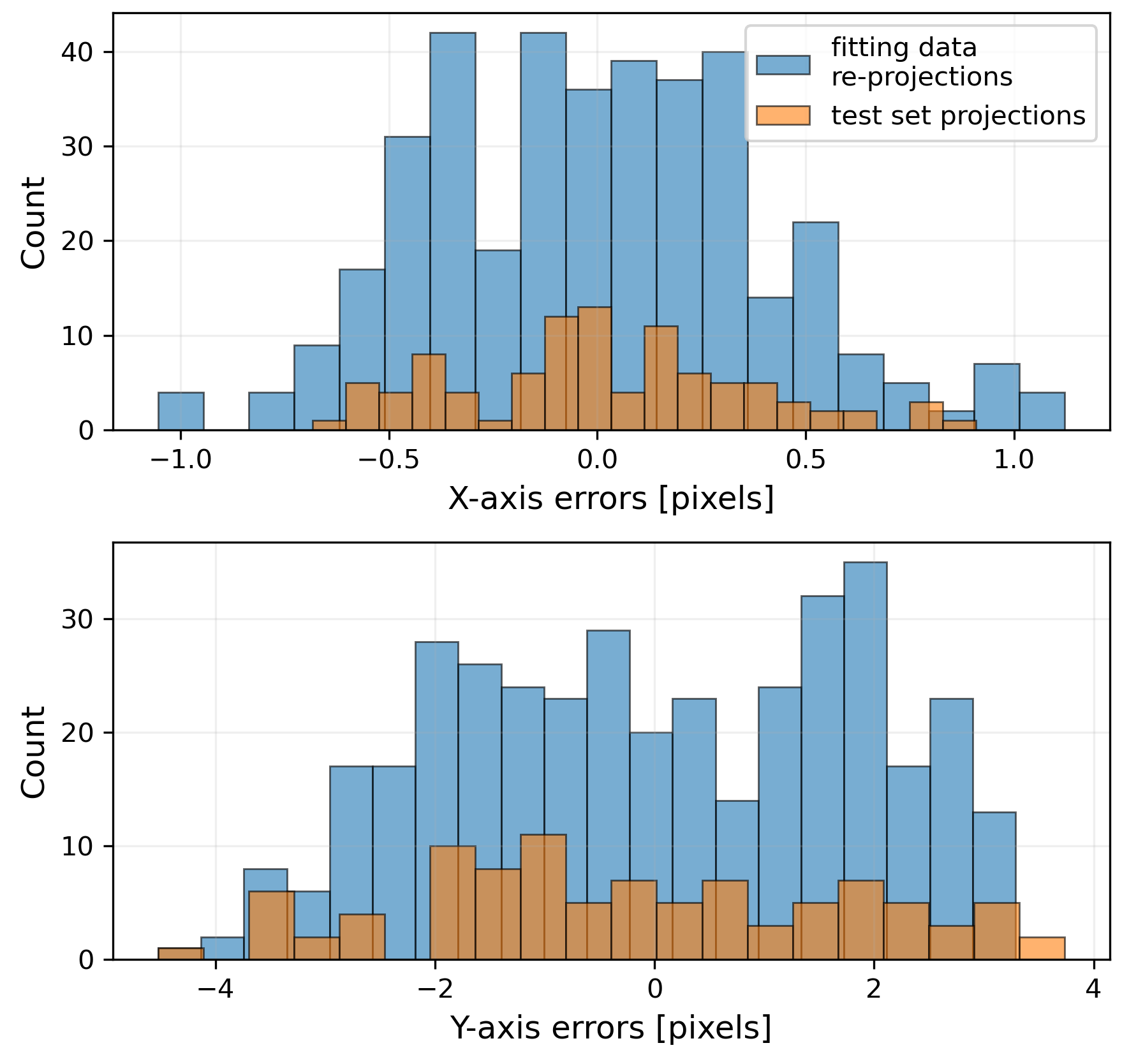}
    \caption{Histograms of residuals from the in-flight geometric calibration model. The larger bars represent the residuals for the data used in model fitting, while the lower bars show the distribution for the independent test set. The top panel corresponds to X-axis errors, and the bottom panel to Y-axis errors.}
    \label{fig:hist_residuals}
\end{figure}

To characterize and correct the optical distortion, a two-dimensional third-order polynomial was fitted to the displacement field between the distorted and undistorted coordinates. The model was fitted using $80\%$ of the data, while the remaining $20\%$ was reserved for testing.  The resulting distortion model was subsequently applied to compute the corrected (undistorted) pixel positions of the testing subset of $20\%$ detections. Therefore, we evaluated the residuals between the predicted and reference WCS positions. The corresponding distributions of the projection errors along both detector axes are presented in Fig.~\ref{fig:hist_residuals}, illustrating the projection errors for the fitting and test data. 

If barrel distortion is not taken into account, the method performs well near the center of the FOV, where position errors remain within a few pixels. However, toward the edges of the field, particularly on the left and right sides, the position errors increase significantly, reaching up to several tens of pixels. The WCS solutions are projected onto the distorted field using both the pre-flight and in-flight calibrations. 

The RMS errors derived from the test set provide an independent estimate of the geometric calibration precision. Table~\ref{tab:rms_distorsion} summarizes this evaluation on both axes due to the distortion corrections obtained from the star field images and from the laboratory checkerboard data (from pre-flight calibration). The data set on which the in-flight distortion model has been fitted was reused to compute the re-projection errors. There is no significant difference between the RMS errors computed for the test set and those of the re-projections. Thus, the dataset was not overfit and we may assume good generalization performance. 

A better match is observed in the case of the in-flight calibration. This is also supported by the fact that the RMS residuals are much larger for the pre-flight calibration, indicating its reduced reliability under in-flight conditions.  Visual comparisons between pre-flight and in-flight distortion corrections can be seen in Fig.~\ref{fig:detections}. There we applied the corrections on the same test image, a star field acquired with a 9-second exposure time.

A slight inclination of the grid can be noticed if we apply the in-flight corrections to correct the checkerboard distortions. The checkerboard pattern appears rotated by a small angle toward the right side as seen in Fig.\ref{fig:checkerboard}. During the pre-flight calibration, it was assumed that the reference checkerboard image was observed face-on. Nonetheless, there are several plausible explanations for these differences, e.g. the checkerboard was not observed exactly face-on or minor geometric effects in the optical alignment that during flight could introduce this small tilt.

\begin{table}[ht]
\centering
\begin{tabular}{lccc}
\hline
\textbf{Calibration dataset}  & \textbf{Star fields} & \textbf{Checkerboard} \\ \hline
RMS$_X$ (px)                                  & 0.36                           & 14.9                  \\
RMS$_Y$ (px)                                   & 1.87                           & 12.3                  \\
RMS (px)                                        & 1.90                           & 19.3                  \\ \hline
\end{tabular}
\caption{Root mean square (RMS) residuals obtained when applying the pre-flight and in-flight geometric corrections on the star fields test set.}
\label{tab:rms_distorsion}
\end{table}

\begin{figure}
    \centering
    \includegraphics[width=\linewidth]{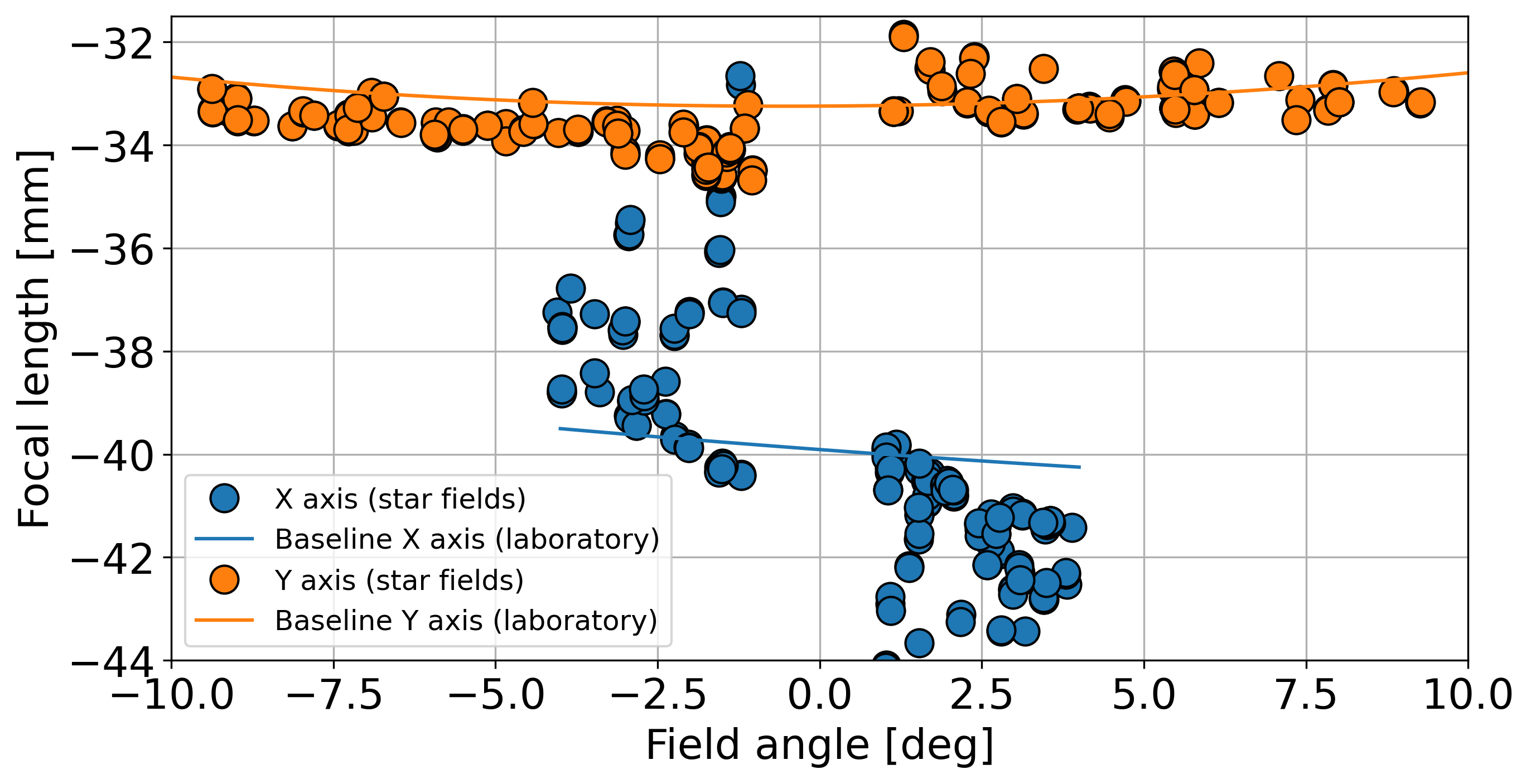}
    \caption{The focal lengths with respect to the field angle on both detector axes. The markers represent the data points determined from the star fields data, whereas the baseline data obtained in the laboratory by the instrument manufacturer is plotted with a solid line. The data points on the bottom part of the plot correspond to the field angles determined for the shorter axis, X.}
    \label{fig:focal_lengths}
\end{figure}

The focal length variations on both axes are derived from the in-flight calibration according to Eq.~\ref{eq:fX}~-~\ref{eq:thY}. For each detection, we calculated the focal length at that position on the detector where the star had been detected. In Fig.~\ref{fig:focal_lengths} is shown the dependence between the focal lengths and the field angles determined for both axes. The Y-axis focal length measurements are in agreement with the laboratory baseline (RMS error of 0.62 mm), with random scatter consistent with measurement uncertainties. However, the X-axis measurements exhibit a different field-angle dependence (RMS error of 2.57 mm), suggesting a change in the optical system. This might be related to the previous observed differences between the pre-flight and in-flight calibrations.

\section{Summary}
\label{sec:conclusions}

We presented a comprehensive in-flight calibration of the HS-H hyperspectral imager aboard ESA's Hera spacecraft, based on observations acquired after the launch in October 2024, including the Mars swing-by in March 2025 and the last observing session in September 2025. The instrument was operated at a wide range of exposure times and illumination conditions. However, HS-H works in space at very low temperatures ($\approx$-12$~^{\circ}C$) compared to the ambient temperatures at which most of the pre-flight calibration experiments had taken place. Thus, the in-flight assessment of HS-H performance is required to update the calibration parameters.

The detector bias structure remained highly stable over the cruise of HS-H. The median bias level fluctuated only between 144~–~156~DN, with no evidence of long-term structural changes. A small number of bad pixels were identified, including three permanently corrupted rows of pixels. The bias distribution and behaviour were stable, and the detector's readout noise was measured to be roughly 8~DN, consistent with expectations. Short-exposure dark frames show dark current to be negligible compared to bias fluctuations, and no correlation with temperature or exposure time was observed for exposures below 100 ms. Longer dark-like frames reveal faint horizontal banding and low-amplitude structures, but their contribution remains minor.

The Earth–Moon observations and stellar exposures confirm that the detector response is linear across the unsaturated regime., whereas measurements from Vega exposures yield FWHM values in excellent agreement with pre-flight expectations (around 1.6~px). Spatial variations of the PSF across the detector spanned between 1.2~and 2.2~px and the simulated PSF models confirm that between 10~–~~26~\% of the total stellar flux falls into the peak pixel as validated by the observed stellar profiles. 

The distorted field of HS-H was assessed using star-field observations acquired at multiple pointings and exposure times. A total of 71 frames were processed, enabling the identification of 312 stellar detections after reliable SNR filtering. These measurements allowed us to quantify the optical distortion across the FoV and verify the camera alignment. The RMS residuals of the distortion model were quantified at 0.36~px on X-axis and 1.87~px on Y-axis.

Using 367 stellar profiles, we derived the first in-flight sensitivity estimate for the reference channel at effective wavelength of 0.806 $\mu$m. The distribution of measured sensitivities is well described by a Gaussian with the mean at $(3030\pm40) \times 10^{11}$ DN / ( J m$^{-2}$ nm$^{-1}$). HS-H spectral measurements agree well with catalog spectra convolved through the instrument response, validating the photometric consistency of the in-flight calibration.

Mars cross–calibration showed the differences in the values of the radiometric constants used in calibration. A first assessment of the in–flight radiometric constants has been made using several standard stars and few regions from Mars surface observed also by CRISM instrument. An extended cross–calibration process is under development and will be published in a separate article. The instrument demonstrated excellent performance under planetary-surface illumination conditions. Comparisons between HS-H reflectance spectra and CRISM data for multiple regions (e.g., Huygens, Schiaparelli, bright terrains) show consistent spectral shapes, but systematic offsets between the two datasets highlight the need to update the radiometric calibration constants.

We recommend that the newly derived in-flight calibration parameters replace the preliminary ground calibration data for all future scientific data processing. The cold space environment changed the detector's thermo-electrical behavior and the optics probably were also affected by the launch procedure. The sole exception is the flat field calibration, which must continue to rely on pre-flight laboratory data as no in-flight flat fields were acquired.

Future efforts will focus on two complementary directions. First, we will perform dedicated simulations of the final stage of the Hera mission, reproducing the geometry, illumination conditions, and surface properties expected during the close-proximity operations at the binary near-Earth asteroid (65803) Didymos. Second, the calibration products presented here will continue to be refined as new in-flight observations become available. Stellar fields, bias, and dark frames acquired throughout the cruise and approach will allow iterative improvement of the radiometric, geometric, and noise-characterization models, ensuring the highest possible accuracy of the HS-H data for the scientific phase of the mission. A calibration version that will be released following the processing of a statistically significant volume of new calibration data, or if any systematic degradation is detected.

\section{Acknowledgements}
We acknowledge the ESA/Hera Flight Control and Flight Dynamics
team for successfully implementing all the proposed requirements.
GP, MP and JdL acknowledge financial support from ESA contract
RFQ/3-18365/23/NL/GLC/my, Spain in response to Tender Action
3-18365 "Hera Investigation Team Instrument Operations and Data
Analysis Preparation". TK acknowledges the Einstitutional support RVO
67985831 of the Institute of Geology of the Czech Academy of Sciences.

\appendix
\section{Noise-free simulation of PSF and stellar profile}
\label{app:psf_sim}

\begin{figure*}
    \centering
    \includegraphics[width=0.98
    \textwidth]{"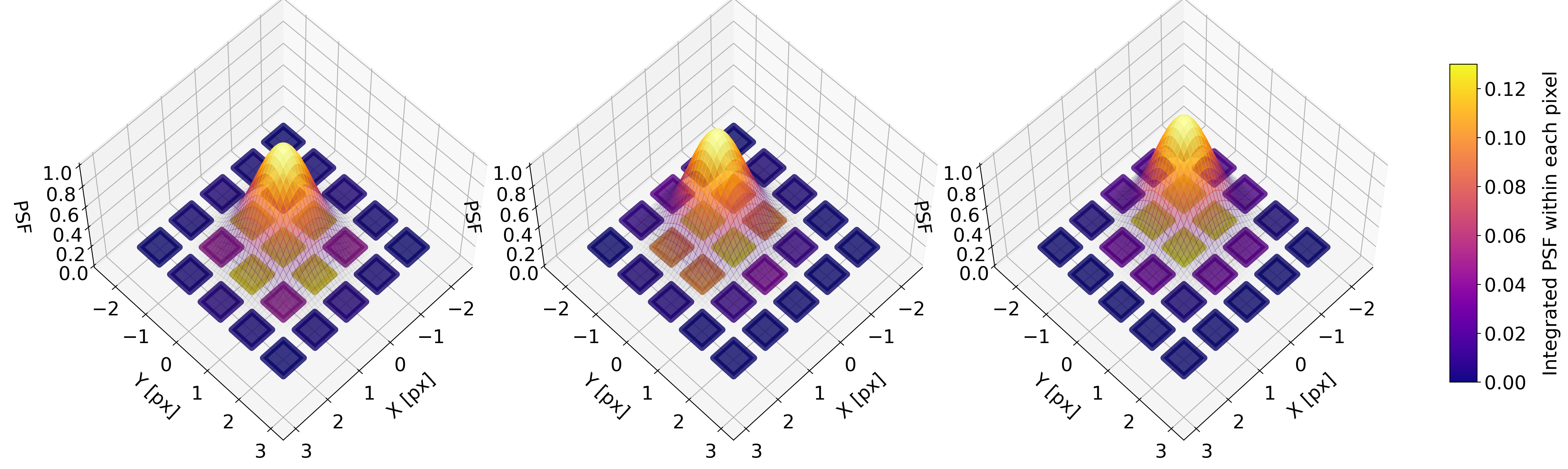"}
    \caption{Simulated PSF flux distribution spread on a detector grid geometry characterized by a 42\% filling factor. Three scenarios illustrate the impact of PSF centroid position on the distribution of the collected signal per pixel: peak centered on a pixel (left), peak between two pixels along one axis (middle), and peak at the intersection of four pixels (right). The detector pixels are color-coded by the fraction of total PSF flux collected (colorbar, right), while the semi-transparent surface represents the normalized 2D Gaussian PSF (FWHM of 1.6~pixels). Gaps between pixels represent inactive detector area. The centroid position significantly affects the flux distribution and total collected signal, with maximum collection efficiency achieved when the PSF is centered on a pixel.}
    \label{fig:psf_dist}
\end{figure*}

Here we present the simulation method we used to quantify the effects of the PSF on the stellar profiles detected by HS-H. One can simulate this type of profile by centering the PSF distribution at different locations inside a pixel area and numerically integrating the PSF distribution in the rectangular domains bound by each pixel taking into account the filling factor of the instrument (42\% for HS-H). A 2D Gaussian PSF with FWHM between 1.2 and 2.2~pixels was modeled, and the fraction of total PSF flux collected by each active detector pixel was determined for three illustrative centroid positions: (i) peak centered on a pixel center, (ii) peak positioned between two pixels along one axis, and (iii) peak positioned at the intersection of four pixels. The results are visualized as 3D surface plots in Fig.\ref{fig:psf_dist}. The model assumes gaps, i.e. inactive detector area, between the squared pixels. These simulations demonstrate that the centroid position significantly affects the spatial flux distribution. However, the total collected signal is slightly affected by the centroid position, varying between 40.5\% and 41.5\% of the total expected flux. 

The results show that 10.8~–~26.6~\% of the flux is distributed to the central pixel, depending on the FWHM and the position of the PSF centroid on the detector. Also, at least 98.5~\% of the flux is collected within a $7 \times 7$ square aperture, whereas 100\% of the flux is contained in a $9 \times 9$ square aperture (with the central pixel containing the peak of the PSF).

The normalized stellar spectrum values, $\hat{j}_\chi$ with $\chi$ from 1 to 25, were evaluated with the help of Planck's radiation law, temperature and spectro-photometric data provided by the HYG catalog (literature spectra were used for a few stars such as Vega, Aldebaran and Wezen). The blackbody radiation curve is convolved using the HS-H transmission functions. Using the catalog data, the effective stellar temperature, $T_{\text{eff}}$, is estimated from the star's color index, B-V, using the empirical Ballesteros' formula:
\begin{equation}
    T_{\text{eff}} = 4600 \ \text{K} \left( \frac{1}{0.92 \cdot \text{(B-V)} + 1.7} + \frac{1}{0.92 \cdot \text{(B-V)} + 0.62} \right)
\end{equation}

The star's radiance spectrum, $B_{\lambda}$, is approximated using the Planck function for the estimated temperature $T_{\text{eff}}$,

\begin{equation}
B_{\lambda}(\lambda, T) = \frac{2 h c^2}{\lambda^5 \left( e^{\frac{h c}{\lambda k T}} - 1 \right)}
\end{equation}

where $h$ is the Planck constant, $c$ is the speed of light, and $k$ is the Boltzmann constant. The blackbody spectrum is scaled to the true absolute flux received at Earth by using the known visual magnitude, $V_{\text{mag}}$. The total flux density in the V-band, $F_V$, is determined using Pogson's ratio relative to the V-band zero-point flux density, $F_{V, \text{ZP}} = 3.67 \times 10^{-11} \text{ W}\cdot\text{m}^{-2}\cdot\text{nm}^{-1}$~\citep{Colina1996Vega},
\begin{equation}
F_{V} = F_{V, \text{ZP}} \cdot 10^{-0.4 \cdot V_{\text{mag}}}
\end{equation}
Then, $B_{\lambda}$ is integrated over the V-band wavelength range to find the flux of the blackbody spectrum in the V-band:
\begin{equation}
F_{\text{BB}, V}/\Omega = \frac{1}{\Delta\lambda_V} \int_{V} B_{\lambda}(\lambda, T) \, d\lambda
\end{equation}
where $\Omega$ is the solid angle within the star is observed. The entire Planck spectrum is scaled by the ratio of the true V-band flux, $F_V$, to the blackbody V-band flux, $F_{\text{BB}, V}$,
\begin{equation}
j_{\lambda} = \frac{F_V}{F_{\text{BB}, V}}  B_{\lambda} \Omega
\label{eq:jlambda}
\end{equation}
The obtained irradiance spectrum $j_{\lambda}$ is convolved with the HS-H filter transfer functions, $\tau_{\chi,\lambda}$,
\begin{equation}
j_\chi = \frac{\int j_{\lambda} \tau_{\chi,\lambda} \lambda  d\lambda}{\int  \tau_{\chi,\lambda} \lambda \, d\lambda} 
\end{equation}
and then normalized to the median channel, $\hat{j}_\chi=j_\chi/j_{13}$. Further, the channel-normalized incoming signal is determined as
\begin{equation}
I_{ref} = \sum_{\mathcal{A}}\frac{I(x,y)}{\hat{j}_{ch(x,y)}\hat{S}_{ch(x,y)}}
\end{equation}
where $\mathcal{A}$ is the extracted aperture area, $(x,y)$ the position on the detector of a pixel inside $\mathcal{A}$, $ch(x,y)$ is the channel corresponding to the pixel at $(x,y)$, and $I(x,y)$ the corresponding DN value. The relative sensitivity constant of channel $ch(x,y)$ is $\hat{S}_{ch(x,y)}$ with $\hat{S}_{12}=1$.

To evaluate the PSF values, we apply a pixel-wise division. For instance, let us take an arbitrary pixel at $(x,y)$. Then, we estimate the PSF value as the ratio between the observed signal within that pixel, normalized to the reference channel, and the channel-normalized incoming signal,
\begin{equation}
    PSF(x,y)=\frac{I(x,y)/I_{ref}}{\hat{j}_{ch(x,y)}\hat{S}_{ch(x,y)}}
\end{equation}

\section{Geometric calibration}

\label{app:processing-geometric}
The right ascension and declination of detected stars were used to determine a tangent-plane projection (TAN) via the Astropy \text{fit\_wcs\_from\_points} routine. The WCS solution provides the pixel scale along each detector axis through the elements of the linear transformation matrix, which convert pixel coordinates to angular units on the sky. 

To assess the variation of the focal length along the two optical axes, we consider the angular projections $(\theta_X,\theta_Y)$ on both axes, measured between the observed position of the stars, $(x_d,y_d)$, and the optical center of the image located at $(x_c,y_c)$. Then, the focal lengths read
\begin{align}
    f_X(\theta_Y)&=\frac{(y_d - y_c)l_{px}}{\tan \theta_Y}& \nonumber \\
    f_Y(\theta_X)&=\frac{(x_d - x_c)l_{px}}{\tan \theta_X}& \label{eq:fX}
\end{align}

The expected FoV without any distortion can be computed directly using the following equations,
\begin{align}
FoV_X &= 2 \tan^{-1}\Big{(}\frac{n_X l_{px}}{2f_Y}\Big{)} \\
FoV_Y &= 2 \tan^{-1}\Big{(}\frac{n_Y l_{px}}{2f_X}\Big{)}
\end{align}
where $n_X=1088$ and $n_Y=2048$ are the number of pixels along each axis. 

The angular projections are easily retrieved with WCS,
\begin{align}
    \theta_X &= (x_u-x_c)s_X& \nonumber \\
    \theta_Y &= (y_u-y_c)s_Y& \label{eq:thY}
\end{align}
Hence, if the distortion mapping $x_d \rightarrow x_u$ is known, the focal length behaviour along that axis can be determined, and vice versa.

The pixel coordinates $(x_d, y_d)$ were extracted from the distorted field as the centroids of the PSF profiles, after normalization using the specific radiometric constant of each spectral channel. The corresponding undistorted positions $(x_u, y_u)$ were determined from the known sky coordinates (RA, Dec) of each star using the WCS solution provided by the SPICE kernels.

The distortion model is fitting two third-order polynomials, $f_x$ and $f_y$, such that 
\begin{align}
    & f_x(x_d,y_d) = \Delta x\\
    &f_y(x_d,y_d) = \Delta y
\end{align}

where $\Delta x = x_u - x_d$ and $\Delta y = y_u - y_d$ represent the distortion corrections along each axis. The polynomial coefficients were estimated through a least-squares fitting procedure using a subset of the matched star detections.

\section{Additional material}
\label{app:additional}

\begin{table*}[h!]
\centering
\begin{tabular}{l c c c c c c c c c} 
\hline 
Star ID & \# & RA & Dec & V mag & B - V & Spectral & $j_{ref}$ & $I_{ref}$ & Sensitivity \\ 
        &    & (h) & (deg) &    &       &  type   & 10$^{-2}$Wm$^{-3}$ & DN/s & DN/(10$^{-2}$Jm$^{-3}$) \\
\hline \\
$\alpha$ Lyr (Vega)       & 54 & 18.615 & 38.783  & 0.03 & -0.001 & A0V   & 1.1808 & 1529$\pm$262 & 3221$\pm$552  \\
$\delta$ CMa (Wezen)      & 27 & 7.139  & -26.393 & 1.83 & 0.671  & F8Ia  & 0.4896 & 550$\pm$94   & 2794$\pm$477  \\
HD 56618                  & 25 & 7.276  & -27.881 & 4.66 & 1.589  & M2III & 0.0782 & 148$\pm$24   & 4713$\pm$770  \\
$\eta$ CMa (Aludra)       & 20 & 7.401  & -29.303 & 2.45 & -0.083 & B5Ia  & 0.1443 & 172$\pm$51   & 2958$\pm$872  \\
$\sigma$ CMa (Unurgunite) & 16 & 7.028  & -27.934 & 3.49 & 1.729  & K4III & 0.2577 & 309$\pm$37   & 2981$\pm$357  \\
$\epsilon$ Tau (Ain)      & 16 & 4.476  & 19.180  & 3.53 & 1.014  & K0III & 0.1370 & 147$\pm$26   & 2673$\pm$480  \\
$\xi$ Pup (Azmidi)        & 15 & 7.821  & -24.859 & 3.34 & 1.218  & G6Ia  & 0.1938 & 217$\pm$27   & 2782$\pm$351  \\
$\alpha$ Tau (Aldebaran)  & 15 & 4.598  & 16.509  & 0.87 & 1.538  & K5III & 2.4606 & 3199$\pm$234 & 3233$\pm$237  \\
$\epsilon$ CMa (Adhara)   & 13 & 6.977  & -28.972 & 1.50 & -0.211 & B2II  & 0.3100 & 286$\pm$91   & 2296$\pm$732  \\
$\gamma$ Tau              & 13 & 4.329  & 15.627  & 3.65 & 0.981  & G8III & 0.1193 & 143$\pm$19   & 2989$\pm$391  \\
$o^{1}$ CMa               & 12 & 6.902  & -24.184 & 3.89 & 1.740  & K3Iab & 0.1799 & 188$\pm$40   & 2602$\pm$547  \\
1 Pup                     & 12 & 7.725  & -28.410 & 4.63 & 1.632  & K5III & 0.0833 & 113$\pm$21   & 3379$\pm$622  \\
HD 56577                  & 11 & 7.276  & -23.315 & 4.83 & 1.601  & K4III & 0.0675 & 90$\pm$10    & 3313$\pm$366  \\
$\delta^{1}$ Tau          & 10 & 4.382  & 17.542  & 3.77 & 0.983  & G8III & 0.1070 & 110$\pm$23   & 2567$\pm$524  \\
$\theta^1$ Tau            & 8  & 4.476  & 15.962  & 3.84 & 0.952  & G7III & 0.0977 & 112$\pm$27   & 2852$\pm$698  \\
$o^{1}$ Ori               & 6  & 4.875  & 14.250  & 4.71 & 1.773  & M3Sv  & 0.0868 & 223$\pm$54   & 6381$\pm$1548 \\
$\omega$ CMa              & 6  & 7.246  & -26.772 & 4.01 & -0.150 & B2IV  & 0.0324 & 60$\pm$12    & 4600$\pm$949  \\
$o^{2}$ CMa               & 6  & 7.050  & -23.833 & 3.02 & -0.077 & B3Ia  & 0.0858 & 86$\pm$27    & 2483$\pm$775  \\ 
\hline
\end{tabular}
\caption{Star identifier, number of detections (\#), coordinates, photometric properties, spectral type, expected spectral irradiance at the HS-H passband, measured count rate, and the resulting sensitivity estimate for each calibration star observed during the in-flight cruise. Data are retrieved from the HYG stellar catalogue.}
\label{tab:stellar_obs}
\end{table*}

\begin{figure*}[ht]
    \centering
    \begin{subfigure}[b]{0.32\textwidth}
        \centering
        \includegraphics[width=\textwidth]{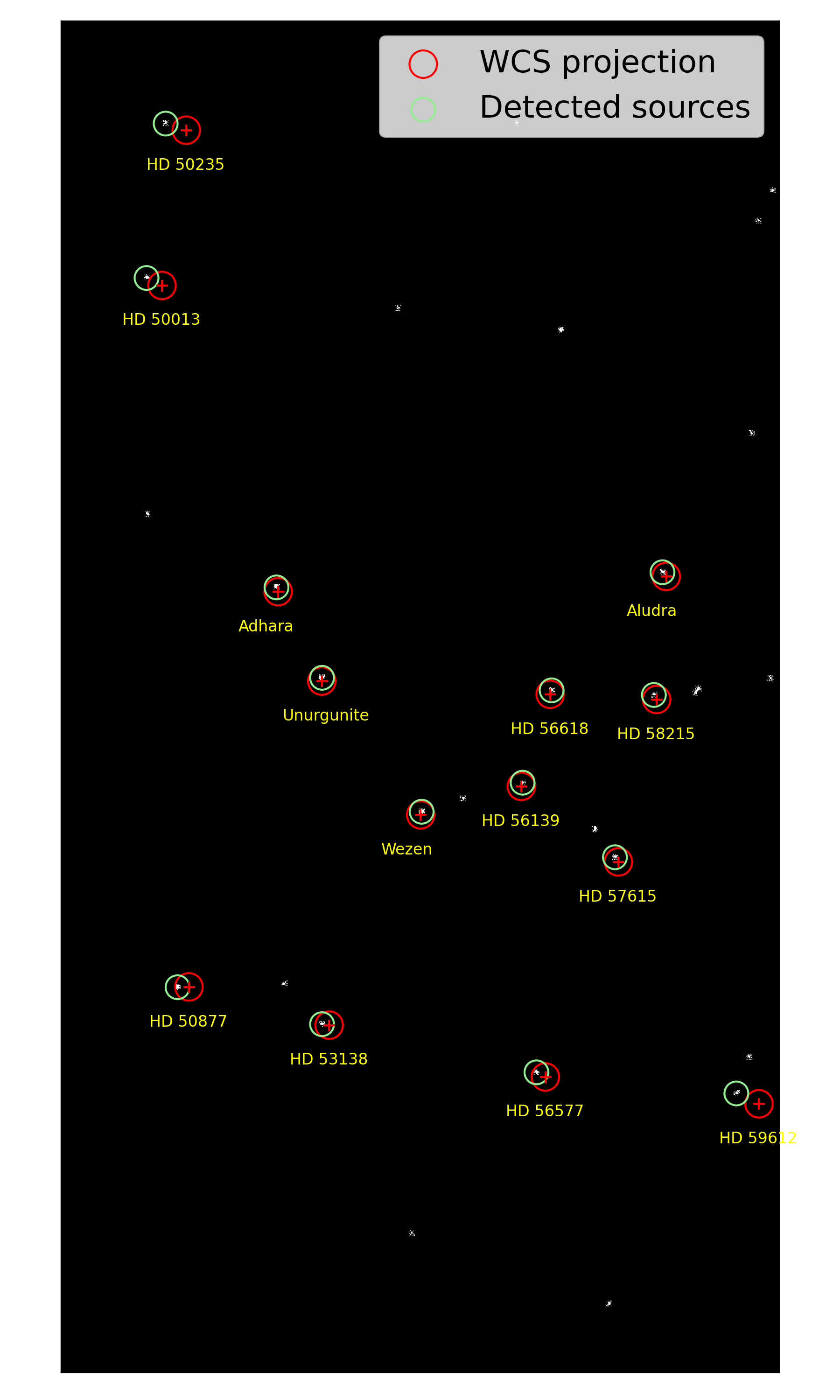}
        \caption{No distortion calibration}
    \end{subfigure}
    \hfill
    \begin{subfigure}[b]{0.32\textwidth}
        \centering
        \includegraphics[width=\textwidth]{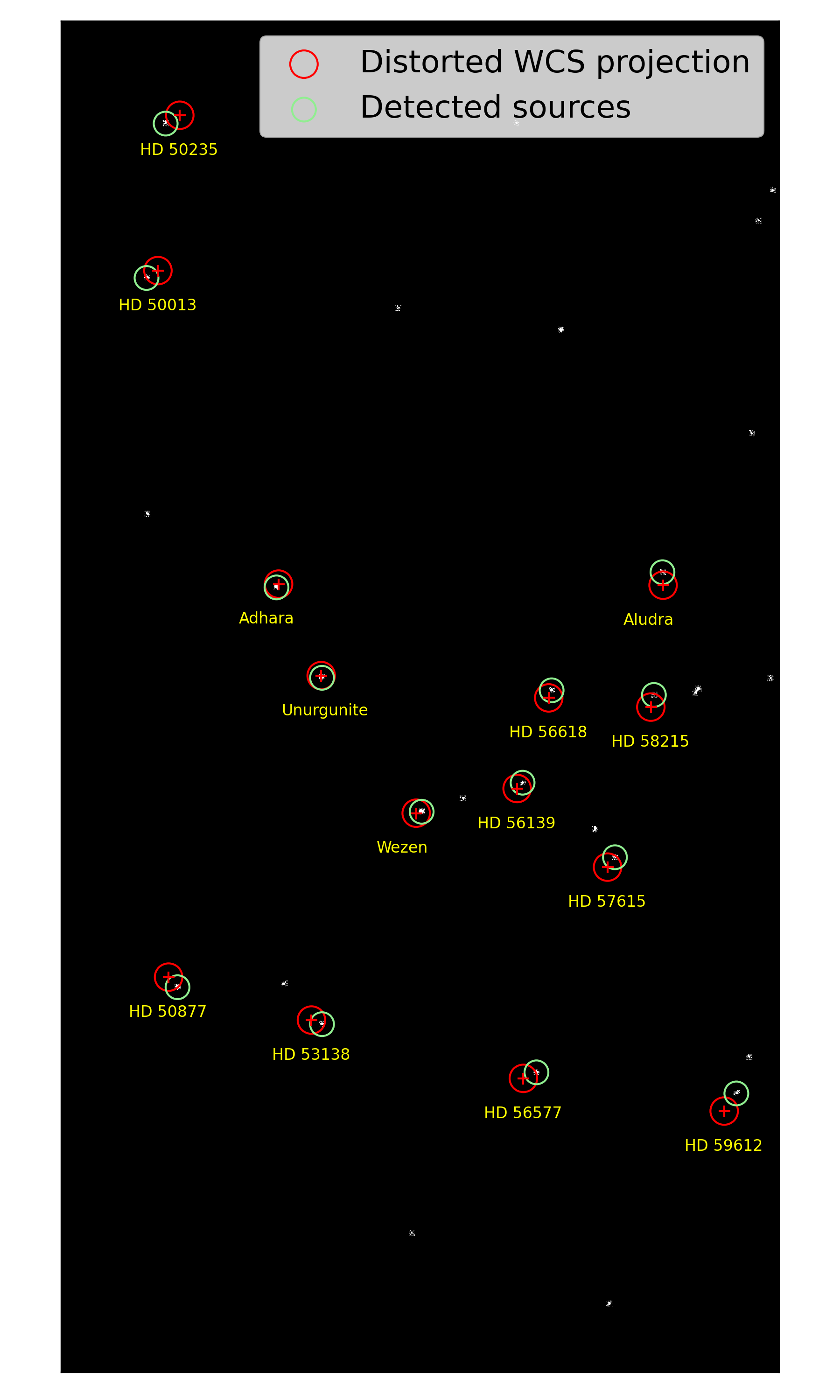}
        \caption{Pre-flight calibration}
    \end{subfigure}
    \hfill
        \begin{subfigure}[b]{0.32\textwidth}
        \centering
        \includegraphics[width=\textwidth]{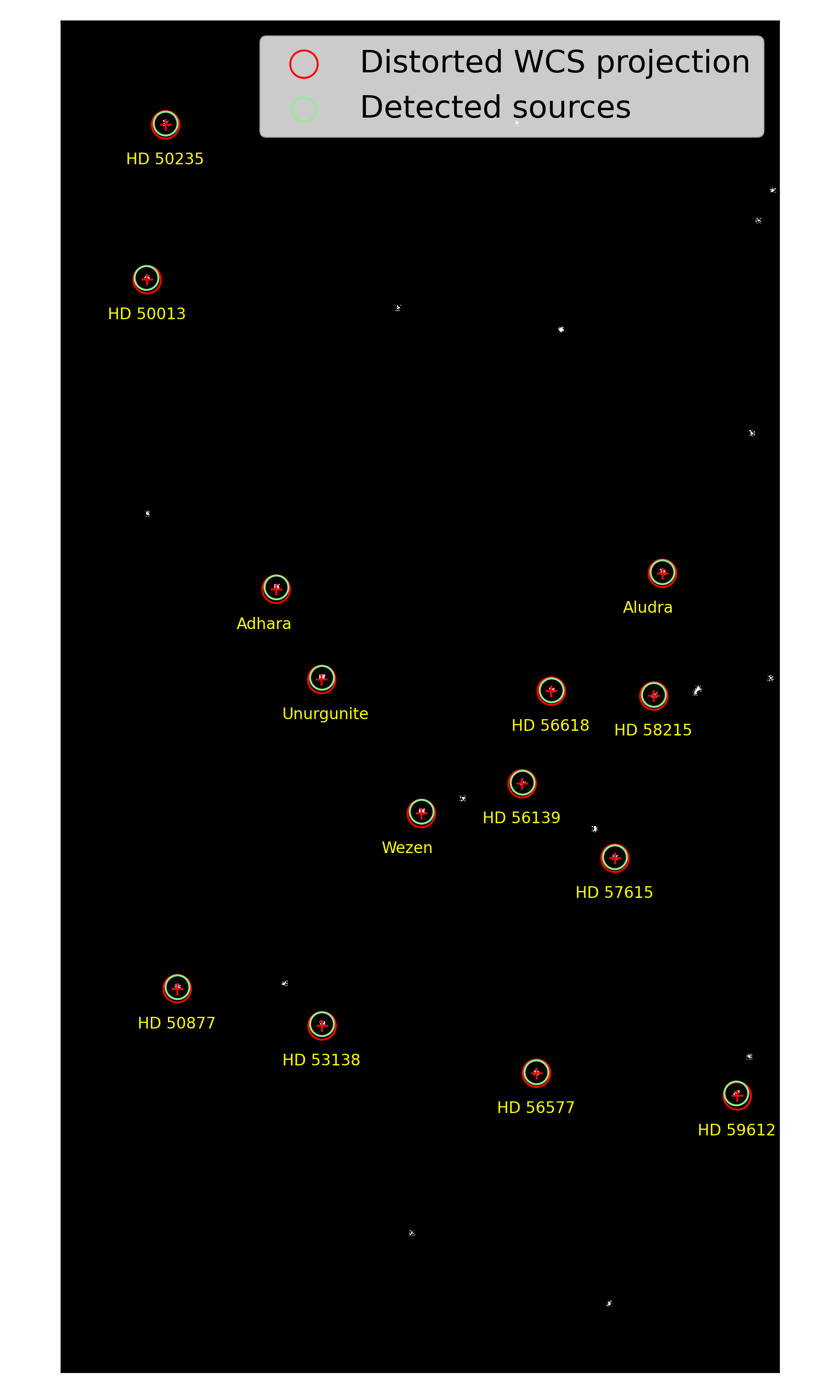}
        \caption{In-flight calibration}
    \end{subfigure}
    \caption{Comparison of WCS projections and detected sources with (a) no distortion calibration, (b) pre-flight calibration, and (c) in-flight calibration. The calibration is applied such that the WCS solutions are projected on the distorted field of HS-H.}
    \label{fig:detections}
\end{figure*}

\begin{figure*}[ht]
    \centering
    \begin{subfigure}[b]{0.32\textwidth}
        \centering
        \includegraphics[width=\textwidth]{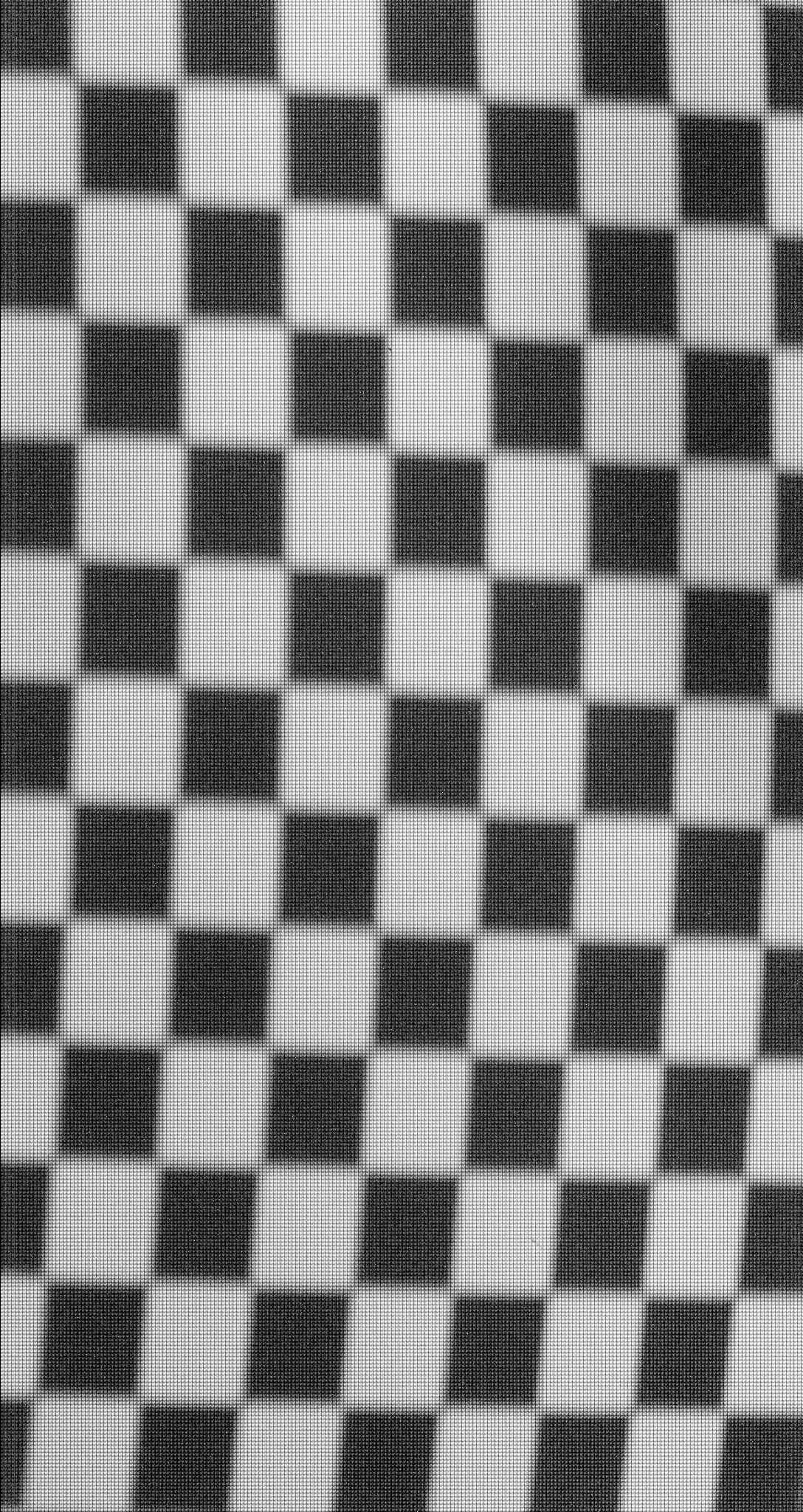}
        \caption{No distortion calibration}
    \end{subfigure}
    \hfill
    \begin{subfigure}[b]{0.32\textwidth}
        \centering
        \includegraphics[width=\textwidth]{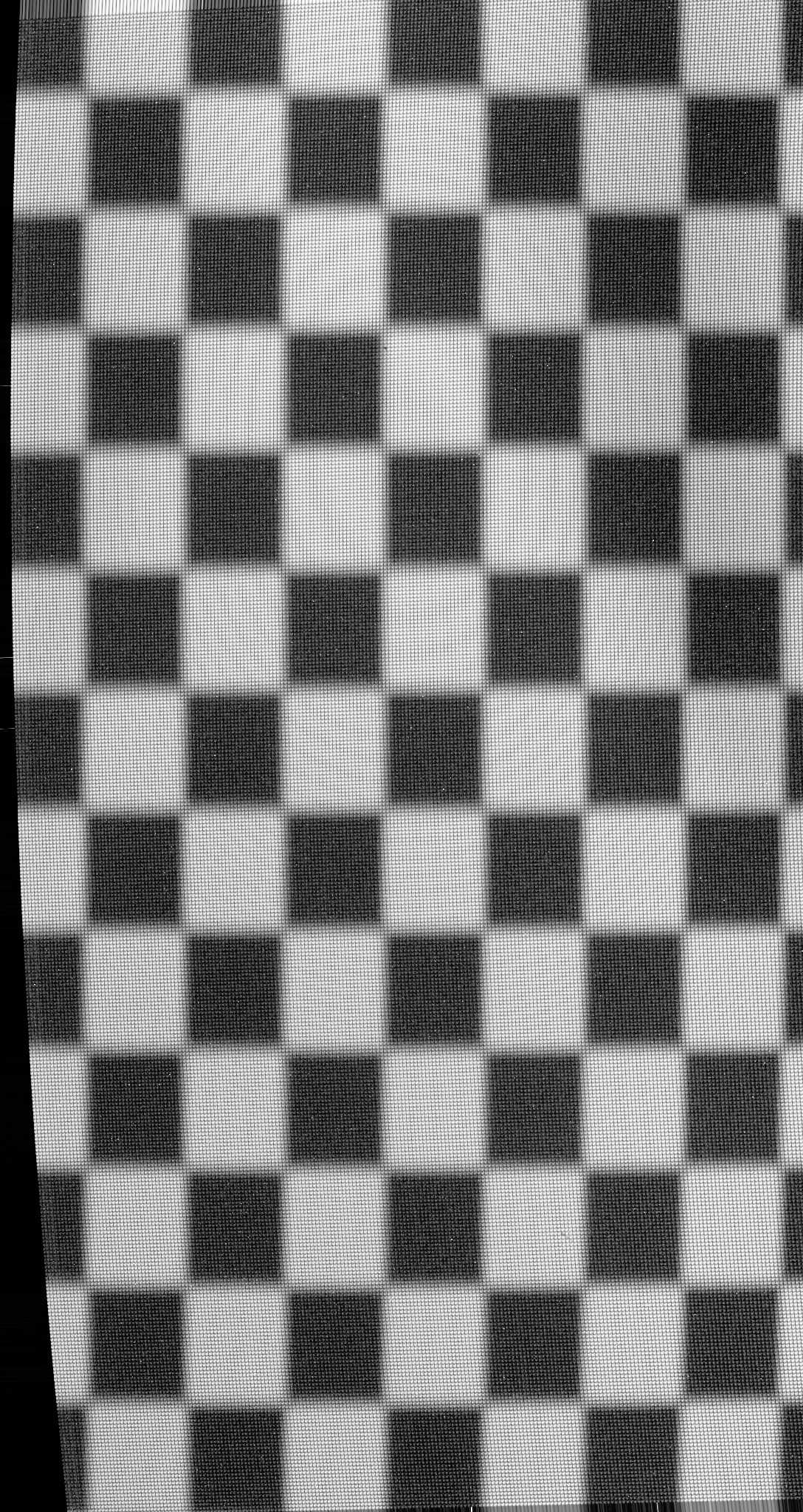}
        \caption{Pre-flight calibration}
    \end{subfigure}
    \hfill
        \begin{subfigure}[b]{0.32\textwidth}
        \centering
        \includegraphics[width=\textwidth]{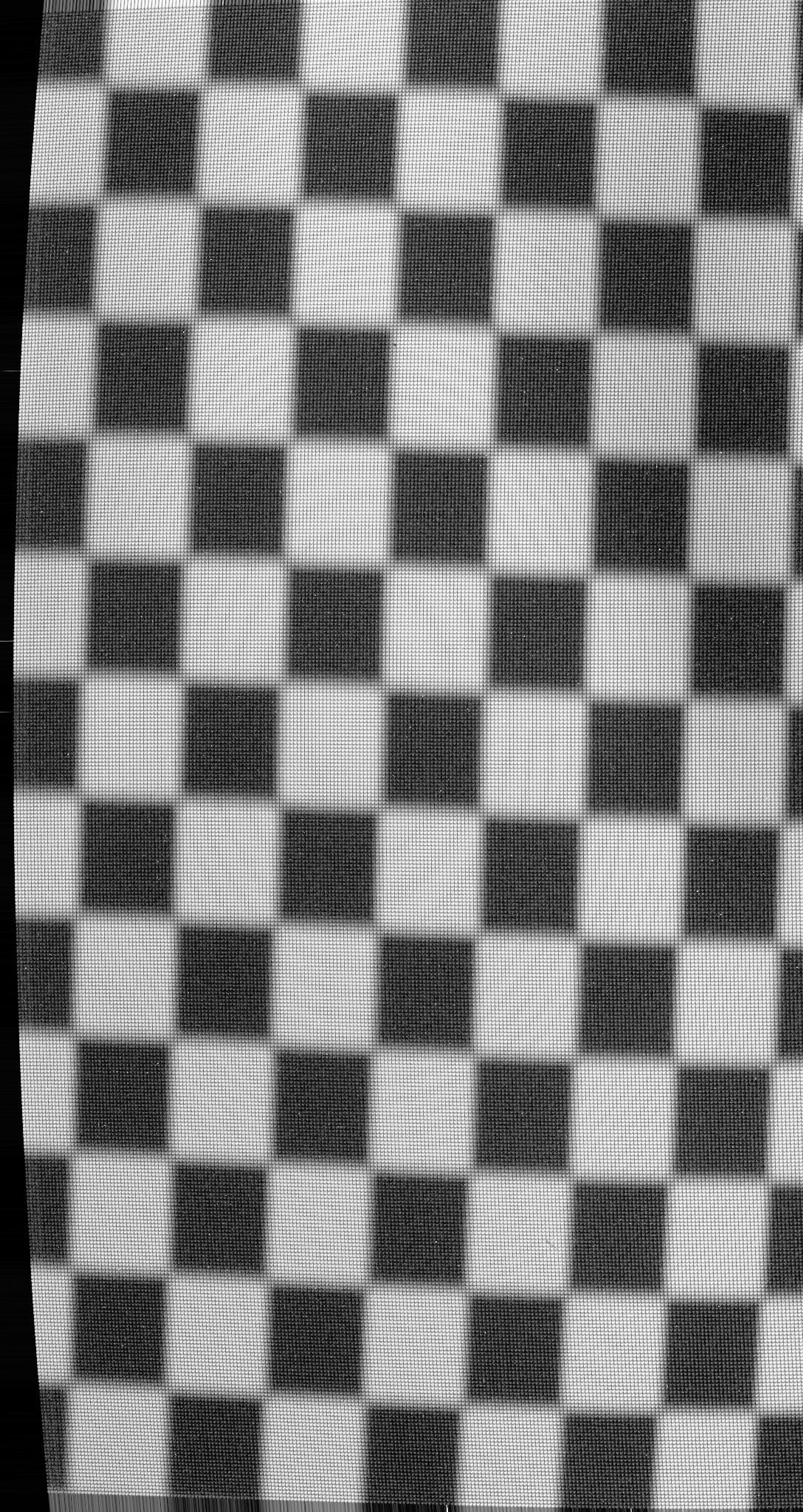}
        \caption{In-flight calibration}
    \end{subfigure}
    \caption{Comparison between checkerboard images after applying (a) no distortion calibration, (b) the pre-flight distortion corrections, and (c)
the in-flight distortion corrections. The calibration is applied to obtain an undistorted field of HS-H.}
    \label{fig:checkerboard}
\end{figure*}
\clearpage

During the preparation of this work the authors used LLMs (ChatGPT,Gemini) in order to perform language corrections. After using this tool/service, the authors reviewed and edited the content as needed and take full responsibility for the content of the published article.

\bibliographystyle{elsarticle-harv} 
\bibliography{hsh.bib}

@ARTICLE{Okada2025,
       author = {{Okada}, Tatsuaki and {Tanaka}, Satoshi and {Sakatani}, Naoya and {Shimaki}, Yuri and {Arai}, Takehiko and {Senshu}, Hiroki and {Demura}, Hirohide and {Sekiguchi}, Tomohiko and {Kouyama}, Toru and {Kanamaru}, Masanori and {Ishizaki}, Takuya and {Furukawa}, Soichiro and {Vilardell-Belles}, Ramon and {Karatekin}, {\"O}zg{\"u}r and {Blommaert}, Joris and {Hera TIRI team}},
        title = "{The Thermal InfraRed Imager on Hera}",
      journal = {Space Science Reviews},
         year = 2025,
        month = nov,
       volume = {221},
       number = {8},
          eid = {104},
        pages = {104},
          doi = {10.1007/s11214-025-01227-w},
       adsurl = {https://ui.adsabs.harvard.edu/abs/2025SSRv..221..104O}
}

@INPROCEEDINGS{Vincent2024,
       author = {{Vincent}, Jean-Baptiste and {Kovacs}, G{\'a}bor and {Nagy}, Bal{\'a}zs and {Preusker}, Frank and {Pajola}, Maurizio and {Kueppers}, Michael and {Michel}, Patrick},
        title = "{The Asteroid Framing Cameras on ESA's Hera mission}",
    booktitle = {European Planetary Science Congress},
         year = 2024,
        month = sep,
          eid = {EPSC2024-445},
        pages = {EPSC2024-445},
          doi = {10.5194/epsc2024-445},
       adsurl = {https://ui.adsabs.harvard.edu/abs/2024EPSC...17..445V}
}

@INPROCEEDINGS{Popescu2025EPSC,
       author = {{Popescu}, Marcel and {de Le{\'o}n}, Julia and {Pantelimon Prodan}, George and {K{\"u}ppers}, Michael and {Kov{\'a}cs}, G{\'a}bor and {Nagy}, Bal{\'a}zs Vince and {Grieger}, Bj{\"o}rn and {Escalante L{\'o}pez}, Alfredo and {Sugita}, Seiji and {Kohout}, Tom{\'a}{\v{s}} and {Korda}, David and {Tatsumi}, Eri and {Lazzarin}, Monica and {Farina}, Andrea and {Poggiali}, Giovanni and {Bickel}, Valentin T. and {Raducan}, Sabina D. and {Licandro}, Javier and {Palomba}, Ernesto and {Michel}, Patrick},
        title = "{Spectral Characterization of the Far Side of Deimos Using the HyperScout-H Instrument Aboard the ESA Hera Spacecraft}",
    booktitle = {EPSC-DPS Joint Meeting 2025},
         year = 2025,
       volume = {2025},
        month = sep,
          eid = {EPSC-DPS2025-1250},
        pages = {EPSC-DPS2025-1250},
          doi = {10.5194/epsc-dps2025-1250},
       adsurl = {https://ui.adsabs.harvard.edu/abs/2025epsc.conf.1250P}
}

@INPROCEEDINGS{Sugita2025EPSC,
       author = {{Sugita}, Seiji and {Nakahara}, Shumpei and {Vincent}, Jean-Baptiste and {Michel}, Patrick and {Kov{\'a}cs}, G{\'a}bor and {Ernst}, Carolyn and {Barnouin}, Olivier and {Miyamoto}, Hirdy and {Kikuchi}, Hiroshi and {Kueppers}, Michael},
        title = "{Geomorphology and Albedo distribution on the Anti-Mars hemisphere of Deimos observed with Asteroid Framing Cameras (AFC) on the Hera Spacecraft}",
    booktitle = {EPSC-DPS Joint Meeting 2025},
         year = 2025,
       volume = {2025},
        month = sep,
          eid = {EPSC-DPS2025-1786},
        pages = {EPSC-DPS2025-1786},
          doi = {10.5194/epsc-dps2025-1786},
       adsurl = {https://ui.adsabs.harvard.edu/abs/2025epsc.conf.1786S}
}

@ARTICLE{1997A&A...323L..49P,
       author = {{Perryman}, M.~A.~C. and {Lindegren}, L. and {Kovalevsky}, J. and {Hoeg}, E. and {Bastian}, U. and {Bernacca}, P.~L. and {Cr{\'e}z{\'e}}, M. and {Donati}, F. and {Grenon}, M. and {Grewing}, M. and {van Leeuwen}, F. and {van der Marel}, H. and {Mignard}, F. and {Murray}, C.~A. and {Le Poole}, R.~S. and {Schrijver}, H. and {Turon}, C. and {Arenou}, F. and {Froeschl{\'e}}, M. and {Petersen}, C.~S.},
        title = "{The HIPPARCOS Catalogue}",
      journal = {Astronomy \& Astrophysics},
         year = 1997,
        month = jul,
       volume = {323},
        pages = {L49-L52},
       adsurl = {https://ui.adsabs.harvard.edu/abs/1997A&A...323L..49P}
}

@article{PROCKTER2002491,
title = {The NEAR shoemaker mission to asteroid 433 eros},
journal = {Acta Astronautica},
volume = {51},
number = {1},
pages = {491-500},
year = {2002},
issn = {0094-5765},
doi = {https://doi.org/10.1016/S0094-5765(02)00098-X},
url = {https://www.sciencedirect.com/science/article/pii/S009457650200098X},
author = {L. Prockter and S. Murchie and A. Cheng and S. Krimigis and R. Farquhar and A. Santo and J. Trombka}
}

@INCOLLECTION{2002aste.book..351C,
       author = {{Cheng}, A.~F.},
        title = "{Near Earth Asteroid Rendezvous: Mission Summary}",
    booktitle = {Asteroids III},
         year = 2002,
       editor = {{Bottke}, Jr., W.~F. and {Cellino}, A. and {Paolicchi}, P. and {Binzel}, R.~P.},
        pages = {351-366},
       adsurl = {https://ui.adsabs.harvard.edu/abs/2002aste.book..351C},
    publisher = {University of Arizona Press, Tucson}
}

@INPROCEEDINGS{2005ASPC..347..491S,
       author = {{Shupe}, D.~L. and {Moshir}, Mehdrdad and {Li}, J. and {Makovoz}, D. and {Narron}, R. and {Hook}, R.~N.},
        title = "{The SIP Convention for Representing Distortion in FITS Image Headers}",
    booktitle = {Astronomical Data Analysis Software and Systems XIV},
         year = 2005,
       editor = {{Shopbell}, P. and {Britton}, M. and {Ebert}, R.},
       series = {Astronomical Society of the Pacific Conference Series},
       volume = {347},
        month = dec,
        pages = {491},
       adsurl = {https://ui.adsabs.harvard.edu/abs/2005ASPC..347..491S}
}

@article{CRISM,
author = {Murchie, S. and Arvidson, R. and Bedini, P. and Beisser, K. and Bibring, J.-P. and Bishop, J. and Boldt, J. and Cavender, P. and Choo, T. and Clancy, R. T. and Darlington, E. H. and Des Marais, D. and Espiritu, R. and Fort, D. and Green, R. and Guinness, E. and Hayes, J. and Hash, C. and Heffernan, K. and Hemmler, J. and Heyler, G. and Humm, D. and Hutcheson, J. and Izenberg, N. and Lee, R. and Lees, J. and Lohr, D. and Malaret, E. and Martin, T. and McGovern, J. A. and McGuire, P. and Morris, R. and Mustard, J. and Pelkey, S. and Rhodes, E. and Robinson, M. and Roush, T. and Schaefer, E. and Seagrave, G. and Seelos, F. and Silverglate, P. and Slavney, S. and Smith, M. and Shyong, W.-J. and Strohbehn, K. and Taylor, H. and Thompson, P. and Tossman, B. and Wirzburger, M. and Wolff, M.},
title = {Compact Reconnaissance Imaging Spectrometer for Mars (CRISM) on Mars Reconnaissance Orbiter (MRO)},
journal = {Journal of Geophysical Research: Planets},
volume = {112},
number = {E5},
pages = {},
doi = {https://doi.org/10.1029/2006JE002682},
year = {2007}
}

@ARTICLE{yale,
       author = {Hoffleit, D. and Warren, Jr., W.H.},
        title = "{The Bright Star Catalog}",
      journal = "{The Bright Star Catalog}",
         year = 1991,
       volume = {5th Revised Edition (Preliminary Version)}
}

@ARTICLE{gliese,
       author = {Wilhelm Gliese and Hartmut Jahreiss},
        title = "{Preliminary Version of the Third Catalogue of Nearby Stars}",
      journal = "Astronomisches Rechen-Institut",
         year = 1991,
       volume = {Heidelberg, Germany}
}

@article{TATSUMI2019153,
title = {Updated inflight calibration of Hayabusa2's optical navigation camera (ONC) for scientific observations during the cruise phase},
journal = {Icarus},
volume = {325},
pages = {153-195},
year = {2019},
issn = {0019-1035},
doi = {https://doi.org/10.1016/j.icarus.2019.01.015},
url = {https://www.sciencedirect.com/science/article/pii/S0019103518304494},
author = {Eri Tatsumi and Toru Kouyama and Hidehiko Suzuki and Manabu Yamada and Naoya Sakatani and Shingo Kameda and Yasuhiro Yokota and Rie Honda and Tomokatsu Morota and Keiichi Moroi and Naoya Tanabe and Hiroaki Kamiyoshihara and Marika Ishida and Kazuo Yoshioka and Hiroyuki Sato and Chikatoshi Honda and Masahiko Hayakawa and Kohei Kitazato and Hirotaka Sawada and Seiji Sugita}
}

@ARTICLE{2014PASP..126..711B,
       author = {{Bohlin}, Ralph C. and {Gordon}, Karl D. and {Tremblay}, P.-E.},
        title = "{Techniques and Review of Absolute Flux Calibration from the Ultraviolet to the Mid-Infrared}",
      journal = {Publications of the Astronomical Society of the Pacific},
         year = 2014,
        month = aug,
       volume = {126},
       number = {942},
        pages = {711},
          doi = {10.1086/677655},
archivePrefix = {arXiv},
       eprint = {1406.1707},
 primaryClass = {astro-ph.IM},
       adsurl = {https://ui.adsabs.harvard.edu/abs/2014PASP..126..711B}
}

@ARTICLE{1977RMxAA...2...71J,
       author = {{Johnson}, H.~L.},
        title = "{An atlas of stellar spectra. I.}",
      journal = {Revista Mexicana de Astronomia y Astrofisica},
         year = 1977,
        month = jan,
       volume = {2},
        pages = {71-170},
       adsurl = {https://ui.adsabs.harvard.edu/abs/1977RMxAA...2...71J}
}

@misc{Rieke2011AbsoluteSpectrum,
  author       = {Rieke, G. H. and Blaylock, M. and Decin, L. and Engelbracht, C. and 
                  Ogle, P. and Avrett, E. and Carpenter, J. and Cutri, R. M. and 
                  Armus, L. and Gordon, K. and Gray, R. O. and Hinz, J. and Su, K. and 
                  Willmer, C. N. A. and Armus, L.},
  title        = {Absolute spectrum of the Sun and Vega for 0.2--30 $\mu$m},
  year         = {2011},
  publisher    = {Centre de Données astronomiques de Strasbourg (CDS)},
  doi          = {10.26093/cds/vizier.51352245},
  note         = {VizieR Online Data Catalog: J/ApJ/728/20}
}

@MISC{vizier,
author = { Ochsenbein, F.},
title = "{ The VizieR database of astronomical catalogues }",
doi = {10.26093/cds/vizier}, 
year = {et al. 2025}
}

@techreport{Colina1996Vega,
  AUTHOR = {Colina, L. and Bohlin, R. and Castelli, F.},
  TITLE = {Absolute Flux Calibrated Spectrum of Vega},
  INSTITUTION = {Space Telescope Science Institute},
  YEAR = {1996},
  NUMBER = {CAL/SCS-008},
  TYPE = {Instrument Science Report},
  URL = {https://www.stsci.edu/instruments/observatory/PDF/scs8.rev.pdf},
  NOTE = {Revised April 22, 1996}
}

@ARTICLE{2022EP&S...74...12K,
       author = {{Kuramoto}, Kiyoshi and {Kawakatsu}, Yasuhiro and {Fujimoto}, Masaki and {Araya}, Akito and {Barucci}, Maria Antonietta and {Genda}, Hidenori and {Hirata}, Naru and {Ikeda}, Hitoshi and {Imamura}, Takeshi and {Helbert}, J{\"o}rn and {Kameda}, Shingo and {Kobayashi}, Masanori and {Kusano}, Hiroki and {Lawrence}, David J. and {Matsumoto}, Koji and {Michel}, Patrick and {Miyamoto}, Hideaki and {Morota}, Tomokatsu and {Nakagawa}, Hiromu and {Nakamura}, Tomoki and {Ogawa}, Kazunori and {Otake}, Hisashi and {Ozaki}, Masanobu and {Russell}, Sara and {Sasaki}, Sho and {Sawada}, Hirotaka and {Senshu}, Hiroki and {Tachibana}, Shogo and {Terada}, Naoki and {Ulamec}, Stephan and {Usui}, Tomohiro and {Wada}, Koji and {Watanabe}, Sei-ichiro and {Yokota}, Shoichiro},
        title = "{Martian moons exploration MMX: sample return mission to Phobos elucidating formation processes of habitable planets}",
      journal = {Earth, Planets and Space},
         year = 2022,
        month = dec,
       volume = {74},
       number = {1},
          eid = {12},
        pages = {12},
          doi = {10.1186/s40623-021-01545-7},
       adsurl = {https://ui.adsabs.harvard.edu/abs/2022EP&S...74...12K}
}

@ARTICLE{1994A&AS..105..311F,
       author = {{Fluks}, M.~A. and {Plez}, B. and {The}, P.~S. and {de Winter}, D. and {Westerlund}, B.~E. and {Steenman}, H.~C.},
        title = "{On the spectra and photometry of M-giant stars}",
      journal = {Astronomy \& Astrophysics Supp.},
         year = 1994,
        month = jun,
       volume = {105},
        pages = {311-336},
       adsurl = {https://ui.adsabs.harvard.edu/abs/1994A&AS..105..311F}
}

@ARTICLE{2024IJRS...45.2488B,
       author = {{Benhadj}, I. and {Livens}, S. and {Esposito}, M. and {Vercruyssen}, N. and {Van Dijk}, C. and {Soukup}, M. and {Zuccaro}, A. and {Maresi}, L.},
        title = "{HyperScout-1 inflight calibration and product validation}",
      journal = {International Journal of Remote Sensing},
         year = 2024,
        month = mar,
       volume = {45},
       number = {7},
        pages = {2488-2513},
          doi = {10.1080/01431161.2024.2331979},
}

@ARTICLE{2024PSJ.....5...49C,
       author = {{Chabot}, Nancy L. and {Rivkin}, Andrew S. and {Cheng}, Andrew F. and {Barnouin}, Olivier S. and {Fahnestock}, Eugene G. and {Richardson}, Derek C. and {Stickle}, Angela M. and {Thomas}, Cristina A. and {Ernst}, Carolyn M. and {Terik Daly}, R. and {Dotto}, Elisabetta and {Zinzi}, Angelo and {Chesley}, Steven R. and {Moskovitz}, Nicholas A. and {Barbee}, Brent W. and {Abell}, Paul and {Agrusa}, Harrison F. and {Bannister}, Michele T. and {Beccarelli}, Joel and {Bekker}, Dmitriy L. and {Bruck Syal}, Megan and {Buratti}, Bonnie J. and {Busch}, Michael W. and {Campo Bagatin}, Adriano and {Chatelain}, Joseph P. and {Chocron}, Sidney and {Collins}, Gareth S. and {Conversi}, Luca and {Davison}, Thomas M. and {DeCoster}, Mallory E. and {Prasanna Deshapriya}, J.~D. and {Eggl}, Siegfried and {Espiritu}, Raymond C. and {Farnham}, Tony L. and {Ferrais}, Marin and {Ferrari}, Fabio and {F{\"o}hring}, Dora and {Fuentes-Mu{\~n}oz}, Oscar and {Gai}, Igor and {Giordano}, Carmine and {Glenar}, David A. and {Gomez}, Edward and {Graninger}, Dawn M. and {Green}, Simon F. and {Greenstreet}, Sarah and {Hasselmann}, Pedro H. and {Herreros}, Isabel and {Hirabayashi}, Masatoshi and {Hus{\'a}rik}, Marek and {Ieva}, Simone and {Ivanovski}, Stavro L. and {Jackson}, Samuel L. and {Jehin}, Emmanuel and {Jutzi}, Martin and {Karatekin}, Ozgur and {Knight}, Matthew M. and {Kolokolova}, Ludmilla and {Kumamoto}, Kathryn M. and {K{\"u}ppers}, Michael and {La Forgia}, Fiorangela and {Lazzarin}, Monica and {Li}, Jian-Yang and {Lister}, Tim A. and {Lolachi}, Ramin and {Lucas}, Michael P. and {Lucchetti}, Alice and {Luther}, Robert and {Makadia}, Rahil and {Mazzotta Epifani}, Elena and {McMahon}, Jay and {Merisio}, Gianmario and {Merrill}, Colby C. and {Meyer}, Alex J. and {Michel}, Patrick and {Micheli}, Marco and {Migliorini}, Alessandra and {Minker}, Kate and {Modenini}, Dario and {Moreno}, Fernando and {Murdoch}, Naomi and {Murphy}, Brian and {Naidu}, Shantanu P. and {Nair}, Hari and {Nakano}, Ryota and {Opitom}, Cyrielle and {Orm{\"o}}, Jens and {Michael Owen}, J. and {Pajola}, Maurizio and {Palmer}, Eric E. and {Palumbo}, Pasquale and {Panicucci}, Paolo and {Parro}, Laura M. and {Pearl}, Jason M. and {Penttil{\"a}}, Antti and {Perna}, Davide and {Petrescu}, Elisabeta and {Pravec}, Petr and {Raducan}, Sabina D. and {Ramesh}, K.~T. and {Ridden-Harper}, Ryan and {Rizos}, Juan L. and {Rossi}, Alessandro and {Roth}, Nathan X. and {Ro{\.z}ek}, Agata and {Rozitis}, Benjamin and {Ryan}, Eileen V. and {Ryan}, William H. and {S{\'a}nchez}, Paul and {Santana-Ros}, Toni and {Scheeres}, Daniel J. and {Scheirich}, Peter and {Senel}, Cem Berk and {Snodgrass}, Colin and {Soldini}, Stefania and {Souami}, Damya and {Statler}, Thomas S. and {Street}, Rachel and {Stubbs}, Timothy J. and {Sunshine}, Jessica M. and {Tan}, Nicole J. and {Tancredi}, Gonzalo and {Tinsman}, Calley L. and {Tortora}, Paolo and {Tusberti}, Filippo and {Walker}, James D. and {Waller}, C. Dany and {W{\"u}nnemann}, Kai and {Zannoni}, Marco and {Zhang}, Yun},
        title = "{Achievement of the Planetary Defense Investigations of the Double Asteroid Redirection Test (DART) Mission}",
      journal = {The Planetary Science Journal},
         year = 2024,
        month = feb,
       volume = {5},
       number = {2},
          eid = {49},
        pages = {49},
          doi = {10.3847/PSJ/ad16e6},
       adsurl = {https://ui.adsabs.harvard.edu/abs/2024PSJ.....5...49C}
}

@ARTICLE{2023Natur.616..457C,
       author = {{Cheng}, Andrew F. and {Agrusa}, Harrison F. and {Barbee}, Brent W. and {Meyer}, Alex J. and {Farnham}, Tony L. and {Raducan}, Sabina D. and {Richardson}, Derek C. and {Dotto}, Elisabetta and {Zinzi}, Angelo and {Della Corte}, Vincenzo and {Statler}, Thomas S. and {Chesley}, Steven and {Naidu}, Shantanu P. and {Hirabayashi}, Masatoshi and {Li}, Jian-Yang and {Eggl}, Siegfried and {Barnouin}, Olivier S. and {Chabot}, Nancy L. and {Chocron}, Sidney and {Collins}, Gareth S. and {Daly}, R. Terik and {Davison}, Thomas M. and {DeCoster}, Mallory E. and {Ernst}, Carolyn M. and {Ferrari}, Fabio and {Graninger}, Dawn M. and {Jacobson}, Seth A. and {Jutzi}, Martin and {Kumamoto}, Kathryn M. and {Luther}, Robert and {Lyzhoft}, Joshua R. and {Michel}, Patrick and {Murdoch}, Naomi and {Nakano}, Ryota and {Palmer}, Eric and {Rivkin}, Andrew S. and {Scheeres}, Daniel J. and {Stickle}, Angela M. and {Sunshine}, Jessica M. and {Trigo-Rodriguez}, Josep M. and {Vincent}, Jean-Baptiste and {Walker}, James D. and {W{\"u}nnemann}, Kai and {Zhang}, Yun and {Amoroso}, Marilena and {Bertini}, Ivano and {Brucato}, John R. and {Capannolo}, Andrea and {Cremonese}, Gabriele and {Dall'Ora}, Massimo and {Deshapriya}, Prasanna J.~D. and {Gai}, Igor and {Hasselmann}, Pedro H. and {Ieva}, Simone and {Impresario}, Gabriele and {Ivanovski}, Stavro L. and {Lavagna}, Mich{\`e}le and {Lucchetti}, Alice and {Epifani}, Elena M. and {Modenini}, Dario and {Pajola}, Maurizio and {Palumbo}, Pasquale and {Perna}, Davide and {Pirrotta}, Simone and {Poggiali}, Giovanni and {Rossi}, Alessandro and {Tortora}, Paolo and {Zannoni}, Marco and {Zanotti}, Giovanni},
        title = "{Momentum transfer from the DART mission kinetic impact on asteroid Dimorphos}",
      journal = {Nature},
         year = 2023,
        month = apr,
       volume = {616},
       number = {7957},
        pages = {457-460},
          doi = {10.1038/s41586-023-05878-z},
archivePrefix = {arXiv},
       eprint = {2303.03464},
 primaryClass = {astro-ph.EP},
       adsurl = {https://ui.adsabs.harvard.edu/abs/2023Natur.616..457C}
}

@ARTICLE{2024Natur.627..505D,
       author = {{Dotto}, E. and {Deshapriya}, J.~D.~P. and {Gai}, I. and {Hasselmann}, P.~H. and {Mazzotta Epifani}, E. and {Poggiali}, G. and {Rossi}, A. and {Zanotti}, G. and {Zinzi}, A. and {Bertini}, I. and {Brucato}, J.~R. and {Dall'Ora}, M. and {Della Corte}, V. and {Ivanovski}, S.~L. and {Lucchetti}, A. and {Pajola}, M. and {Amoroso}, M. and {Barnouin}, O. and {Campo Bagatin}, A. and {Capannolo}, A. and {Caporali}, S. and {Ceresoli}, M. and {Chabot}, N.~L. and {Cheng}, A.~F. and {Cremonese}, G. and {Fahnestock}, E.~G. and {Farnham}, T.~L. and {Ferrari}, F. and {Gomez Casajus}, L. and {Gramigna}, E. and {Hirabayashi}, M. and {Ieva}, S. and {Impresario}, G. and {Jutzi}, M. and {Lasagni Manghi}, R. and {Lavagna}, M. and {Li}, J. -Y. and {Lombardo}, M. and {Modenini}, D. and {Palumbo}, P. and {Perna}, D. and {Pirrotta}, S. and {Raducan}, S.~D. and {Richardson}, D.~C. and {Rivkin}, A.~S. and {Stickle}, A.~M. and {Sunshine}, J.~M. and {Tortora}, P. and {Tusberti}, F. and {Zannoni}, M.},
        title = "{The Dimorphos ejecta plume properties revealed by LICIACube}",
      journal = {Nature},
         year = 2024,
        month = mar,
       volume = {627},
       number = {8004},
        pages = {505-509},
          doi = {10.1038/s41586-023-06998-2},
       adsurl = {https://ui.adsabs.harvard.edu/abs/2024Natur.627..505D}
}

@ARTICLE{2023PSJ.....4..214R,
       author = {{Rivkin}, Andrew S. and {Thomas}, Cristina A. and {Wong}, Ian and {Rozitis}, Benjamin and {de Le{\'o}n}, Julia and {Holler}, Bryan and {Milam}, Stefanie N. and {Howell}, Ellen S. and {Hammel}, Heidi B. and {Arredondo}, Anicia and {Brucato}, John R. and {Epifani}, Elena M. and {Ieva}, Simone and {La Forgia}, Fiorangela and {Lucas}, Michael P. and {Lucchetti}, Alice and {Pajola}, Maurizio and {Poggiali}, Giovanni and {Sunshine}, Jessica N. and {Trigo-Rodr{\'\i}guez}, Josep M.},
        title = "{Near to Mid-infrared Spectroscopy of (65803) Didymos as Observed by JWST: Characterization Observations Supporting the Double Asteroid Redirection Test}",
      journal = {The Planetary Science Journal},
         year = 2023,
        month = nov,
       volume = {4},
       number = {11},
          eid = {214},
        pages = {214},
          doi = {10.3847/PSJ/ad04d8},
archivePrefix = {arXiv},
       eprint = {2310.11168},
 primaryClass = {astro-ph.EP},
       adsurl = {https://ui.adsabs.harvard.edu/abs/2023PSJ.....4..214R}
}

@ARTICLE{2024PSJ.....5...35M,
       author = {{Moskovitz}, Nicholas and {Thomas}, Cristina and {Pravec}, Petr and {Lister}, Tim and {Polakis}, Tom and {Osip}, David and {Kareta}, Theodore and {Ro{\.z}ek}, Agata and {Chesley}, Steven R. and {Naidu}, Shantanu P. and {Scheirich}, Peter and {Ryan}, William and {Ryan}, Eileen and {Skiff}, Brian and {Snodgrass}, Colin and {Knight}, Matthew M. and {Rivkin}, Andrew S. and {Chabot}, Nancy L. and {Ayvazian}, Vova and {Belskaya}, Irina and {Benkhaldoun}, Zouhair and {Berte{\c{s}}teanu}, Daniel N. and {Bonavita}, Mariangela and {Bressi}, Terrence H. and {Brucker}, Melissa J. and {Burgdorf}, Martin J. and {Burkhonov}, Otabek and {Burt}, Brian and {Contreras}, Carlos and {Chatelain}, Joseph and {Choi}, Young-Jun and {Daily}, Matthew and {de Le{\'o}n}, Julia and {Ergashev}, Kamoliddin and {Farnham}, Tony and {Fatka}, Petr and {Ferrais}, Marin and {Geier}, Stefan and {Gomez}, Edward and {Greenstreet}, Sarah and {Gr{\"o}ller}, Hannes and {Hergenrother}, Carl and {Holt}, Carrie and {Hornoch}, Kamil and {Hus{\'a}rik}, Marek and {Inasaridze}, Raguli and {Jehin}, Emmanuel and {Khalouei}, Elahe and {Eluo}, Jean-Baptiste Kikwaya and {Kim}, Myung-Jin and {Krugly}, Yurij and {Ku{\v{c}}{\'a}kov{\'a}}, Hana and {Ku{\v{s}}nir{\'a}k}, Peter and {Larsen}, Jeffrey A. and {Lee}, Hee-Jae and {Lejoly}, Cassandra and {Licandro}, Javier and {Longa-Pe{\~n}a}, Pen{\'e}lope and {Mastaler}, Ronald A. and {McCully}, Curtis and {Moon}, Hong-Kyu and {Morrell}, Nidia and {Nath}, Arushi and {Oszkiewicz}, Dagmara and {Parrott}, Daniel and {Phillips}, Liz and {Popescu}, Marcel M. and {Pray}, Donald and {Prodan}, George Pantelimon and {Rabus}, Markus and {Read}, Michael T. and {Reva}, Inna and {Roark}, Vernon and {Santana-Ros}, Toni and {Scotti}, James V. and {Tatara}, Taiyo and {Thirouin}, Audrey and {Tholen}, David and {Troianskyi}, Volodymyr and {Tubbiolo}, Andrew F. and {Villa}, Katelyn},
        title = "{Photometry of the Didymos System across the DART Impact Apparition}",
      journal = {The Planetary Science Journal},
         year = 2024,
        month = feb,
       volume = {5},
       number = {2},
          eid = {35},
        pages = {35},
          doi = {10.3847/PSJ/ad0e74},
archivePrefix = {arXiv},
       eprint = {2311.01971},
 primaryClass = {astro-ph.EP},
       adsurl = {https://ui.adsabs.harvard.edu/abs/2024PSJ.....5...35M}
}

@ARTICLE{2022PSJ.....3..175P,
       author = {{Pravec}, P. and {Thomas}, C.~A. and {Rivkin}, A.~S. and {Scheirich}, P. and {Moskovitz}, N. and {Knight}, M.~M. and {Snodgrass}, C. and {de Le{\'o}n}, J. and {Licandro}, J. and {Popescu}, M. and {Thirouin}, A. and {F{\"o}hring}, D. and {Chandler}, C.~O. and {Oldroyd}, W.~J. and {Trujillo}, C.~A. and {Howell}, E.~S. and {Green}, S.~F. and {Thomas-Osip}, J. and {Sheppard}, S.~S. and {Farnham}, T.~L. and {Mazzotta Epifani}, E. and {Dotto}, E. and {Ieva}, S. and {Dall'Ora}, M. and {Kokotanekova}, R. and {Carry}, B. and {Souami}, D.},
        title = "{Photometric Observations of the Binary Near-Earth Asteroid (65803) Didymos in 2015-2021 Prior to DART Impact}",
      journal = {The Planetary Science Journal},
         year = 2022,
        month = jul,
       volume = {3},
       number = {7},
          eid = {175},
        pages = {175},
          doi = {10.3847/PSJ/ac7be1},
       adsurl = {https://ui.adsabs.harvard.edu/abs/2022PSJ.....3..175P}
}

@ARTICLE{2006AdSpR..37..178D,
       author = {{de Le{\'o}n}, J. and {Licandro}, J. and {Duffard}, R. and {Serra-Ricart}, M.},
        title = "{Spectral analysis and mineralogical characterization of 11 olivine pyroxene rich NEAs}",
      journal = {Advances in Space Research},
         year = 2006,
        month = jan,
       volume = {37},
       number = {1},
        pages = {178-183},
          doi = {10.1016/j.asr.2005.05.074},
       adsurl = {https://ui.adsabs.harvard.edu/abs/2006AdSpR..37..178D}
}

@ARTICLE{2024PSJ.....5...17S,
       author = {{Scheirich}, Peter and {Pravec}, Petr and {Meyer}, Alex J. and {Agrusa}, Harrison F. and {Richardson}, Derek C. and {Chesley}, Steven R. and {Naidu}, Shantanu P. and {Thomas}, Cristina and {Moskovitz}, Nicholas A.},
        title = "{Dimorphos Orbit Determination from Mutual Events Photometry}",
      journal = {The Planetary Science Journal},
         year = 2024,
        month = jan,
       volume = {5},
       number = {1},
          eid = {17},
        pages = {17},
          doi = {10.3847/PSJ/ad12cf},
archivePrefix = {arXiv},
       eprint = {2403.02804},
 primaryClass = {astro-ph.EP},
       adsurl = {https://ui.adsabs.harvard.edu/abs/2024PSJ.....5...17S}
}

@ARTICLE{2024Icar..41816138P,
       author = {{Pravec}, P. and {Meyer}, A.~J. and {Scheirich}, P. and {Scheeres}, D.~J. and {Benson}, C.~J. and {Agrusa}, H.~F.},
        title = "{Rotational lightcurves of Dimorphos and constraints on its post-DART impact spin state}",
      journal = {Icarus},
         year = 2024,
        month = aug,
       volume = {418},
          eid = {116138},
        pages = {116138},
          doi = {10.1016/j.icarus.2024.116138},
       adsurl = {https://ui.adsabs.harvard.edu/abs/2024Icar..41816138P}
}

@ARTICLE{2024NatAs...8..445R,
       author = {{Raducan}, S.~D. and {Jutzi}, M. and {Cheng}, A.~F. and {Zhang}, Y. and {Barnouin}, O. and {Collins}, G.~S. and {Daly}, R.~T. and {Davison}, T.~M. and {Ernst}, C.~M. and {Farnham}, T.~L. and {Ferrari}, F. and {Hirabayashi}, M. and {Kumamoto}, K.~M. and {Michel}, P. and {Murdoch}, N. and {Nakano}, R. and {Pajola}, M. and {Rossi}, A. and {Agrusa}, H.~F. and {Barbee}, B.~W. and {Syal}, M. Bruck and {Chabot}, N.~L. and {Dotto}, E. and {Fahnestock}, E.~G. and {Hasselmann}, P.~H. and {Herreros}, I. and {Ivanovski}, S. and {Li}, J. -Y. and {Lucchetti}, A. and {Luther}, R. and {Orm{\"o}}, J. and {Owen}, M. and {Pravec}, P. and {Rivkin}, A.~S. and {Robin}, C.~Q. and {S{\'a}nchez}, P. and {Tusberti}, F. and {W{\"u}nnemann}, K. and {Zinzi}, A. and {Epifani}, E. Mazzotta and {Manzoni}, C. and {May}, B.~H.},
        title = "{Physical properties of asteroid Dimorphos as derived from the DART impact}",
      journal = {Nature Astronomy},
         year = 2024,
        month = apr,
       volume = {8},
        pages = {445-455},
          doi = {10.1038/s41550-024-02200-3},
       adsurl = {https://ui.adsabs.harvard.edu/abs/2024NatAs...8..445R}
}

@ARTICLE{2023PSJ.....4..229P,
       author = {{Polishook}, David and {DeMeo}, Francesca E. and {Burt}, Brian J. and {Thomas}, Cristina A. and {Rivkin}, Andrew S. and {Sanchez}, Juan A. and {Reddy}, Vishnu},
        title = "{Near-IR Spectral Observations of the Didymos System: Daily Evolution Before and After the DART Impact Indicates that Dimorphos Originated from Didymos}",
      journal = {The Planetary Science Journal},
         year = 2023,
        month = dec,
       volume = {4},
       number = {12},
          eid = {229},
        pages = {229},
          doi = {10.3847/PSJ/ad08ae},
archivePrefix = {arXiv},
       eprint = {2311.00421},
 primaryClass = {astro-ph.EP},
       adsurl = {https://ui.adsabs.harvard.edu/abs/2023PSJ.....4..229P}
}

@ARTICLE{2022PSJ.....3..183I,
       author = {{Ieva}, Simone and {Mazzotta Epifani}, E. and {Perna}, D. and {Dall'Ora}, M. and {Petropoulou}, V. and {Deshapriya}, J.~D.~P. and {Hasselmann}, P.~H. and {Rossi}, A. and {Poggiali}, G. and {Brucato}, J.~R. and {Pajola}, M. and {Lucchetti}, A. and {Ivanovski}, S.~L. and {Palumbo}, P. and {Della Corte}, V. and {Zinzi}, A. and {Rivkin}, A.~S. and {Thomas}, C.~A. and {de Le{\'o}n}, J. and {Dotto}, E. and {Amoroso}, M. and {Bertini}, I. and {Capannolo}, A. and {Cotugno}, B. and {Cremonese}, G. and {Di Tana}, V. and {Gai}, I. and {Impresario}, G. and {Lavagna}, M. and {Meneghin}, A. and {Miglioretti}, F. and {Modenini}, D. and {Pirrotta}, S. and {Simioni}, E. and {Simonetti}, S. and {Tortora}, P. and {Zannoni}, M. and {Zanotti}, G.},
        title = "{Spectral Rotational Characterization of the Didymos System prior to the DART Impact}",
      journal = {The Planetary Science Journal},
         year = 2022,
        month = aug,
       volume = {3},
       number = {8},
          eid = {183},
        pages = {183},
          doi = {10.3847/PSJ/ac7f34},
       adsurl = {https://ui.adsabs.harvard.edu/abs/2022PSJ.....3..183I}
}

@ARTICLE{2023Natur.616..443D,
       author = {{Daly}, R. Terik and {Ernst}, Carolyn M. and {Barnouin}, Olivier S. and {Chabot}, Nancy L. and {Rivkin}, Andrew S. and {Cheng}, Andrew F. and {Adams}, Elena Y. and {Agrusa}, Harrison F. and {Abel}, Elisabeth D. and {Alford}, Amy L. and {Asphaug}, Erik I. and {Atchison}, Justin A. and {Badger}, Andrew R. and {Baki}, Paul and {Ballouz}, Ronald-L. and {Bekker}, Dmitriy L. and {Bellerose}, Julie and {Bhaskaran}, Shyam and {Buratti}, Bonnie J. and {Cambioni}, Saverio and {Chen}, Michelle H. and {Chesley}, Steven R. and {Chiu}, George and {Collins}, Gareth S. and {Cox}, Matthew W. and {DeCoster}, Mallory E. and {Ericksen}, Peter S. and {Espiritu}, Raymond C. and {Faber}, Alan S. and {Farnham}, Tony L. and {Ferrari}, Fabio and {Fletcher}, Zachary J. and {Gaskell}, Robert W. and {Graninger}, Dawn M. and {Haque}, Musad A. and {Harrington-Duff}, Patricia A. and {Hefter}, Sarah and {Herreros}, Isabel and {Hirabayashi}, Masatoshi and {Huang}, Philip M. and {Hsieh}, Syau-Yun W. and {Jacobson}, Seth A. and {Jenkins}, Stephen N. and {Jensenius}, Mark A. and {John}, Jeremy W. and {Jutzi}, Martin and {Kohout}, Tomas and {Krueger}, Timothy O. and {Laipert}, Frank E. and {Lopez}, Norberto R. and {Luther}, Robert and {Lucchetti}, Alice and {Mages}, Declan M. and {Marchi}, Simone and {Martin}, Anna C. and {McQuaide}, Maria E. and {Michel}, Patrick and {Moskovitz}, Nicholas A. and {Murphy}, Ian W. and {Murdoch}, Naomi and {Naidu}, Shantanu P. and {Nair}, Hari and {Nolan}, Michael C. and {Orm{\"o}}, Jens and {Pajola}, Maurizio and {Palmer}, Eric E. and {Peachey}, James M. and {Pravec}, Petr and {Raducan}, Sabina D. and {Ramesh}, K.~T. and {Ramirez}, Joshua R. and {Reynolds}, Edward L. and {Richman}, Joshua E. and {Robin}, Colas Q. and {Rodriguez}, Luis M. and {Roufberg}, Lew M. and {Rush}, Brian P. and {Sawyer}, Carolyn A. and {Scheeres}, Daniel J. and {Scheirich}, Petr and {Schwartz}, Stephen R. and {Shannon}, Matthew P. and {Shapiro}, Brett N. and {Shearer}, Caitlin E. and {Smith}, Evan J. and {Steele}, R. Joshua and {Steckloff}, Jordan K. and {Stickle}, Angela M. and {Sunshine}, Jessica M. and {Superfin}, Emil A. and {Tarzi}, Zahi B. and {Thomas}, Cristina A. and {Thomas}, Justin R. and {Trigo-Rodr{\'\i}guez}, Josep M. and {Tropf}, B. Teresa and {Vaughan}, Andrew T. and {Velez}, Dianna and {Waller}, C. Dany and {Wilson}, Daniel S. and {Wortman}, Kristin A. and {Zhang}, Yun},
        title = "{Successful kinetic impact into an asteroid for planetary defence}",
      journal = {Nature},
         year = 2023,
        month = apr,
       volume = {616},
       number = {7957},
        pages = {443-447},
          doi = {10.1038/s41586-023-05810-5},
archivePrefix = {arXiv},
       eprint = {2303.02248},
 primaryClass = {astro-ph.EP},
       adsurl = {https://ui.adsabs.harvard.edu/abs/2023Natur.616..443D}
}

@ARTICLE{michel2022psj,
       author = {{Michel}, Patrick and {K{\"u}ppers}, Michael and {Bagatin}, Adriano Campo and {Carry}, Benoit and {Charnoz}, S{\'e}bastien and {de Leon}, Julia and {Fitzsimmons}, Alan and {Gordo}, Paulo and {Green}, Simon F. and {H{\'e}rique}, Alain and {Juzi}, Martin and {Karatekin}, {\"O}zg{\"u}r and {Kohout}, Tomas and {Lazzarin}, Monica and {Murdoch}, Naomi and {Okada}, Tatsuaki and {Palomba}, Ernesto and {Pravec}, Petr and {Snodgrass}, Colin and {Tortora}, Paolo and {Tsiganis}, Kleomenis and {Ulamec}, Stephan and {Vincent}, Jean-Baptiste and {W{\"u}nnemann}, Kai and {Zhang}, Yun and {Raducan}, Sabina D. and {Dotto}, Elisabetta and {Chabot}, Nancy and {Cheng}, Andy F. and {Rivkin}, Andy and {Barnouin}, Olivier and {Ernst}, Carolyn and {Stickle}, Angela and {Richardson}, Derek C. and {Thomas}, Cristina and {Arakawa}, Masahiko and {Miyamoto}, Hirdy and {Nakamura}, Akiko and {Sugita}, Seiji and {Yoshikawa}, Makoto and {Abell}, Paul and {Asphaug}, Erik and {Ballouz}, Ronald-Louis and {Bottke}, William F. and {Lauretta}, Dante S. and {Walsh}, Kevin J. and {Martino}, Paolo and {Carnelli}, Ian},
        title = "{The ESA Hera Mission: Detailed Characterization of the DART Impact Outcome and of the Binary Asteroid (65803) Didymos}",
      journal = {The Planetary Science Journal},
         year = 2022,
        month = jul,
       volume = {3},
       number = {7},
          eid = {160},
        pages = {160},
          doi = {10.3847/PSJ/ac6f52},
       adsurl = {https://ui.adsabs.harvard.edu/abs/2022PSJ.....3..160M}
}

@ARTICLE{2006Sci...312.1341S,
       author = {{Saito}, J. and {Miyamoto}, H. and {Nakamura}, R. and {Ishiguro}, M. and {Michikami}, T. and {Nakamura}, A.~M. and {Demura}, H. and {Sasaki}, S. and {Hirata}, N. and {Honda}, C. and {Yamamoto}, A. and {Yokota}, Y. and {Fuse}, T. and {Yoshida}, F. and {Tholen}, D.~J. and {Gaskell}, R.~W. and {Hashimoto}, T. and {Kubota}, T. and {Higuchi}, Y. and {Nakamura}, T. and {Smith}, P. and {Hiraoka}, K. and {Honda}, T. and {Kobayashi}, S. and {Furuya}, M. and {Matsumoto}, N. and {Nemoto}, E. and {Yukishita}, A. and {Kitazato}, K. and {Dermawan}, B. and {Sogame}, A. and {Terazono}, J. and {Shinohara}, C. and {Akiyama}, H.},
        title = "{Detailed Images of Asteroid 25143 Itokawa from Hayabusa}",
      journal = {Science},
         year = 2006,
        month = jun,
       volume = {312},
       number = {5778},
        pages = {1341-1344},
          doi = {10.1126/science.1125722},
       adsurl = {https://ui.adsabs.harvard.edu/abs/2006Sci...312.1341S}
}

@ARTICLE{2019Sci...364..268W,
       author = {{Watanabe}, S. and {Hirabayashi}, M. and {Hirata}, N. and {Hirata}, Na. and {Noguchi}, R. and {Shimaki}, Y. and {Ikeda}, H. and {Tatsumi}, E. and {Yoshikawa}, M. and {Kikuchi}, S. and {Yabuta}, H. and {Nakamura}, T. and {Tachibana}, S. and {Ishihara}, Y. and {Morota}, T. and {Kitazato}, K. and {Sakatani}, N. and {Matsumoto}, K. and {Wada}, K. and {Senshu}, H. and {Honda}, C. and {Michikami}, T. and {Takeuchi}, H. and {Kouyama}, T. and {Honda}, R. and {Kameda}, S. and {Fuse}, T. and {Miyamoto}, H. and {Komatsu}, G. and {Sugita}, S. and {Okada}, T. and {Namiki}, N. and {Arakawa}, M. and {Ishiguro}, M. and {Abe}, M. and {Gaskell}, R. and {Palmer}, E. and {Barnouin}, O.~S. and {Michel}, P. and {French}, A.~S. and {McMahon}, J.~W. and {Scheeres}, D.~J. and {Abell}, P.~A. and {Yamamoto}, Y. and {Tanaka}, S. and {Shirai}, K. and {Matsuoka}, M. and {Yamada}, M. and {Yokota}, Y. and {Suzuki}, H. and {Yoshioka}, K. and {Cho}, Y. and {Tanaka}, S. and {Nishikawa}, N. and {Sugiyama}, T. and {Kikuchi}, H. and {Hemmi}, R. and {Yamaguchi}, T. and {Ogawa}, N. and {Ono}, G. and {Mimasu}, Y. and {Yoshikawa}, K. and {Takahashi}, T. and {Takei}, Y. and {Fujii}, A. and {Hirose}, C. and {Iwata}, T. and {Hayakawa}, M. and {Hosoda}, S. and {Mori}, O. and {Sawada}, H. and {Shimada}, T. and {Soldini}, S. and {Yano}, H. and {Tsukizaki}, R. and {Ozaki}, M. and {Iijima}, Y. and {Ogawa}, K. and {Fujimoto}, M. and {Ho}, T.-M. and {Moussi}, A. and {Jaumann}, R. and {Bibring}, J.-P. and {Krause}, C. and {Terui}, F. and {Saiki}, T. and {Nakazawa}, S. and {Tsuda}, Y.},
        title = "{Hayabusa2 arrives at the carbonaceous asteroid 162173 Ryugu{\textemdash}A spinning top-shaped rubble pile}",
      journal = {Science},
         year = 2019,
        month = apr,
       volume = {364},
       number = {6437},
        pages = {268-272},
          doi = {10.1126/science.aav8032},
       adsurl = {https://ui.adsabs.harvard.edu/abs/2019Sci...364..268W}
}

@article{Michel2025,
  author = {Michel, P. and K{\"u}ppers, M. and Fitzsimmons, A. and others},
  title = {The Hera Space Mission in the Context of Small Near-Earth Asteroid Missions in the Past, Present and Future},
  journal = {Space Science Reviews},
  volume = {221},
  pages = {70},
  year = {2025},
  doi = {10.1007/s11214-025-01195-1}
}

@article{Russell2007Dawn,
  author       = {Russell, C. T. and Capaccioni, F. and Coradini, A. and others},
  title        = {Dawn Mission to Vesta and Ceres},
  journal      = {Earth, Moon, and Planets},
  year         = {2007},
  volume       = {101},
  number       = {1-2},
  pages        = {65--91},
  doi          = {10.1007/s11038-007-9151-9}
}

@article{Mainzer2017,
author = {Mainzer, A.},
title = {The future of planetary defense},
journal = {Journal of Geophysical Research: Planets},
volume = {122},
number = {4},
pages = {789-793},
doi = {https://doi.org/10.1002/2017JE005318},
url = {https://agupubs.onlinelibrary.wiley.com/doi/abs/10.1002/2017JE005318},
eprint = {https://agupubs.onlinelibrary.wiley.com/doi/pdf/10.1002/2017JE005318},
year = {2017}
}

@article{Baum2018AsteroidRisk,
  author       = {Baum, Seth D.},
  title        = {Uncertain Human Consequences in Asteroid Risk Analysis and the Global Catastrophe Threshold},
  journal      = {Natural Hazards},
  year         = {2018},
  volume       = {94},
  number       = {2},
  pages        = {759--775},
  doi          = {10.1007/s11069-018-3419-4},
  note         = {Available at SSRN: https://ssrn.com/abstract=3218342}
}

@ARTICLE{2022PSJ.....3..158A,
       author = {{Agrusa}, Harrison F. and {Ferrari}, Fabio and {Zhang}, Yun and {Richardson}, Derek C. and {Michel}, Patrick},
        title = "{Dynamical Evolution of the Didymos-Dimorphos Binary Asteroid as Rubble Piles following the DART Impact}",
      journal = {The Planetary Science Journal},
         year = 2022,
        month = jul,
       volume = {3},
       number = {7},
          eid = {158},
        pages = {158},
          doi = {10.3847/PSJ/ac76c1},
archivePrefix = {arXiv},
       eprint = {2207.06995},
 primaryClass = {astro-ph.EP},
       adsurl = {https://ui.adsabs.harvard.edu/abs/2022PSJ.....3..158A}
}

@article{Richardson2024DidymosDynamicalState,
  author       = {Richardson, Derek C. and Agrusa, Harrison F. and Barbee, Brent and Cueva, Rachel H. and Ferrari, Fabio and Jacobson, Seth A. and Makadia, Rahil and Meyer, Alex J. and Michel, Patrick and Nakano, Ryota},
  title        = {The Dynamical State of the Didymos System before and after the {DART} Impact},
  journal      = {The Planetary Science Journal},
  year         = {2024},
  volume       = {5},
  number       = {8},
  pages        = {182},
  doi          = {10.3847/PSJ/ad62f5}
}

@article{Popescu2025hyperscout,
  author = {Popescu, M. M. and de Le{\'o}n, J. and Prodan, G. P. and others},
  title = {HyperScout-H: The Hyperspectral Imager for the ESA Hera Mission},
  journal = {Space Science Reviews},
  volume = {221},
  pages = {112},
  year = {2025},
  doi = {10.1007/s11214-025-01237-8}
}

@ARTICLE{2001AJ....122.2118B,
       author = {{Bohlin}, R.~C. and {Dickinson}, M.~E. and {Calzetti}, D.},
        title = "{Spectrophotometric Standards from the Far-Ultraviolet to the Near-Infrared: STIS and NICMOS Fluxes}",
      journal = {The Astronomical Journal},
         year = 2001,
        month = oct,
       volume = {122},
       number = {4},
        pages = {2118-2128},
          doi = {10.1086/323137},
       adsurl = {https://ui.adsabs.harvard.edu/abs/2001AJ....122.2118B}
}

@ARTICLE{1993yCat.6039....0K,
       author = {{Kurucz}, R.~L.},
        title = "{VizieR Online Data Catalog: Model Atmospheres (Kurucz, 1979)}",
 journal = {VizieR On-line Data Catalog},
         year = 1993,
        month = oct,
          eid = {VI/39},
       adsurl = {https://ui.adsabs.harvard.edu/abs/1993yCat.6039....0K}
}

@article{verissimostructural,
author = {Veríssimo, Gonçalo and Graça, Guilherme and Alves, Matias and Miguéns, Gonçalo},
year = {2026},
month = {01},
pages = {},
title = {Structural Analysis of LISAT-01 CubeSat},
doi = {10.13140/RG.2.2.11796.36481}
}

@article{Takashashi:95,
author = {Haruo Takashashi},
journal = {Appl. Opt.},
number = {4},
pages = {667--675},
publisher = {Optica Publishing Group},
title = {Temperature stability of thin-film narrow-bandpass filters produced by ion-assisted deposition},
volume = {34},
month = {Feb},
year = {1995},
url = {https://opg.optica.org/ao/abstract.cfm?URI=ao-34-4-667},
doi = {10.1364/AO.34.000667},
}

@inproceedings{Gilmore2002SpacecraftTC,
  title={Spacecraft Thermal Control Handbook, Volume I: Fundamental Technologies},
  author={Gilmore, David},
  year={2002},
  url={https://api.semanticscholar.org/CorpusID:125023491}
}

@techreport{osti_5671682,
  author       = {Berk, A and Bernstein, L S and Robertson, D C},
  title        = {MODTRAN: a moderate resolution model for LOWTRAN. Technical report, 12 May 1986-11 May 1987},
  institution  = {Spectral Sciences, Inc., Burlington, MA (USA)},
  url          = {https://www.osti.gov/biblio/5671682},
  place        = {United States},
  year         = {1987},
  month        = {07}}

@ARTICLE{2004SoEn...76..423G,
       author = {{Gueymard}, Christian A.},
        title = "{The sun's total and spectral irradiance for solar energy applications and solar radiation models}",
      journal = {Solar Energy},
         year = 2004,
        month = apr,
       volume = {76},
       number = {4},
        pages = {423-453},
          doi = {10.1016/j.solener.2003.08.039},
       adsurl = {https://ui.adsabs.harvard.edu/abs/2004SoEn...76..423G}
}

@inproceedings{Mikkonen2014VerificationOC,
  title={Verification of camera module conducted immunity},
  author={Jussi Ville Mikkonen},
  year={2014},
  url={https://api.semanticscholar.org/CorpusID:114335724}
}

@article{Shao2023TheIO,
  title={The Impact of Bias Row Noise to Photometric Accuracy: Case Study Based on a Scientific CMOS Detector},
  author={Li Shao and Hu Zhan and Chao Liu and Haonan Chi and Qiuyan Luo and Huaipu Mu and Wenzhong Shi},
  journal={Research in Astronomy and Astrophysics},
  year={2023},
  volume={24},
  url={https://api.semanticscholar.org/CorpusID:266435571}
}

@article{Ono1996,
  author = {Ono, A. and Sakuma, F. and Arai, K. and Yamaguchi, Y. and Fujisada, H. and Slater, P. N. and Thome, K. J. and Palluconi, F. D. and Kieffer, H. H.},
  title = {Preflight and In-Flight Calibration Plan for ASTER},
  journal = {Journal of Atmospheric and Oceanic Technology},
  volume = {13},
  number = {2},
  pages = {321--335},
  year = {1996},
  doi = {10.1175/1520-0426(1996)013<0321:PAIFCP>2.0.CO;2}
}

@article{BenMoussa_2013,
   title={On-Orbit Degradation of Solar Instruments},
   volume={288},
   ISSN={1573-093X},
   url={http://dx.doi.org/10.1007/s11207-013-0290-z},
   DOI={10.1007/s11207-013-0290-z},
   number={1},
   journal={Solar Physics},
   publisher={Springer Science and Business Media LLC},
   author={BenMoussa, A. and Gissot, S. and Schühle, U. and Del Zanna, G. and Auchère, F. and Mekaoui, S. and Jones, A. R. and Walton, D. and Eyles, C. J. and Thuillier, G. and Seaton, D. and Dammasch, I. E. and Cessateur, G. and Meftah, M. and Andretta, V. and Berghmans, D. and Bewsher, D. and Bolsée, D. and Bradley, L. and Brown, D. S. and Chamberlin, P. C. and Dewitte, S. and Didkovsky, L. V. and Dominique, M. and Eparvier, F. G. and Foujols, T. and Gillotay, D. and Giordanengo, B. and Halain, J. P. and Hock, R. A. and Irbah, A. and Jeppesen, C. and Judge, D. L. and Kretzschmar, M. and McMullin, D. R. and Nicula, B. and Schmutz, W. and Ucker, G. and Wieman, S. and Woodraska, D. and Woods, T. N.},
   year={2013},
   month=apr, pages={389–434} }

@article{SHI20243993,
title = {Vibration isolation methods in spacecraft: A review of current techniques},
journal = {Advances in Space Research},
volume = {73},
number = {8},
pages = {3993-4023},
year = {2024},
issn = {0273-1177},
doi = {https://doi.org/10.1016/j.asr.2024.01.020},
url = {https://www.sciencedirect.com/science/article/pii/S0273117724000462},
author = {H.T Shi and Musa Abubakar and X.T. Bai and Zhong Luo}
}

@inproceedings{Esposito2019,
  author       = {Esposito, Marco and Zuccaro Marchi, Alessandro},
  title        = {In-orbit demonstration of the first hyperspectral imager for nanosatellites},
  booktitle    = {Proceedings of SPIE — International Conference on Space Optics \& Applications (ICSO)},
  year         = {2019},
  doi          = {10.1117/12.2535991},
  url          = {https://doi.org/10.1117/12.2535991},
  publisher    = {SPIE}
}

@misc{ESA_HERA_Operational_SPICE_2024,
  author       = {{European Space Agency}},
  title        = {Hera Operational SPICE Kernel Dataset},
  year         = {2024},
  howpublished = {Data set},
  version      = {1.0},
  publisher    = {European Space Agency},
  doi          = {10.57780/esa-k25x2cv},
  url          = {https://doi.org/10.57780/esa-k25x2cv}
}

@article{Acton1996NAIF,
  author  = {Acton, Charles H.},
  title   = {Ancillary Data Services of NASA's Navigation and Ancillary Information Facility},
  journal = {Planetary and Space Science},
  volume  = {44},
  number  = {1},
  pages   = {65--70},
  year    = {1996},
  doi     = {10.1016/0032-0633(95)00107-7}
}

@article{sturm1992silicon,
  title={Silicon temperature measurement by infrared absorption: fundamental processes and doping effects},
  author={Sturm, James C. and Reaves, C. M.},
  journal={IEEE Transactions on Electron Devices},
  volume={39},
  number={1},
  pages={81--88},
  year={1992},
  publisher={IEEE}
}

@article{nguyen2014temperature,
  title={Temperature dependence of the band-band absorption coefficient in crystalline silicon from photoluminescence},
  author={Nguyen, Hieu T. and Macdonald, Daniel and Rougieux, Fiacre E.},
  journal={Journal of Applied Physics},
  volume={115},
  number={4},
  pages={043710},
  year={2014},
  publisher={American Institute of Physics}
}

@inproceedings{velichko2019intrinsic,
  title={Intrinsic {Si} Quantum Efficiency, Sensitivity, and Other Parameters Temperature Dependence for {BSI} Image Sensors},
  author={Velichko, Sergey and Gravelle, Bob and Tekleab, Daniel and Guidash, Michael and Johnson, Scott and Oh, Minseok and Chang, Hung Chih},
  booktitle={2019 International Image Sensor Workshop (IISW)},
  month={June},
  year={2019},
  address={Snowbird, UT, USA},
  publisher={International Image Sensor Society},
  doi={10.60928/hrhu-uj3d},
  url={https://doi.org/10.60928/hrhu-uj3d}
}

@inproceedings{10.1117/12.476792,
author = {Stephan Maestre and Pierre Magnan and Francis Lavernhe and Franck Corbiere},
title = {{Hot carriers effects and electroluminescence in the CMOS photodiode active pixel sensors}},
volume = {5017},
booktitle = {Sensors and Camera Systems for Scientific, Industrial, and Digital Photography Applications IV},
editor = {Nitin Sampat and Ricardo J. Motta and Morley M. Blouke and Nitin Sampat and Ricardo J. Motta},
organization = {International Society for Optics and Photonics},
publisher = {SPIE},
pages = {59 -- 67},
year = {2003},
doi = {10.1117/12.476792},
URL = {https://doi.org/10.1117/12.476792}
}

@book{10.1088/978-0-7503-1272-1,
author = {Simoen, Eddy and Claeys, Cor},
title = {Random Telegraph Signals in Semiconductor Devices},
publisher = {IOP Publishing},
year = {2016},
series = {2053-2563},
isbn = {978-0-7503-1272-1},
url = {https://doi.org/10.1088/978-0-7503-1272-1},
doi = {10.1088/978-0-7503-1272-1}
}

@misc{apergis2025highprecisionphotometryscientificcmos,
      title={High-Precision Photometry with a scientific CMOS Camera: I Lab Testing of the Marana camera}, 
      author={Ioannis Apergis and Daniel Bayliss and Leonidas Asimakoulas and Paul Chote and James McCormac and Morgan A. Mitchell and Sam Gill and Philip G. Steen and Peter Wheatley},
      year={2025},
      eprint={2510.14484},
      archivePrefix={arXiv},
      primaryClass={astro-ph.IM},
      url={https://arxiv.org/abs/2510.14484}, 
}

@article{Alarcon_2023,
doi = {10.1088/1538-3873/acd04a},
url = {https://doi.org/10.1088/1538-3873/acd04a},
year = {2023},
month = {may},
publisher = {The Astronomical Society of the Pacific},
volume = {135},
number = {1047},
pages = {055001},
author = {Alarcon, Miguel R. and Licandro, Javier and Serra-Ricart, Miquel and Joven, Enrique and Gaitan, Vicens and de Sousa, Rebeca},
title = {Scientific CMOS Sensors in Astronomy: IMX455 and IMX411},
journal = {Publications of the Astronomical Society of the Pacific}
}

\end{document}